\pdfpageattr {/Group << /S /Transparency /I true /CS /DeviceRGB>>}
\documentclass[aps,reprint,prb,superscriptaddress,longbibliography,floatfix]{revtex4-2}
\usepackage[version=3]{mhchem}
\usepackage{graphicx}
\usepackage{fancyref}
\usepackage{soul}
\usepackage{amssymb}
\usepackage[normalem]{ulem}
\usepackage{amsmath}
\usepackage{commath}
\usepackage{blindtext}
\usepackage{siunitx}
\usepackage{miller}
\usepackage{xcolor}
\usepackage{float}
\usepackage{lipsum}
\usepackage{bm}
\usepackage{ulem}
\usepackage{physics}
\makeatletter
\let\Hy@backout\@gobble
\makeatother
\usepackage{titlesec}
\setcitestyle{super}
\usepackage{subcaption}
\usepackage{ragged2e}
\DeclareCaptionJustification{justified}{\justifying}
\begin{document}
\title{Dynamics of 2D Material Membranes}
\author{Peter G. Steeneken}
    \email[Correspondence email address: ]{p.g.steeneken@tudelft.nl}
    \affiliation{Precision and Microsystems Engineering  Department, Delft University of Technology, Mekelweg 2, 2628 CD Delft, The Netherlands}
    \affiliation{Kavli Institute of Nanoscience, Delft University of Technology, P.O. Box 5046, 2600 GA Delft, The Netherlands}
\author{Robin J. Dolleman}
    \affiliation{Second institute of Physics, RWTH Aachen University, 52074, Aachen, Germany}
\author{Dejan Davidovikj}
\affiliation{Kavli Institute of Nanoscience, Delft University of Technology, P.O. Box 5046, 2600 GA Delft, The Netherlands}
\author{Farbod Alijani}
\affiliation{Precision and Microsystems Engineering  Department, Delft University of Technology, Mekelweg 2, 2628 CD Delft, The Netherlands}
\author{Herre S.J. van der Zant}
\affiliation{Kavli Institute of Nanoscience, Delft University of Technology, P.O. Box 5046, 2600 GA Delft, The Netherlands}
\email[Correspondence email address: ]{h.s.j.vanderzant@tudelft.nl}
\date{\today} 

\begin{abstract}
The dynamics of suspended two-dimensional (2D) materials has received increasing attention during the last decade, yielding new techniques to study and interpret the physics that governs the motion of atomically thin layers. This has led to insights into the role of thermodynamic and nonlinear effects as well as the mechanisms that govern dissipation and stiffness in these resonators. In this review, we present the current state-of-the-art in the experimental study of the dynamics of 2D membranes. The focus will be both on the experimental measurement techniques and on the interpretation of the physical phenomena exhibited by atomically thin membranes in the linear and nonlinear regimes. We will show that resonant 2D membranes have emerged both as sensitive probes of condensed matter physics in ultrathin layers, and as sensitive elements to monitor small external forces or other changes in the environment. New directions for utilizing suspended 2D membranes for material characterization, thermal transport, and gas interactions will be discussed and we conclude by outlining the challenges and opportunities in this upcoming field.
\end{abstract}


\maketitle
~
\newpage
~
\newpage

\tableofcontents

\section{Introduction}

The exfoliation of a single layer of graphite~\cite{novoselov2005two}, and the demonstration of the unique properties of graphene~\cite{novoselov2004electric,novoselov2005two,zhang2005experimental,nair2008fine}, marked the start of an era where atomically thin crystalline materials can be studied and used for next generation devices~\cite{geim2007rise,dhanabalan20172d,jayakumar20182d,gupta2015recent,macha20192d,lemme2020nanoelectromechanical}. Soon afterwards, this also led to the first mechanically movable structures of one atom thick graphene, which were shown to operate at high resonance frequencies in the MHz range~\cite{Bunch2007}. Since then, the research field, focusing on the study of the dynamics of atomically thin membranes, has grown steadily. Many different 2D materials have been explored~\cite{Castellanos2013,lee2013high}, the control over their mechanical actuation and detection has increased~\cite{Chen2009,DollemanCalibration,song2014graphene,Weber2014} and the understanding of the link between high-frequency mechanics, material properties and physical interactions 
has improved~\cite{Barton2012,DollemanThermal,Farbod2017,Banafsheh2017,inoue2017resonance,Dolleman2016A}. 

\begin{figure*}[htb]
	\includegraphics[]{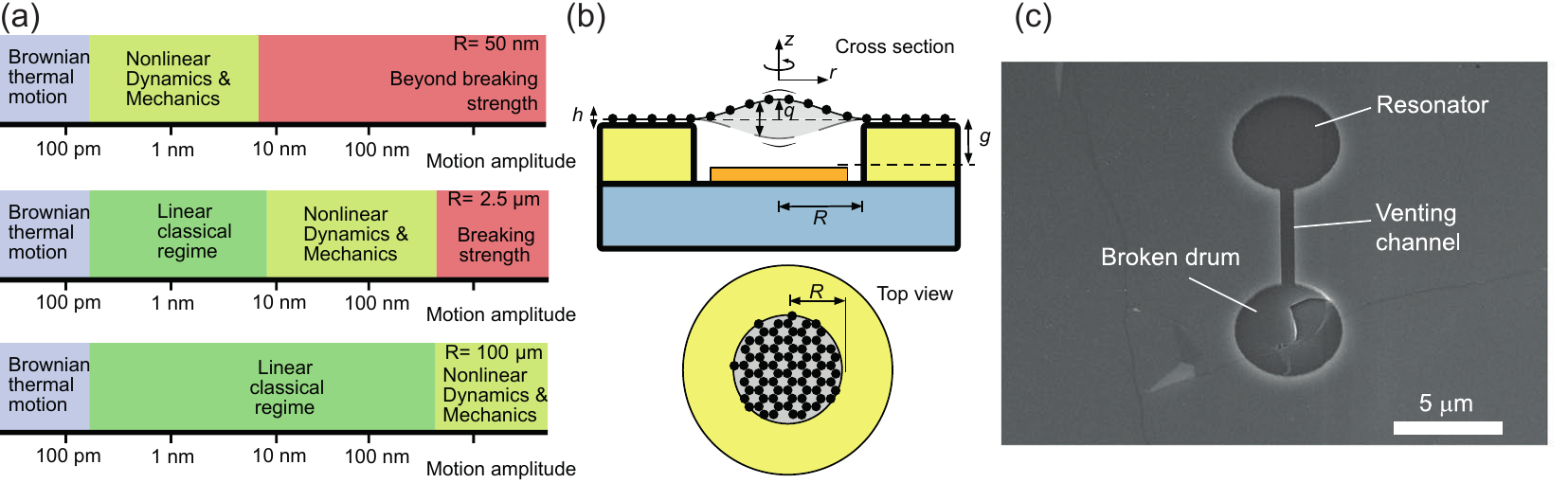}
	\caption{(a) Schematic indicating ranges of dynamic motion exhibited by circular graphene membranes of different radii $R$ at room temperature. The $x$-axis indicates the membrane's maximum out-of-plane amplitude of the fundamental resonance mode. Large membranes with a radius of 100~$\mu$m exhibit linear motion over a large range of out-of-plane amplitudes. When the membrane radius reduces to 2.5~$\mu$m, the linear region shrinks and the membrane exhibits nonlinear response at amplitudes of less than 10~nm. For very small membranes with $R<50$~nm, the linear regime disappears completely, and the Brownian thermal forces will drive the drum into the nonlinear regime. Details of the determination of the ranges can be found in appendix~\ref{AppendixA}. (b) Schematic of a circular graphene membrane resonating in its fundamental mode with center displacement $q(t)$. The cross section in the top panel indicates the thickness $h$, membrane radius $R$ and gap distance $g$. The bottom panel shows the suspended part of the graphene in a top view sketch. (c) Scanning electron microscope image of a circular single-layer graphene resonator. Reprinted from Ref.~\onlinecite{Dolleman2018} licensed under CC BY 4.0.	\label{fig:introduction}}
\end{figure*}

The atomic thickness and high aspect-ratio of suspended membranes of 2D materials result in large differences between their mechanical response in the in-plane and out-of-plane directions. They are extremely flexible out-of-plane as a consequence of their small thickness, yet very stiff within the plane due to their high Young's modulus~\cite{Lee2008}. The ultra-thin nature of 2D membranes thus brings unique mechanical features that are not easily attainable in their macroscopic counterparts. 
First of all, their flexibility results in low and tunable stiffness, making them highly sensitive to forces~\cite{Weber2016}. As a result, already at low actuation forces, the nonlinear regime is reached~\cite{Eichler2011,Castellanos2013}. This makes 2D material membranes excellent probes for studying a variety of nonlinear dynamic effects, including mode-coupling~\cite{Samanta2015,liu2015optical,DeAlba2016, schwarz2016deviation}, high resonance-frequency-tunability~\cite{Chen2009,Chen2013}, parametric~\cite{Prasad2017,Dolleman2018} and internal resonances~\cite{guttinger2017energy,kecskekler2020enhanced}.  Second, their mass is extremely small, which increases resonance frequencies yielding high sensitivity in sensing applications~\cite{sakhaee2008applications,atalaya2010nanomechanical,Dolleman2016A,dolleman2019high,lemme2020nanoelectromechanical}.
 Third, their high surface-to-volume-ratio makes them very sensitive to their environment. For example, their membrane dynamics is highly responsive to gases in the environment~\cite{Dolleman2016A,lee2014air,roslon2020graphene} and to thermal fluctuations~\cite{dolleman2019high}. Finally, the large stiffness difference between the in-plane and out-of-plane directions, results in unique properties via out-of-plane wrinkles and ripples~\cite{Fasolino2007,ahmadpoor2017perspective,los2017mechanics,ackerman2016anomalous} and nonequilibrium thermodynamics of flexural and in-plane phonons~\cite{Dolleman2020A}. This interplay between thermal properties and out-of-plane mechanical motion is particularly strong, such that at room temperature the thermal 'Brownian' forces in the undriven regime lead to significant motion amplitudes of the order of the thickness~\cite{Davidovikj2016}. Figure~\ref{fig:introduction}(a) illustrates the different regimes of motion for a circular graphene drum, highlighting the increasing importance of Brownian and nonlinear dynamics when scaling down membrane radius. In fact, it shows that for graphene membrane radii below $R=50$~nm, the linear regime disappears, and thermal fluctuations at room temperature drive the membrane motion into the nonlinear regime as discussed in appendix~\ref{AppendixA}.

In this review we will discuss and describe, from an experimental point of view, the progress that has been made in the study of the physics and dynamics of 2D material resonators, with a particular emphasis on graphene as a model system. We will provide insight into both the underlying concepts and the measurement techniques, which build on know-how from the fields of micro and nanoelectromechanical systems (MEMS and NEMS)\cite{Senturia2001, Schmid2016}, and which are crucial for detecting motion at high-frequencies and small displacements down to the picometer regime. We will not focus on motion in the quantum regime~\cite{poot2012mechanical}, Raman phonon excitations at THz frequencies~\cite{Paillet2018}, static mechanical properties~\cite{Castellanos2015,Akinwande2017} of 2D materials nor on applications~\cite{lemme2020nanoelectromechanical,hu2020resonant}, for which we refer the reader to the provided references. The review aims both at giving an introduction to new researchers in the field, providing them with relevant information and references on common methodologies, as well as providing an overview for experts active in this emerging field, by including recent developments and new research directions.

To understand the dynamics of 2D membranes, we start in section~\ref{sec:eqom} by introducing the equations of motion as central reference for describing the forces and motion. Then in the subsequent sections, techniques for detecting the motion of 2D membranes are discussed (Sec.~\ref{readout}), and the different types of actuation methods for driving the membranes in motion are summarized (Sec.~\ref{sec:actuation}). 
Next, the solutions of the equation of motion are outlined, in both the linear and the nonlinear regime (Sec.~\ref{sec:linreg} and \ref{sec:nonlsoln}). For each of these regimes the types of solutions that can occur are discussed, followed by a subsection where the different terms in the equation of motion are related to the underlying physics, both from a theoretical and an experimental point of view. We continue with outlining the emerging research direction that uses the link between dynamics and physics of 2D materials to quantify physical and material related parameters (Sec.~\ref{sec:physlin} and \ref{sec:orignonl}).
In section~\ref{sec:physint} this concept - linking dynamics to underlying physics - is taken a step further. It deals with the use of dynamics of 2D resonators for the study of respectively electromagnetic order, external forces, gas flows and thermodynamics. Finally, we discuss and conclude with some open research questions and future directions in this exciting field.

\subsection{General equations of motion} 
\label{sec:eqom}
The dynamics of 2D material membranes is governed by their equations of motion (EOM). In general, the motion of a flat ultrathin membrane (Fig.~\ref{fig:introduction}(b)) can be described by a time-dependent displacement vector field $\mathbf{r}(x,y,t)$, where for small-amplitude out-of-plane motion, the in-plane motion can be neglected 
so that only the out-of-plane displacement function $w(x,y,t)$ in the $z$-direction is of importance. The motion can be expanded in terms of the linear eigenmodes $\phi_i(x,y)$ of the membrane $w(x,y,t)=\sum_i^N q_i(t) \phi_i(x,y)$, where $q_i(t)$ are defined as the $N$ time dependent generalized coordinates and $i$ is the mode number as can be derived from classical mechanics~\cite{geradin2014mechanical,Steeneken2007comsol}. We will choose to normalize the eigenmodes $\phi_i(x,y)$ to have a maximum absolute value of 1, such that $q_i(t)$ represents the maximum deflection of mode $i$. In the linear free vibration case, the motion has a sinusoidal time dependence, that is $q_i(t) = \Re q_{i,0} e^{i\omega_i t}$, where $\omega_i$ is the mode's angular eigenfrequency and $|q_{i,0}|$ the amplitude. The total motion of the membrane is therefore a superposition of the different eigenmodes, where the generalized coordinates $q_{i}(t)$ describe the motion of the points of maximum deflection for each of the modes. For convenience we order the coefficients $i$ on ascending eigenfrequency, such that $i=1$ corresponds to the fundamental mode of the membrane. 

This eigenmode decomposition allows obtaining a set of $N$ coupled EOMs, in terms of the $N$ generalized coordinates $q_i$ as follows~\cite{amabili2008nonlinear}:  
\begin{equation}
\begin{aligned}
\label{eq:motion}
    m_1 \ddot{q_1} + c_1 \dot{q_1} + k_1 q_1+  &F_{nl,1}( q_j, \dot{q_j}, \dots) \\ 
    =  &F_{\rm ext,1}(q_j, \dot{q}_j,\dots, t)-k_{{\rm p},1}(t) q_1 , \\
    \dots & \\
     m_i \ddot{q_i} + c_i \dot{q_i} + k_i q_i+ &F_{nl,i}(q_j,\dot{q_j}, \dots) & \\
     = &F_{\rm ext,i}(q_j,  \dot{q}_j, \dots, t)  -k_{{\rm p},i}(t) q_i , \\
       \mathrm{for~all~} & i,j= 1 \dots N.
\end{aligned}     
\end{equation}
\noindent
In these equations, the terms $m_i$, $c_i$ and $k_i$ describe the mode-dependent linear modal mass, damping coefficient and linear stiffness, respectively. All nonlinear membrane forces that are intrinsic to the membrane itself, e.g. due to material and geometric nonlinearities, are described by the term $F_{nl,i}$. On the right side of the EOM there is the external forcing term $F_{\rm ext,i}(q_i,\dot{q}_i,t)$, which captures the externally applied forces on the membrane, that can depend on time, position and membrane speed, and which can, as we will see later on, also introduce nonlinear effects. Finally there are the parametric terms $k_{{\rm p},i}(t) q_i$, which might be also categorized as part of $F_{\rm ext,i}$, but are specified separately to emphasize their significance.
We emphasize that Eq.~(\ref{eq:motion}) is constructed such that all force terms that are intrinsic to the mechanical resonator itself are on the left side of the equal sign and all other terms are on the right side, even though this separation is not always easily made, for instance when the material properties or membrane tension are modulated externally. Each of the following sections will focus on specific terms in these equations of motion, discussing both their physical origins and their effect on the membrane dynamics.

\color{black}

\begin{figure*}[htb]
	\centering
	\includegraphics[]{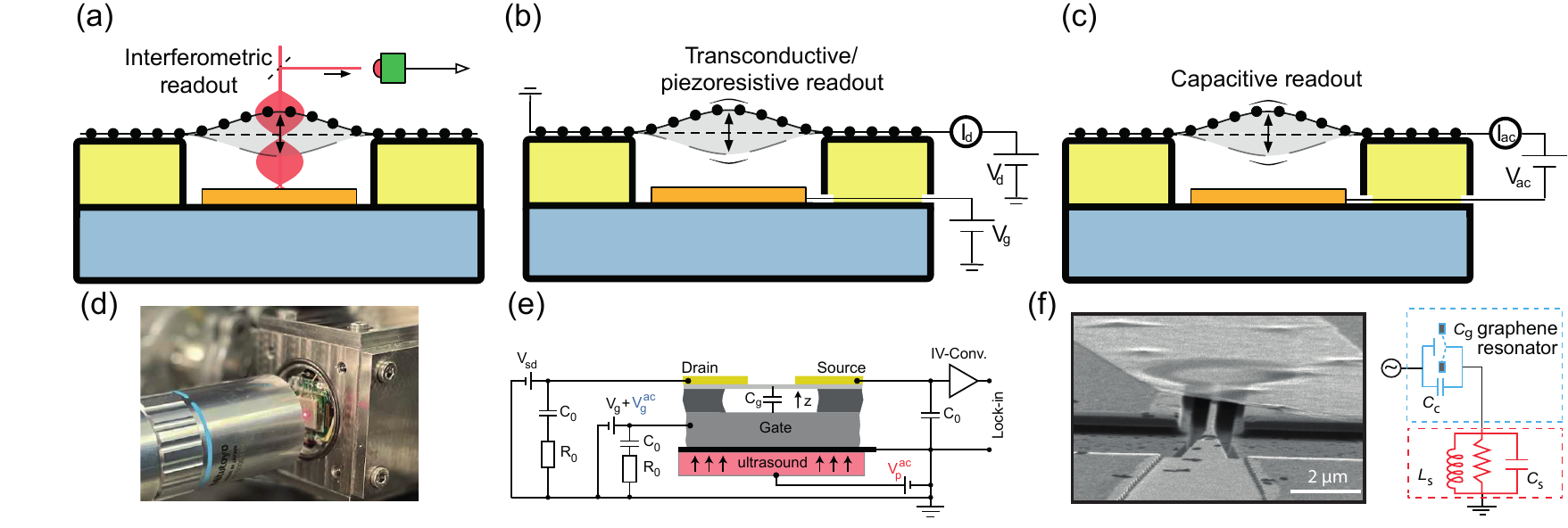}
	\caption{Readout methods. (a) Interferometric readout: the intensity of reflected laser light is modulated by the motion of the membrane. (b) Transconductive/piezoresistive readout: the resistance of the membrane depends on deflection. (c) Capacitive readout: the capacitance between the membrane and the gate electrode depends on membrane position. (d) Photograph of an experimental setup using interferometric readout. The sample is mounted behind a window in a vacuum chamber; the laser is focused on the sample through the objective in front. (e) Image of a graphene resonator suspended between a source (S) and drain (D) above a local gate (LG) whose motion is read out using transconductance. Reprinted with permission from Ref.~\onlinecite{Verbiest2018}, copyright (2018) American Chemical Society. (f) False-coloured electron microscope image of a few-layer graphene resonator suspended over a gate electrode. The gate electrode is coupled to a microwave cavity, allowing motion detection through the capacitance. Reprinted with permission from Ref.~\onlinecite{singh2016negative}, copyright (2016) American Physical Society.  
	}
	\label{fig:readout}
\end{figure*}

\section{Readout methods}
\label{readout}

For studying the dynamics of 2D materials, readout methods for measuring motion $w(x,y,t)$ and actuation methods for driving the membrane, via the terms $F_{\rm ext,i}(t)$ and $k_{{\rm p},i}(t)$) in the EOM, are essential. Due to the high frequencies, small amplitudes and small size of 2D material resonators, accurate readout is challenging. Moreover, several conventional mechanical engineering actuation and detection methods like modal hammers and accelerometers~\cite{ewins2009modal} are too large or invasive to apply. This has rendered contactless optical and electronic readout and actuation techniques to be most effective; notable exceptions are atomic force microscope (AFM) based detection of the dynamic motion of graphene membranes~\cite{Garcia2008} and base excitation methods (section~\ref{base}). 

In this section~\ref{readout} we will discuss the most important dynamic readout and detection methods for 2D materials. These methods convert the position or velocity of the membrane into an electrical signal that is subsequently analyzed by measurement equipment such as network, lock-in and/or spectrum analyzers. Mostly,the out-of-plane motion of the membrane $w(x,y,t)$ is measured, since the in-plane dynamic motion is usually much smaller and more difficult to detect, although techniques like Raman~\cite{zhang2020dynamicallyenhanced} and piezoresistive~\cite{Smith2013} readout are able to probe it. Figure~\ref{fig:readout} lists the three main readout methods for studying the dynamics of 2D material membranes that will be discussed in the following subsections: optical, transconductive and capacitive readout.

\subsection{Optical readout}
\label{optreadout}
 The first studies of the dynamics of 2D materials were performed using the interferometric technique~\cite{Bunch2007} shown in Figs.~\ref{fig:readout}(a) and (d), where a laser is reflected from the Fabry-Perot cavity formed by the semi-transparent 2D material and the underlying reflective substrate.
In the presence of a reflecting substrate, a standing wave electric field intensity $I(z)$ is obtained from the superposition of the incoming and reflected optical wave as illustrated by the red sinusoidal waves in Fig.~\ref{fig:readout}(a). When the graphene moves through this standing wave, it absorbs light proportional to $I(z)$ and thus modulates the reflected light beam. In addition, modulation also arises from interference between the light reflected from the graphene and from the substrate, but since the reflectivity of graphene is very low, this contribution is relatively small. The highest motion sensitivity is achieved when the membrane resides at a distance from the substrate with maximum slope in the optical field intensity $\frac{{\rm d}I(z)}{{\rm d}z}$, which can be calculated using standard techniques if optical material properties and geometry are known~\cite{Blake2007, Davidovikj2016}. 

Besides this type of interferometric readout, other optical detection techniques have been developed, such as the recent demonstration of a Michelson interferometer setup~\cite{Singh2018}, which has the advantage that it requires just a free-hanging membrane and not a reflective substrate behind it, and that the distance between the different paths in this Michelson set-up can be adjusted for calibration purposes. On the other hand, a drawback is that this technique is more sensitive for relative vibrations between the arms and requires more careful alignment. Furthermore, a balanced homodyne technique has been demonstrated to probe the phase fluctuations of the light reflected from a graphene membrane~\cite{schwarz2016deviation}. Laser Doppler Vibrometry (LDV) has also been used for  characterization of graphene membrane dynamics using optical interferometry~\cite{zhou2013electrostatic,grady2014low,Akbari2020}. Another interesting development is the use of Raman spectroscopy to determine the dynamically induced strain in the membrane, allowing one to obtain information on the in-plane strain, in addition to the out-of-plane motion~\cite{zhang2020dynamicallyenhanced}. 

\subsection{Transconductive readout}
\label{transreadout}
Transconductive mechanical readout methods detect motion via changes in the electrical resistance or conductance of the suspended 2D material. In this section we discuss both conduction variations due to a motion-induced change in the electric gate field~\cite{Nathanson1967} that causes changes in carrier density, as well as strain-induced changes in the resistivity via the piezoresistive material properties of the membrane.
In a transconductive readout scheme, shown in Figs. \ref{fig:readout}(b) and (e), a constant current $I_d$ runs through the 2D material. When the resistance of the 2D material is displacement-dependent $R(q_i)$, and the membrane moves, the voltage across the material is modulated as $V_d=R(q_i) I_d\approx (R_0 + \frac{{\rm d}R}{{\rm d}q_i} q_i)I_d$. The position dependence of the resistance $\frac{{\rm d}R}{{\rm d}q_i}$ can be the result of the semiconducting nature of the material in a field-effect transistor geometry, where for a constant voltage $V_g$ on a bottom gate-electrode, the motion of the membrane causes a variation of the electric field 
that causes a variation in the charge carrier density in the semiconducting or semi-metallic 2D membrane, thereby changing its resistance~\cite{Chen2009, Verbiest2018}. 
A second effect that can cause the resistance to change is the piezoresistive effect. When the membrane moves out of the flat equilibrium position, its strain and lattice spacing increases, which causes the resistance of the material to change~\cite{Smith2013, fan2019graphene, manzeli2019self}. Piezoresistance can either be caused by strain-induced geometrical changes in the conductor, or by changes in the material's band-structure that cause the charge carriers to move to bands with different carrier mobilities. 

\subsection{Capacitive readout}
\label{capreadout}
Another form of electrical readout is the capacitive method (Fig.~\ref{fig:readout}(c)). In this configuration an $ac$ current is driven through the capacitor formed by the suspended membrane and the gate electrode, yielding a time-dependent gate capacitance, $C_g(t)=\int_A \varepsilon_0{\rm d}A/(g+w(x,y,t))$, in the parallel plate configuration ($g \ll R$), where the capacitance is integrated over the membrane area $A$ and $\varepsilon_0$ is the permittivity of vacuum. When the membrane center displacement $q_i$ changes, this can be detected as a change in the impedance of the capacitor $Z_C=1/(i\omega C_g(q_i))=V_{\rm ac}/I_{\rm ac}$. This change, for a 1 nm displacement of a $5~\mu$m diameter membrane, is~\cite{Davidovikj2017} typically only 2~aF. Such small capacitance changes are challenging to detect at low frequencies, at which $Z_C$ is very high, and are therefore more conveniently captured at GHz range frequencies at which impedances are lower~\cite{Herfst2012}, see also the example in Fig.~\ref{fig:readout}(f).

The main advantage of capacitive readout compared to transconductive schemes is that the capacitance only depends on the membrane geometry and is to a large extent independent of the material properties or contamination on the membrane~\cite{Davidovikj2017}; a disadvantage is that parasitic capacitances, e.g. due to electrical interconnects, are generally much larger than the membrane capacitance changes themselves and complicate accurate readout. 
Nevertheless, capacitive readout has been successfully applied to measure slow deflections of a single graphene drum~\cite{Davidovikj2017}, large capacitive graphene sensor arrays~\cite{siskins2020sensitive, berger2017} and fast capacitance changes in MEMS devices~\cite{Herfst2012}.

\subsection{Mixing techniques}

Although in some works the high-frequency electrical signals from 2D material resonators described in sections~\ref{transreadout} and \ref{capreadout} have successfully been measured directly ~\cite{xu2010radio}, this can be challenging in practice, because the high-impedance of the sample causes the motional signal to be small, whereas the parasitic cross-talk from the driving voltage is large. Distinguishing the small motional signal on the large background parasitic signal of the same frequency is difficult~\cite{sazonova2006tunable,sazonova2004tunable2}. 
This problem can be mitigated using down-mixing schemes that convert the signal to another frequency that is far away from the parasitic cross-talk signal. To down-mix transconductive readout signals~\cite{knobel2003nanometre,bargatin2005sensitive,verbiest2016tunable,Chen2009}, the membrane conductance $G$ is modulated by the motion at a frequency $\omega_m$ and a modulated bias voltage $V_{\rm sd}$ at frequency $\omega_c$ is applied between the source and drain. The resistance modulation causes the current through the sample $I = V_{\rm sd}G$ to consist of the product of the two sinusoidal functions, which results in a low-frequency mixing term in the current at frequency $\Delta \omega=|\omega_c-\omega_m|$. In principle, $\Delta \omega$ can be arbitrarily low and the technique can even be applied in the {\it dc} domain, meaning that no high-frequency measurement equipment is needed to read out the signal~\cite{rosenblatt2005mixing}. Typically, values of $\Delta \omega$ are in the 0.1-10 kHz range to avoid low-frequency noise.  Several works have demonstrated this downmixing technique in 2D materials resonators~\cite{Verbiest2018,verbiest2020tunable}. A potential drawback of mixing techniques is that the sideband signal may cause cross-talk and may also actuate the drum~\cite{Chen2009}, in particular when $\Delta \omega < \frac{\omega_m}{2Q}$.

For capacitive radio-frequency (RF) readout of 2D membranes, a similar mixing technique can be used. Essentially, the scheme resembles techniques used in the cavity optomechanics community~\cite{bothner2020cavity}, since in both cases an electromagnetic (EM) wave is stored in an EM cavity resonance, whose EM resonance frequency $\omega_{\rm EM}$ is modulated by the motion of a mirror or capacitor plate. When an RF input signal with frequency $\omega_{c}$ close to $\omega_{\rm EM}$ is sent into this optical filter, it will be amplitude-modulated by the movement of the mirror at frequency $\omega_m$, resulting in mixed output signals at $\omega_{c} \pm \omega_m$. For a sufficiently high electromagnetic wave intensity, the mirror will also be actuated by the radiation pressure forces of the optical field, leading to optomechanical couplings that are essential in the field of quantum optomechanics. 
This approach has been successfully carried out with 2D material membranes, albeit at low temperatures using zero-loss superconducting transmission lines~\cite{Singh2014, Weber2014} and side-band resolved detection. Due to the higher losses of the transmission lines, this type of readout is difficult to apply at room temperature. 
Moreover, since 2D material membranes have not reached the ultrahigh mechanical and optical quality factors of optical cavities made out of materials like high-tension silicon nitride~\cite{norte2016mechanical}, they are presently less attractive to the quantum optomechanics community.

\subsection{Position dependent readout and mode-shapes}
\label{sec:modeshapereadout}
With a relatively localized probe, like a laser beam, the point at which the motion amplitude is measured can be laterally scanned over the drum, either by moving the spot position or the sample. By analyzing the position-dependent amplitude and phase of the membrane motion, the complete motion $w(x,y,t)$ can be measured. By monitoring amplitude and phase of the resonance peaks as a function of position (Fig.~\ref{fig:mechanical}a,c), the mode-shapes can be determined\cite{Garcia2008, Wang2014,Davidovikj2016}.
In case the motion consists of a superposition of eigenmodes, with multiple nonzero generalized coordinates $q_i(t)$, a projection procedure~\cite{Banafsheh2018} can be applied to decompose the measured motion $w(x,y,t)$ into modal participation factors~\cite{geradin2014mechanical} and generalized coordinates $q_i(t)$.

It is important to note that none of the previously described readout methods can be used to measure the membrane deflection at exactly one point. Instead, the readout signal is typically a weighed average of the deflection or speed over a certain area of the membrane. For optical readout that area is determined by the area of the optical focal spot and for transconductive or capacitive readout, this area depends on the area of the gate electrode below the membrane. The effect of averaging caused by the readout method, can be accounted for in so-called reduced order parameter models~\cite{Steeneken2007comsol, geradin2014mechanical}, and can significantly affect the relative peak heights in the frequency response spectrum. In some cases, for example if half of the membrane moves up and the other half moves down, the averaged motion over the whole membrane can even add up to zero, making a mode invisible or of very small amplitude, depending on the mode-shape and the measurement or electrode position~\cite{Davidovikj2016}.
When analyzing the dynamics of 2D materials, it is therefore of importance to be aware of the position dependence of the applied readout method. Position dependent characterisation of membrane dynamics and mode-shapes can provide useful additional information on membrane characteristics and imperfections (Fig.~\ref{fig:mechanical}b), that is hard to determine when characterising the motion at a single point or with a single electrode.

\subsection{Readout calibration}
\label{calibration}
Relating the output signal of the readout system to the actual amplitude of the membrane is not trivial and requires accurate calibration. In some cases calibration is of less interest, for example if the topic of study is the resonance frequency or $Q$-factor of the membrane. However, in other cases, like the study of the nonlinear dynamics, knowledge of the exact amplitude is essential. 
Three main methods of amplitude calibration have been discussed in literature. The first one is based on the measurement of the thermal Brownian motion of a harmonic oscillator in the undriven situation, which according to the equipartition theorem corresponds to an energy per mode of $\frac{1}{2}k_i \langle q_i^2\rangle  + \frac{1}{2}m_i \langle \dot{q_i}^2\rangle =m_i \omega_i^2 \langle q_i^2\rangle = k_{\rm B} T$. Using this equation, the measured voltage can be converted to a displacement $q_i$ if the temperature $T$ and the effective modal mass $m_i$ or stiffness $k_i$ are known or can be estimated from the geometry and material parameters of the structure~\cite{Hauer2013, Davidovikj2016}.  It should be noted that such estimations can be risky, especially for monolayers, e.g. because the mass can significantly deviate from the theoretical value due to contamination or because the stiffness is affected by tension variations and wrinkles; see also Sec.~\ref{sec:mass} and \ref{sec:stiffness}.

The second reported calibration method is based on fitting the resonance frequency versus gate voltage curve by a theoretical curve that has the mass-density and tension as fit parameters~\cite{Chen2009}. This method is based on the electrostatic reduction of the spring constant as discussed in sections~\ref{sec:elstact} and \ref{sec:feedback}, and uses a model for the electrostatic force and the expected membrane deflection to determine the deflection amplitude.
The third method is based on using the optical wavelength as a measuring rod, driving the membrane to large amplitudes, and analysing the harmonics generated by the nonlinearities of the optical readout method to calibrate the motion~\cite{DollemanCalibration}. The advantage of this method is that it does not require knowledge about the mass nor the mechanical properties of the membrane. The origin of these readout nonlinearities are discussed in the next subsection.

\subsection{Readout nonlinearities and other artifacts}

When driving the membranes to large amplitudes, besides mechanical nonlinearities, that will be discussed in section~\ref{nonlineardynamics}, the readout voltage response function $V_{\rm out}(q_i)$ can also become nonlinear, such that higher harmonics are generated. When the response function is well-known, measurement of the higher-harmonics can be used to correct the output signal for nonlinearities in the response function and in combination with the calibration methods discussed in the previous section, determine the time dependent position~\cite{DollemanCalibration}. Especially when studying the nonlinear dynamics of 2D material membranes, assessing the importance of these nonlinear readout effects is important to distinguish intrinsic mechanical nonlinearities described by the equation of motion (\ref{eq:motion}) from nonlinearities caused by the readout mechanism.

We conclude the section on readout by noting that for every readout method, effects of the readout on the actually measured motion should be avoided. For that reason it should be verified that variations in the laser power and electrical readout currents do not significantly affect the measured motion, via effects like membrane heating that shift the resonance frequency, or via feedback mechanisms in the actuation that will be discussed later. Also, care must be taken that spurious signals in the readout system, due to instrumentation noise or cross-talk, are minimized as much as possible to enable accurate readout. 

\color{black}

\section{Actuation methods}
\label{sec:actuation}
\begin{figure}[ht]
	\centering
	\includegraphics[scale=1]{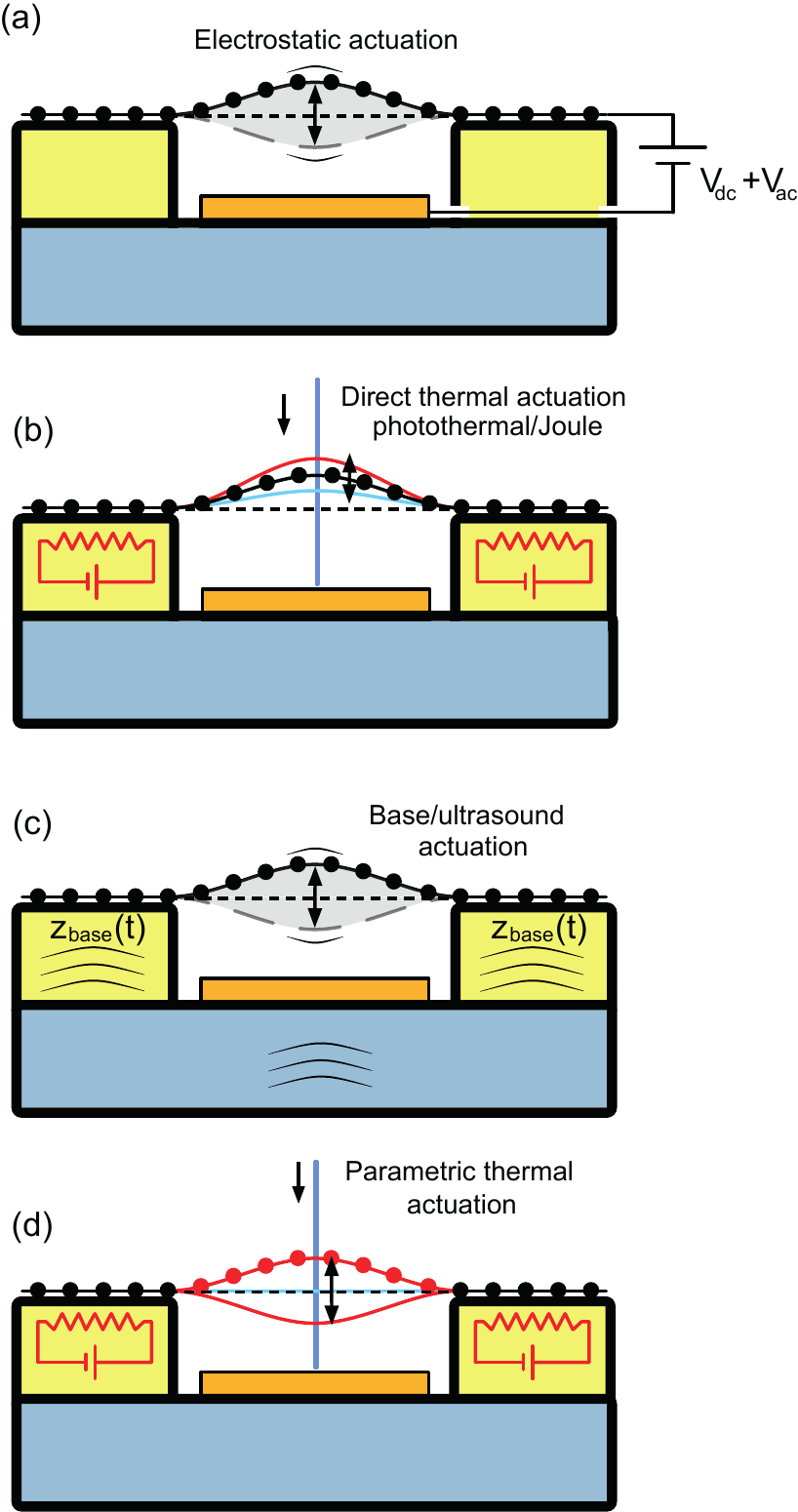}
	\caption{Actuation mechanisms for 2D material membranes. (a) Electrostatic actuation. (b) Optothermal actuation and thermal actuation by resistive heating. (c) Base excitation by high frequency vibrations of the substrate. (d) Several mechanisms leading to parametric actuation by stiffness modulation.
	}
	\label{fig:actuation}
\end{figure}

In  Eq. (1), external actuation can either be applied directly via the term $F_{\rm ext,i}$ or parametrically via the term $k_{{\rm p},i}$ that modulates the stiffness. Preferably, actuation should not be done by making mechanical contact to the suspended structure, since adhesion forces significantly alter the membrane shape and tension, and the mass of the contacting structure significantly alters the dynamics.
Actuation mechanisms that will be discussed in this section focus therefore on contactless methods, including electrostatic actuation, thermal actuation and base excitation. Finally, we discuss the effect of feedback forces and methods to parametrically actuate 2D membranes.

\subsection{Electrostatic actuation}
\label{sec:elstact}
When a voltage $V_{\rm act}(t)$ is applied between the suspended 2D material and a gate electrode (Fig.~\ref{fig:actuation}(a)), a time and position dependent electrostatic membrane pressure is generated with this functional form:
\begin{equation}
\label{elforce}
    P_{\rm ext}(w, t)=\frac{-\varepsilon_0 (V_{\rm act}(t)-V_{\rm off})^2}{2(g-w(x,y,t))^2}.
\end{equation}
\noindent
As indicated by the minus sign in this equation, this force is always attractive towards the gate electrode with a quadratic voltage dependence. The intrinsic offset voltage $V_{\rm off}$ is usually zero or close to zero, but can become nonzero in the presence of work-function differences or trapped charges~\cite{Steeneken2004}. By adjusting the voltage $V_{\rm act}$ properly, the effect of this offset can be eliminated; however, if the trapped charge distribution is non-uniform, this is not fully possible~\cite{Rottenberg2007}. In addition to trapped charge, Casimir forces can also generate a permanent downward force. For $g=100$~nm, the Casimir pressure between two perfectly conducting mirrors is equal to that of the electrostatic pressure at $V_{\rm act}-V_{\rm off}=0.17$~V as calculated by Eq.~\ref{elforce}. Since it cannot be avoided, the Casimir force can become a factor limiting the minimum gap distance beyond which the membrane always collapses~\cite{Chudnovsky2016}.


It is often desirable to eliminate nonlinear effects, i.e., to have an electrostatic actuation force that is proportional to an {\it ac} applied voltage $V_{\rm ac}$ and independent of membrane position. Therefore, ideally, the gap size $g$ is small compared to the lateral radius of the drum ($g \ll R$) and the displacements are much smaller than the gap size ($w \ll g$) such that the denominator of Eq. (\ref{elforce}) is almost constant. In that case the electric field lines are parallel to the $z$-axis, and the parallel-plate approximation holds for the capacitance between the 2D material and the bottom electrode. Moreover, to achieve an actuation force on the membrane at the same $ac$ frequency $\omega$ as the driving voltage, often a sum of $dc$ and $ac$ voltages is used with $V_{\rm eff}(t) = V_{\rm act}(t)-V_{\rm off} = V_{\rm dc} + V_{\rm ac} \sin \omega t$, with $V_{\rm dc} \gg V_{\rm ac}$ (see Fig.~\ref{fig:actuation}(a)). 
Using these approximations, quadratic terms in $w$ and $V_{\rm ac}$ can be neglected and equation (\ref{elforce}) becomes:
\begin{equation}
\label{eq:elpressure2}
 P_{\rm ext}(w,t)\approx -(V_{\rm dc}^2 + 2 V_{\rm dc}V_{\rm ac} \sin \omega t) \frac{\varepsilon_0}{2 g^2} (1+2\frac{w}{g}) .
\end{equation}

This equation implies that an electrostatic force generates a static downward force proportional to $V_{\rm dc}^2$, and a sinusoidal driving force proportional to $2 V_{\rm dc}V_{\rm ac}$. The last term proportional to $\frac{w}{g}$ effectively acts as a negative spring constant, and thus reduces the resonance frequencies at large $V^2_{\rm dc}$; this effect is called spring softening and is discussed in more detail in sections~\ref{sec:feedback} and \ref{sec:elstinteractions}. Just as for readout (Sec.~\ref{sec:modeshapereadout}), the effective modal force that drives a certain resonance mode, depends both on the electrode configuration and on the mode shape to be driven and can be calculated by a weighted integral of the pressure of Eq.~(\ref{eq:elpressure2}) over the actuation surface \cite{Steeneken2007comsol}, which is usually the largest for the fundamental mode. Since the electrostatic energy is given by $U_{\rm es}=\frac{1}{2}C_g V^2$, only mode shapes that significantly change $C_g$ can be efficiently excited using electrostatic forces.

There are several additional aspects that one should consider when using this form of actuation. The first one is that since the $ac$ voltage is applied on a high-impedance capacitor (the 2D nanodrum) and the voltage source (e.g. network analyzer) often has a 50~$\mathrm{\Omega}$ output impedance, the $ac$ voltage across the drum $V_{\rm ac}$ will be almost twice as large as if the source would be connected to a 50 Ohm load. The second one is that the geometry, and finite thickness of the electrodes underneath the membrane (e.g. a graphene flake stamped on top of an electrode made of gold with a circular hole in it) can significantly affect~\cite{Davidovikj2016} the electric field lines near the edge of the drum, such that they deviate from the parallel plate approximation, leading to a lower effective electrostatic force than predicted from Eq.~(\ref{eq:elpressure2}). Thirdly, a high resistance of the membrane, in combination with cable and device capacitances to ground, can result in $RC$-times that can diminish the efficiency of the actuation at high frequencies (Sec.~\ref{sec:elstinteractions}). Finally, quantum capacitance effects~\cite{Xia2009} can decrease the efficiency of capacitive readout and actuation of membranes.

\subsection{Optothermal and electrothermal actuation}
\label{sec:thermalact}
Due to their low heat capacitance and high thermal conductivity, suspended 2D materials can be heated very rapidly and efficiently, either by absorbing optical power~\cite{Bunch2007}, by resistive electrical Joule heating of the membrane itself or via a resistive heating ring by which the membrane is suspended~\cite{DavidovikjHeaters}. Although electrothermal actuation of a large graphene/PMMA heterostructure membrane has been demonstrated~\cite{al2017dynamic}, to our knowledge high-frequency electrothermal actuation of resonances by Joule heating in a freestanding single 2D material membrane has not been demonstrated. 
For both optothermal and electrothermal actuation, the basic principle is shown in Fig.~\ref{fig:actuation}(b). The black line in the figure is the initial state and when the 2D material membrane is heated, it thermally expands (assuming a positive thermal expansion coefficient) and moves upward to the red position; on the other hand, when it cools, it contracts and moves to the blue position. In the figure, the time-dependent heating power is provided by the blue power-modulated laser, or by the red resistive Joule heaters near the suspension points of the membrane. The cooling of the membrane occurs via heat transfer towards the substrate, surrounding gas, or by radiation. Even though the heat capacitance of the membrane is low, the temperature change will not occur instantaneously. This delay between heating power, membrane temperature $T$ and thermal expansion can have interesting effects on the membrane dynamics, as will be discussed in more detail in section~\ref{thermalcharacterization}. The linear equation of motion of a mode in the presence of an effective thermal expansion coefficient $\alpha_i$, a mode-dependent device parameter, can be written as:
\begin{equation}\label{eq:thermalexcitation}
    m_i \ddot{q_i} + c_i \dot{q_i} +  k_i (q_i-\alpha_i T(t))=0.
\end{equation}
It is important to note that the initial membrane shape (represented in black in Fig.~\ref{fig:actuation}(b)) has been given an intentional offset from the flat position. The reason for this is that if the membrane would be perfectly flat, a temperature change would not result in an out-of-plane deflection. 
A consequence of this is that both the magnitude and the direction of the resulting out-of-plane thermal expansion forces depend on the magnitude and direction of the initial deformation. If the initial deformation is upward, the membrane will move upward upon heating, and vice-versa for downward initial deflection. Although the physical origin for these initial deformation related effects has not been clarified, potential causes could be fabrication induced wrinkles, buckling, edge-adhesion, electrostatic or Casimir forces. A large variation in the magnitude and direction of the thermal expansion force between different CVD graphene drums, made by the same fabrication procedure, was recently observed~\cite{Dolleman2020A}, suggesting that this thermal expansion force is a very sensitive function of the membrane properties and geometrical imperfections. The exact determination of the effective thermal expansion coefficient $\alpha_i$ is therefore complex and can only be estimated from measurements using the calibration methods discussed in section~\ref{calibration} in combination with fits or models for the linear membrane parameters. 

It is usually desirable to have a force that is at the same frequency as the {\it ac} driving voltage $V_{\rm ac}$. However, for resistive Joule heating it is known that heating power in the membrane follows $P \propto V_{\rm act}^2/R$, so similar to electrostatic actuation a voltage $V_{\rm act}=V_{\rm dc} + V_{\rm ac} \sin \omega t$, with $V_{\rm ac} \ll V_{\rm dc}$ is used to ensure that there is a component in $\Delta T$ at frequency $\omega$ proportional to $V_{\rm ac}$. Also for linear opto-thermal modulation, to get a force of the same frequency of the driving voltage, the time-dependent laser power should be modulated on top of a {\it dc} background power $P_{\rm L}(t)=P_{\rm dc}+P_{\rm ac} \sin \omega t$, with $P_{\rm ac}\ll P_{\rm dc}$. This modulation is usually done by running a constant {\it dc} current, $I_{\rm L}$, through a diode laser, while modulating the voltage around a certain bias point $V_{\rm L}(t)=V_{\rm dc}+V_{\rm ac} \sin \omega t$, noting that the laser's optical power is proportional to $P_{\rm L}(t)=I_{\rm L}\times V_{\rm L}(t)$. It is important to inspect the $I_{\rm L}-V_{\rm L}$ curve of the laser diode, to ensure that it is sufficiently linear to prevent nonlinear terms in the actuation force to appear. For opto-thermal actuation, it should be noted that position dependent actuation forces can emerge, because the optical field intensity and optical absorption of the standing wave formed by the actuation laser is position dependent as discussed in section~\ref{parametricact}, causing feedback forces that can affect damping and resonance frequency and can even lead to self-oscillation (Sec.~\ref{sec:feedback}).

Besides direct thermal actuation, parametric thermal actuation by tension modulation is also possible as will be discussed in section~\ref{parametricact}. Finally, we mention that instead of opto-thermal actuation a modulated laser can also excite the membrane by radiation pressure of light, given by the ratio of the light intensity with the speed of light, $P_{\rm rad}=I_{\rm rad}/c$. However, due to the relatively high optical absorption and low reflectivity of graphene membranes, thermal expansion forces tend to exceed radiation pressure forces. Other 2D materials might offer better opportunities for demonstrating radiation pressure actuation.

\subsection{Base actuation}
\label{base}
Base excitation is a method where a piezoelectric resonator, or other type of shaker, is mounted below the substrate to excite the substrate with the membrane sinusoidally (see Fig.~\ref{fig:actuation}(c)). Acoustic waves flow through the entire substrate and excite the resonant membrane at its edges (base). The simplest model for base excitation in the absence of damping is a mass at position $q_i$ that is connected via a spring to a base at position $q_b(t)$ that moves in time.
The equation of motion in that case is given by:
\begin{equation}\label{eq:baseexcitation}
    m_i \ddot{q_i} + c_i \dot{q_i} + k_i (q_i-q_b)=0.
\end{equation}
\noindent
This equation can be rewritten, such that it is identical to Eq.~(\ref{eq:motion}) with an effective base excitation force $F_{\rm ext,bi}=k_i q_b(t)$. Like for other types of excitation, the amplitude of the resonator at resonance, is a factor $Q_i$ higher than at low frequencies, such that at resonance $|q_i|=Q_i |q_b|$. Using SiN membranes with integrated graphene membranes, base excitation was generated around 4 MHz and was indeed observed~\cite{Singh2018} to result in motion amplification by a factor $Q$. In another work, also off-resonant base excitation was used to move graphene with respect to the base and detected using transconductive readout~\cite{Verbiest2018}, thus functioning as an ultrasound detector.

When using a resonant actuation element, like a piezoelectric resonator, for driving the base actuation, it is important to note that when actuating at constant voltage amplitude, both resonances of the membrane and resonances of the actuation element will be observed in the motion $q_{i}(t)$. Another point to note is that in Eq.~(\ref{eq:baseexcitation}) it is assumed that the mass and stiffness of the base are infinitely much larger than the mass and stiffness of the 2D membrane, such that the membrane motion does not affect the motion of the base. If this assumption does not hold anymore, the combined membrane-base systems needs to be analyzed using coupled equations of motion for base and membrane~\cite{Singh2018, Zanette2018}.

\subsection{Parametric actuation}
\label{parametricact}
Instead of direct actuation, where the force $F_{\rm ext,i}(t)$ only depends on time, it is also possible to externally excite motion by force terms of the form $-k_{{\rm p},i}(t) q_i$, which are the product of an externally modulated time-dependent stiffness $k_{{\rm p},i}(t)$ and the membrane position $q_i(t)$. This parametric actuation term can originate from special (nonlinear) terms in the excitation force, like~\cite{turner1998five} the term proportional to $V_{\rm ac}\frac{w}{g}\sin \omega t $ in Eq.~(\ref{eq:elpressure2}), but can also be generated by physical modulation of the linear mass, damping and stiffness parameters in the equation of motion: $m_i(t)$, $c_i(t)$ or $k_i(t)$. For example, when heating a 2D membrane with a modulated laser, its tension reduces when it thermally expands, and since the stiffness is proportional to the tension, the stiffness will be modulated proportionally to the laser-induced temperature change~\cite{Zalalutdinov2001b,Aubin2004,Dolleman2018,Mathew2016,Prasad2017}: 
\begin{equation}\label{eq:thermalparametricexcitation}
    m_i \ddot{q_i} + c_i \dot{q_i} +  (k_i + k_{{\rm p}T,i}(T(t))) q_i=0.
\end{equation}
The modulated stiffness can be rewritten as a parametric force term $-k_{{\rm p},i}(t) q_i$ in Eq.~(\ref{eq:motion}), with $k_{{\rm p},i}(t) = k_{{\rm p}T,i}(T(t))$.

This type of tension modulation, illustrated in Fig. ~\ref{fig:actuation}(d), is especially efficient in 2D materials, since their temperature can be more efficiently opto-thermally modulated at high frequencies than bulk materials due to their small thickness.
Parametric terms in the equation of motion can result in interesting effects like parametric oscillation (also called parametric resonance) and noise squeezing or amplification as will be discussed in section~\ref{sec:parametricres}. Finally, it should be noted that when a parametric force term is accompanied by a constant static offset $q_{\rm off}$ that might be caused be a static force or fabrication imperfection in the system, the parametric force term $-k_{{\rm p},i}(t) (q_i(t)-q_{\rm off})$ consists of a parametric and direct actuation force term ($k_{{\rm p},i}(t)q_{\rm off}$).

\subsection{Feedback forces}
\label{sec:feedback}

In addition to the parametric terms discussed in the previous section, that depend on time and position, the actuation force $F_{\rm ext,i}$ also contains terms that depend on the position or speed of the membrane but not on time, which are called feedback or back-action forces. These force terms can originate from an external feedback system, that e.g. measures a displacement $w$ and accordingly applies a force $F_{\rm ext,i}(w)$, but they can also originate from the intrinsic physics of the actuation (Sec.~\ref{sec:actuation}) or physical interactions of the membrane with its environment (Sec.~\ref{sec:physint}).
In its simplest form $F_{\rm ext,i}=k_{\rm fb} q_i +c_{\rm fb} \dot{q}_i$, the feedback force is linearly proportional to the position and velocity of the membrane . The feedback force can be merged with the left side of Eq.~(\ref{eq:motion}), resulting in modified stiffness and damping~\cite{Miller2018} terms, $k_{\rm eff,i}=k_i-k_{\rm fb}$ and $c_{\rm eff,i}=c_i-c_{\rm fb}$. Linear feedback terms thus provide a route to tune the damping, stiffness and resonance frequency of the system:
\begin{equation}
    \label{eq:omegatuning}
    \omega_{\rm eff,i}^2 = \omega_i^2 - \frac{k_{\rm fb}}{m_i}, 
\end{equation}
\begin{equation}
\label{eq:qtuning}
    \frac{1}{Q_{\rm eff,i}} = \frac{1}{Q_{i}} - \frac{c_{\rm fb}}{\omega_i m_i}.
\end{equation}

By external control of the delay between force and position, the feedback force can be brought in-phase with either position or velocity~\cite{Steeneken2011}, either enhancing or diminishing $\omega_{\rm eff,i}$ or $Q_{\rm eff,i}$, thus providing a route for tuning the resonators characteristics. For $1/Q_{\rm eff,i} < 0$ it can even result in self-sustained oscillations (Sec. \ref{sec:brownian}). Nonlinear feedback terms can result in even more complex behaviour as will be discussed in Sec.~\ref{sec:nonlsoln}.

\section{Vibration in the linear regime}
\label{sec:linreg}

In this section, we will consider the dynamics of 2D material membranes that follows from the linear terms in the equation of motion (Eq.~\ref{eq:motion}) under (i) free, (ii) driven and (iii) feedback conditions. In the first subsection we will look at the solutions of the EOM and in the second subsection at the underlying physics that governs the values of the linear coefficients $m_i$, $c_i$ and $k_i$, and their theoretical and experimental determination. 

\subsection{Linear dynamic motion}
\label{sec:sollineom}
For small displacements $q_i$, the terms of quadratic and higher order in $q_i$ and $\dot{q_i}$ in Eq.~(\ref{eq:motion}) become negligible compared to the linear terms, such that the equation of motion is linear. Although the solutions of the linear equation of motion are well known and discussed in textbooks on dynamics~\cite{geradin2014mechanical}, we quickly review them here for completeness, before focusing on the specific mechanisms that determine the linear dynamics of 2D materials.

\begin{figure*}[ht]
    \centering
    \includegraphics{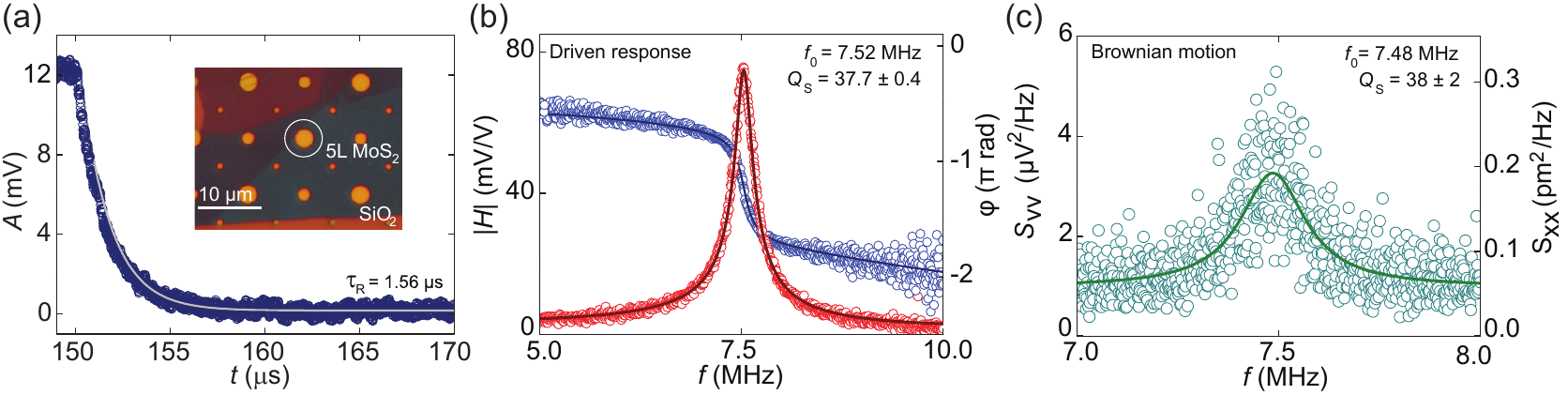}
    \caption{Linear vibration measured on a 5 layer MoS$_2$ drum; the motion is actuated opto-thermally and read out using Fabry-Perot interferometry. (a) Free vibration: measurement of the amplitude as a function of time after the optical drive is switched off at $t = 150$ $\mu$s. The decay of the amplitude is exponential in line with Eq.~\ref{eq:qringdown}. The inset shows an optical image of the drum that was measured.  (b) Driven motion: amplitude and phase of the driven response. A fit to the response using Eq.~\ref{eq:FRF} gives a resonance frequency of 7.52 MHz and a Q factor of 37.7. (c) Measured power spectrum of the undriven membrane (same as in (b), showing a peak due to Brownian motion. A Lorentzian fit to the data results in a resonance frequency of 7.48 MHz and a Q factor of 38. Reprinted with permission from Ref.~\onlinecite{leeuwen2014}, copyright (2014) AIP Publishing LLC. }
    \label{fig:linearmotion}
\end{figure*}
\subsubsection{Free vibration}
\label{sec:freevib}
When the forcing terms $F_{\rm ext,i}(t)$ and $k_{{\rm p},i}(t)$ in Eq.~(\ref{eq:motion}) are zero, the system exhibits free vibrations that are the solutions of the well-known harmonic oscillator equation:
\begin{equation}
\label{eq:freelineareom}
    m_i \ddot{q}_i + c_i \dot{q}_i + k_i q_i = 0 .
\end{equation}
If we plug in a trial solution $q_i(t) = q_{i,0} e^{\lambda_i t}$, we obtain:
\begin{equation}
    \lambda_i^2 m_i  + \lambda_i c_i  + k_i  = 0.
\end{equation}
Solving this quadratic equation for $\lambda_i$, and taking the small damping approximation $c_i \ll \sqrt{4 k_i m_i}$ we find the underdamped solutions for $\lambda_i$:
\begin{equation}
\label{eq:lambdapm}
    \lambda_{i\pm} = -\frac{c_i}{2 m_i} \pm i\sqrt{\frac{k_i}{m_i}-\frac{c_i^2}{4 m_i^2}} \approx -\frac{\omega_i}{2 Q_i} \pm i \omega_i,
\end{equation}
where $\omega_i = \sqrt{k_i/m_i}$ is the (natural) resonance frequency of mode $i$, with corresponding quality factor $Q_i=\sqrt{k_i m_i}/c_i \approx |\Im \lambda_\pm/2 \Re \lambda_\pm|$. The imaginary exponent represents the fast oscillatory part of the motion and the negative real exponent represents the slowly decaying envelope of the motion, which is called ringdown (see Fig.~\ref{fig:linearmotion}(a)). There are two solutions $q_{i\pm}(t)$ to the equation of motion, of which a superposition with suitable coefficients $q_{i,0\pm}$  satisfies the initial conditions for position and speed: 
\begin{equation}
\label{eq:qringdown}
    q_{i\pm}(t) = q_{i,0\pm} e^{-\frac{-\omega_i}{2 Q_i} t} e^{\pm i\omega_i t}.
\end{equation} 
If the damping is increased such that $Q_i<1/2$, the square-root in Eq.~(\ref{eq:lambdapm}) changes sign, such that the system becomes overdamped.

\subsubsection{Driven motion}
\label{sec:drivenmotion}

For the linear differential equation of motion, any superposition of solutions is again a solution of the differential equation. Any periodic driving force $F_{\rm ext,i}(t)$ can be written as a Fourier sum of sinusoidal functions, so if we find the solution for a sinusoidal driving force with complex amplitude $F_{\rm ext,i} (\omega)$ and frequency $\omega$, we can construct the solution for any waveform, and $F_{\rm ext,i} (\omega)$ is the Fourier transform of $\sqrt{2\pi}F_{\rm ext,i}(t)$. For the linear driven case, Eq.~(\ref{eq:motion}) reads:
\begin{equation}\label{eq:linear-harmonic-osc}
   m_i \ddot{q_i} + c_i \dot{q_i} + k_i q_i = F_{\rm ext,i}(\omega) e^{i \omega t}.
\end{equation}
The steady-state solutions are of the form $q_i(t)=  q_i(\omega) e^{i \omega t}$, and the frequency response function FRF$(\omega)$ equals $q_i(\omega)/F_{\rm ext,i}(\omega)=(-\omega^2 m_i+ i\omega c_i + k_i)^{-1}$. Its magnitude $|\rm FRF(\omega)|$, which is also called its compliance, is displayed in Figs.~\ref{fig:mechanical}(c) and \ref{fig:hardening}(a), and obeys the equation:
\begin{equation}\label{eq:FRF}
    \left|\frac{q_i(\omega)}{F_{\rm ext,i}(\omega)}\right|=\frac{1}{\sqrt {(k_i- m_i\omega^2)^2+ (c_i \omega)^2}}.
\end{equation}
 A peak in the magnitude is found when the membrane is driven at its resonance frequency ($\omega=\omega_i$); the full-width-half-maximum (FWHM) of the peak in $|{\rm {FRF}(\omega)}|^2$ is $\Delta \omega_{FWHM}=\omega_i/Q_i$ (in the small damping limit) and defines the linewidth $\Gamma = \omega_{FWHM}/(2\pi)$ . The phase angle, by which the motion lags behind with respect to the driving force, is $\phi_i(\omega)=\arctan \left(\frac{-\Im{\rm{FRF}}}{\Re{\rm{FRF}}}\right)$. Near the resonance frequency, it changes abruptly from zero to $\pi$, and equals $\phi_i=\pi/2$ at resonance. Note that the lineshape of FRF$(\omega)$ is not exactly identical to a Lorentzian. An example of a measured resonance peak is shown in Fig.~\ref{fig:linearmotion}(b). 

When measuring the response of a membrane as a function of driving frequency, one can find multiple peaks at different frequencies $\omega_i$, of which the fundamental mode typically shows the largest amplitude, because the quality factor and the weighting integrals for actuation and readout (Sec.~\ref{sec:modeshapereadout}, \ref{sec:elstact}) tend to be largest when all points on the membrane move up and down in phase. Some of the higher modes may be degenerate, i.e., have the same resonance frequency, although in practical experiments they often split and become non-degenerate when symmetry is broken by deviations from the ideal membrane shape such as wrinkles~\cite{Davidovikj2016}. 

\subsubsection{Brownian motion cooling, amplification and oscillation}
\label{sec:brownian}
Even without intentionally applying an external driving force, the membrane moves due to thermal or quantum fluctuations. Although macroscopic mechanical resonators have been brought to the quantum ground state~\cite{Cleland2010}, this has not yet been achieved for 2D material resonators. We therefore focus here on the stochastic (random) thermal or Brownian motion forces that drive the mode, $F_{\rm ext,i}=F_{\rm th}$, which are sometimes called Langevin forces. They are a consequence of the thermal coupling of the resonator to the environment via for example the random collisions of molecules in the gas surrounding it, or via the phonons in the substrate that couple to the membrane at the clamping points.

These thermomechanical forces have a white (frequency independent) magnitude and random phase, such that in Eq.~(\ref{eq:linear-harmonic-osc}), $\langle |F_{\rm ext,i}(\omega)|^2\rangle =4 c_i k_{\rm B} T\times BW$, where $BW$ is the bandwidth in Hz over which the force power spectral density (one-sided, so only integrated over positive frequencies) is integrated and where the brackets $\langle \rangle $ indicate the expected value or long-time average. $T$ is the ambient temperature and $k_{\rm B}$ is the Boltzmann constant. The thermal fluctuations lead to an amplitude that obeys $\langle q_i^2\rangle =k_{\rm B} T/k_i$ for all resonance modes, as follows from the equipartition theorem~\cite{Hauer2013} (see also appendix~\ref{AppendixA}). This relationship can also be used to relate the measured signal to the actual motion amplitude  (amplitude calibration), provided that $T$ and $k_i$ are known (see also Sec.~\ref{calibration}). An example of a measured power spectral density due to Brownian motion is shown in Fig.~\ref{fig:linearmotion}(c). 

As mentioned in section~\ref{sec:feedback}, the damping coefficient of a mode can be affected by feedback forces. Interestingly, this effect can be used to change the effective temperature of the mode as follows: in the presence of velocity proportional feedback forces $F_{\rm ext,i}=c_{\rm fb} \dot{q}_i$, the effective damping coefficient of the mode is altered to a value $c_{\rm eff}=c_{\rm i}-c_{\rm fb}$ without affecting the thermo-mechanical noise force $\langle |F_{\rm ext,i}(\omega)|^2\rangle $ (which only depends on the intrinsic damping $c_i$). Thus by increasing $c_{\rm eff}$ the motion can be damped, reducing $\langle q_i^2\rangle $, such that it moves stochastically as if it were in a system without feedback at a lower temperature $T_{\rm eff}$. This type of feedback can thus be used~\cite{Barton2012,DeAlba2016,Steeneken2011} for cooling (lowering) the effective temperature $T_{\rm eff}$ of a mode, below the ambient temperature $T$. 
Along a similar fashion, the effective damping $c_{\rm eff}$ can also be reduced, increasing the effective quality factor $Q_{\rm eff}=\frac{\sqrt{k_i}{m_i}}{c_{\rm eff}}$ and amplifying the Brownian motion until the effective damping becomes negative $c_{\rm eff} < 0$ such that it reaches the threshold for self-sustained oscillation; beyond this threshold the motion amplitude is amplified up to a level where it is limited by nonlinear effects. Since the frequency of this so-called limit-cycle is close to $\omega_i$, this kind of oscillation behaviour can be used as a clock and has been reported in graphene membranes with opto-thermal~\cite{Barton2012,Samer2017} and electronic feedback~\cite{Chen2013}.

\subsection{Physical parameters determining the linear dynamics}
\label{sec:physlin}
In conventional MEMS (microelectromechanical systems) devices material properties are usually accurately known, and fabrication induced mechanical stresses are usually uniform over a wafer (the substrate on top of which the devices are fabricated) and can be characterised with dedicated test structures~\cite{Senturia1997}.  This makes it possible to accurately estimate the modal mass and stiffness of MEMS resonators from their geometry. In 2D material membranes, however, it is much more challenging to determine these parameters since  i)  the material properties are more difficult to measure because conventional characterisation techniques fail, and ii) there is large variability and non-uniformity of the suspended 2D material parameters caused by non-reproducible material growth and device fabrication methods. As a consequence, there is a large variation in literature values of the relevant material and device parameters, that are important to predict and understand the dynamics of 2D material membranes. However, once fabricated, several of these parameters can be deduced from the measured static and dynamic motion, as will be outlined in this section.

The resonance frequency and quality factor can be extracted relatively easily from experiments by fitting Eq.~(\ref{eq:FRF}) to the data and the amplitude can be determined with some more effort using the calibration methods described in Sec.~\ref{calibration}. 
With the amplitude, resonance frequency, and quality factor determined accurately, the remaining coefficients  in Eq.~(\ref{eq:FRF}) can be determined provided that the force can also be determined either from models or measurements. An example of this procedure is given in~\cite{Chen2009}, where the mass increase of a graphene membrane due to a small amount of pentacene (a small aromatic hydrocarbon molecule) is studied.

An important point in determining the actual values of the physical parameters is to realize that e.g. the modal mass coefficient $m_i$ is not the actual mass of the system. Numerical coefficients, which depend on mode shape, measurement position and resonator geometry, that relate the actual mass to $m_i$ need to be calculated. This can be done by evaluating the kinetic and potential energy in the resonator~\cite{amabili2008nonlinear, Hauer2013,Steeneken2007comsol}. 
For a circular drum, the derivation is given in appendix~\ref{AppendixB} and results in a simplified equation of motion that resembles Eq.~(\ref{eq:motion}). Analytical expressions are found for modal mass $m_i$, modal stiffness $k_i$, the effective force coefficients and also the nonlinear terms that will be discussed in the next section. 
Specifically, for the fundamental mode of a circular drum one finds $m_1=0.2695 m$ and $k_1=4.8967 n_0$, where $m = \rho h \pi R^2$ is the total mass of the membrane ($\rho$ is the mass density of the membrane material) and $n_0$ its initial tension. For more complicated device geometries, finite element methods can be more convenient to determine the mode-shapes. Using these mode-shapes, the stiffness and mass coefficients can be determined by integrating elastic and kinetic energy over the device volume~\cite{Steeneken2007comsol}. If the damping mechanism is known, the modal quality factor can even be estimated using finite elements for support losses~\cite{Steeneken2007comsol} and thermo-elastic damping~\cite{Duwel2006} (see section~\ref{sec:Q}). In the following three subsections we will first discuss experimental determination of the modal mass ($m_i$) and stiffness ($k_i$), followed by a subsection on determining dissipation ($c_i$) characterized by the quality factor. 

\subsubsection{Modal mass}
\label{sec:mass}

In principle, the modal mass of a resonance mode can be determined experimentally from the relation $m_i=k_i/\omega_i^2$ if the resonance frequency and modal stiffness are known. Although the resonance frequency $\omega_i$ is straightforward to measure, the stiffness mainly depends on the pretension in the membrane, a parameter that is difficult to measure or control directly as will be discussed in the next subsection. One way to circumvent this problem is to vary the pretension and extract the modal stiffness from the resulting change in resonance frequency. Along this line experimental estimates~\cite{Bunch2008, Chen2009, Lee2018} of the modal mass $m_i$ of a graphene resonator were made by fitting the relation between resonance frequency and applied 
pressure curve, where the pressure was applied either by electrostatic forces~\cite{Chen2009, Banafsheh2017} or by a gas pressure~\cite{Bunch2008, Lee2018}.
The method relies on the fact that the modal stiffness $k_i(\Delta P)$ and pretension in a membrane change with the applied pressure difference $\Delta P$ across the membrane. 

By fitting the resulting $\omega_i (\Delta P)$ measurement both $m_i$ and $k_i$ can be inferred, noting that the model should include tension changes and for electrostatic pressure also include electrostatic softening effects~\cite{Banafsheh2017} (see Sec.~\ref{sec:elstact}, \ref{sec:elstinteractions} and Eq.~(\ref{eq:Ffit})).

However, in practice models for $k_i$ and its pressure dependence can be inaccurate due to the presence of wrinkles, tension non-uniformity or other imperfections. To avoid this problem, mass determination based on the squeeze-film effect can be applied~\cite{Dolleman2016A}, which uses the fact that the resonance frequency of a graphene membrane, that is close to a counter electrode at a distance $g$, depends on the ambient gas pressure $P_{a}$ by the relation $\omega_i^2(P_{a})-\omega_i^2=\frac{P_{a}}{g \rho h}$ (see Eq.~(\ref{eq:sqz_freqshift})). From this relation the mass density per area $\rho h$ can be obtained directly by measuring the resonance frequency versus ambient pressure curve. Using this technique the mass $m_i$ of a 31 layer graphene resonator was determined~\cite{Dolleman2016A}. 

In Table~\ref{tab:mass}, we list the experimental mass density of several single-layer graphene, MoS$_2$ and WSe$_2$ devices reported in literature. Although errors can be expected since the mode shape can be uncertain in some cases, it is clear that consistently a larger mass is measured than theoretically expected. The cleanest devices that are closest to the theoretical values are either produced by mechanical exfoliation without transfer polymers involved~\cite{Bunch2008}, or by cleaning the device using Ohmic heating~\cite{Morell2016}. The table includes methods where electrostatic softening or tensioning are used; these two are differentiated, since electrostatic softening is independent of the mode-shape, whereas  electrostatic tensioning does depend on the mode shape.

\begin{table*}[] 
\caption{Table with mass density $\rho h$ determined in several single--layer 2D material membranes using different methods; the deviation indicates the ratio between the measured and theoretically calculated mass. \label{tab:mass} }
\begin{tabular}{l|l|l|l|l}
Authors & Material & Method & Measured $\rho h$ (kg/m$^2$) & Deviation \\ \hline 
  \citeauthor{Bunch2008} \cite{Bunch2008}     &   Exfoliated graphene       &  Tension (gas)  &    $\num{9.6e-7} $   &  1.3    \\\hline 
 \citeauthor{Singh2010} \cite{Singh2010}  &    Exfoliated graphene  &    Tension (electrostatic)      &   $\num{5.7e-6}$ & 7.4       \\\hline 
  \citeauthor{Barton2012} \cite{Barton2012}     &   CVD graphene       & Tension (electrostatic) & \num{3.54e-6}  & 4.6 \\
  &     &    & \num{2.23e-6} & 2.9 \\ \hline
   \citeauthor{Song2012} \cite{Song2012}  & Exfoliated graphene & Electrostatic softening & $\num{7.47e-6 }$ & 9.7 \\ \hline
   \citeauthor{Chen2009} \cite{Chen2009} & Exfoliated graphene & Tension (electrostatic) & $\num{3.6e-6}$ & 4.7 \\ 
    & Annealed by Ohmic heating & Tension (electrostatic) & $\num{1.6e-6}$ & 2.1 \\ \hline
    \citeauthor{Singh2018} \cite{Singh2018} & CVD graphene & Tension (electrostatic) & $\num{2.2e-5}$ & 29 \\ \hline
    \citeauthor{DeAlba2016} \cite{DeAlba2016} & CVD graphene & Tension (electrostatic) & $\num{8.4e-6}$ & 11 \\
  &  &  & $\num{7.3e-6}$ & 9.5 \\ \hline
  \citeauthor{Morell2016} \cite{Morell2016} & WSe$_2$ & Electrostatic softening & $\num{15.6e-6} $ & 1.3 \\ \hline
   \citeauthor{manzeli2019self} \cite{manzeli2019self} & CVD MoS$_2$ & Tension (electrostatic)& $\num{3.8e-6} $ & 1.16 \\ 
    &  &  &  $\num{2.3e-5}$ & 7.1 \\ \hline
\end{tabular}
\end{table*}

\subsubsection{Modal stiffness} 
\label{sec:stiffness}
\begin{figure*}[ht]
    \centering
    \includegraphics{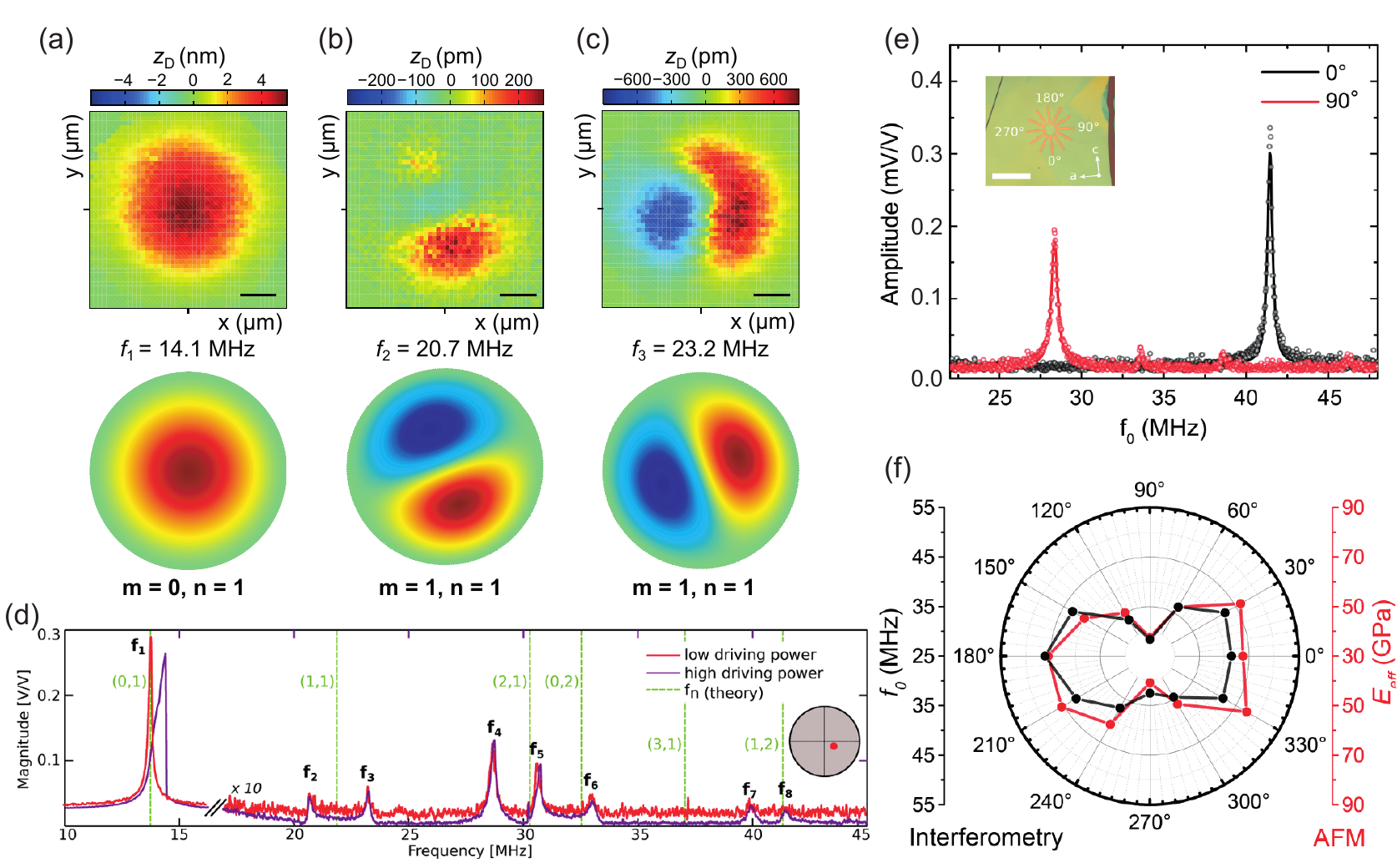}
    \caption{(a) Characterization of the first mode shape of a circular graphene drum with a diameter of 5~$\mu$m and thickness of 5 nm (scale bar is 1~$\mu$m). The bottom panel shows the mode shape predicted by theory (see Appendix~\ref{AppendixB}). (b) and (c) Second and third mode shape, showing larger deviations compared to theory due to nonuniformities in the tension of the drum. (d) Peaks in the measured mechanical response of the drum in Figs. (a)--(c) deviate from the theoretical peaks (green dashed lines) due to nonuniform tension. (e) Measurement of the resonance frequency of a rectangular shaped anisotropic As$_2$S$_3$ resonator, showing its frequency depends on the crystal orientation. Inset: optical image of a star-like cavity used to charactize the anisoptropy (scale bar is 4~$\mu$m). e) Resonance frequency and Young's modulus dependence on the angular orientation of the As$_2$S$_3$ resonator. Figs. (a--d) are reprinted with permission from Ref.~\onlinecite{Davidovikj2016}, copyright (2016) American Chemical Society. Figs. (e--f) are reprinted from Ref.~\onlinecite{siskins2019highly} licensed under CC-BY-NC-ND.
    \label{fig:mechanical}}
\end{figure*}

The stiffness of thin structures is determined both by their bending rigidity, and by their tensile stress. If one of these effects dominates, the structure is in the plate or membrane limit respectively. The dynamics in these limits are discussed in appendix \ref{AppendixB} in sections \ref{sec:membranedynamics} and \ref{sec:platedynamics}. When increasing the thickness of the structure, a transition from the membrane to the plate limit occurs\cite{Castellanos2013}, which can be observed in the resonance frequencies and their ratios (appendix~\ref{sec:plateundertension}). For thicker flakes the bending rigidity can be calculated from the material's Young's modulus, Poisson's ratio and thickness, but in the membrane limit the stiffness is mainly determined by the pretension.

The pretension in 2D membranes is hard to control, because it depends strongly on the fabrication method. As a result the modal stiffness of 2D resonators is difficult to predict from models and requires experimental determination. One route for this is the measurement of the modal mass by one of the techniques from the previous subsection and subsequent calculation of the modal stiffness using the relation $k_i=m_i \omega_i^2$. However, often the methodologies discussed in the previous section are not easily applied, or the models on which they rely are inaccurate due to device imperfections as discussed below. A second route, thermomechanical motion for determination of stiffness, is closely related to calibration, since if the measurement system is well-calibrated (section~\ref{calibration}), such that $q_i$ is known, the modal stiffness $k_i$ can be determined from the thermal motion using the relation $\frac{1}{2}k_{\rm B} T = \frac{1}{2}k_i \langle q_i^2\rangle $. A third method is characterization of the membrane's pretension by AFM or Raman methods and analytical or finite element method (FEM) calculation of the modal dynamic stiffness. Many studies have focused on AFM and Raman spectroscopy for studying the tension and stiffness of suspended 2D materials and their uniformity~\cite{Castellanos2015, Akinwande2017, Colangelo2019, Nicholl2017, Sonntag2019, zhang2020dynamicallyenhanced}. 

In addition to tension variation, some 2D materials naturally exhibit large mechanical anisotropy in their bending rigidity, which in the plate limit can significantly modify their mode shapes and resonance frequencies with respect to those of isotropic materials (see Figs.~\ref{fig:mechanical}(e)-(f))~\cite{wang2016resolving,siskins2019highly}. 
It is much more difficult to determine modal stiffness and mode-shapes in the presence of these tension uniformities or material anisotropies and building a better understanding of these effects is an active topic of research. For this purpose mode shapes need to be measured, using methods~\cite{Davidovikj2016} discussed in Sec.~\ref{sec:modeshapereadout}, and shown in Fig.~\ref{fig:mechanical}(a--c). Here, the nonuniform tension in the resonator, 
causes the modeshape of the second mode to become asymmetric (Figs.~\ref{fig:mechanical}(b) and (c)). Since all mode shapes are affected by this nonuniform tension, deviations in all other resonance frequencies of the membrane are present as well (Fig.~\ref{fig:mechanical}(d)).

\begin{figure*}[ht]
    \centering
    \includegraphics{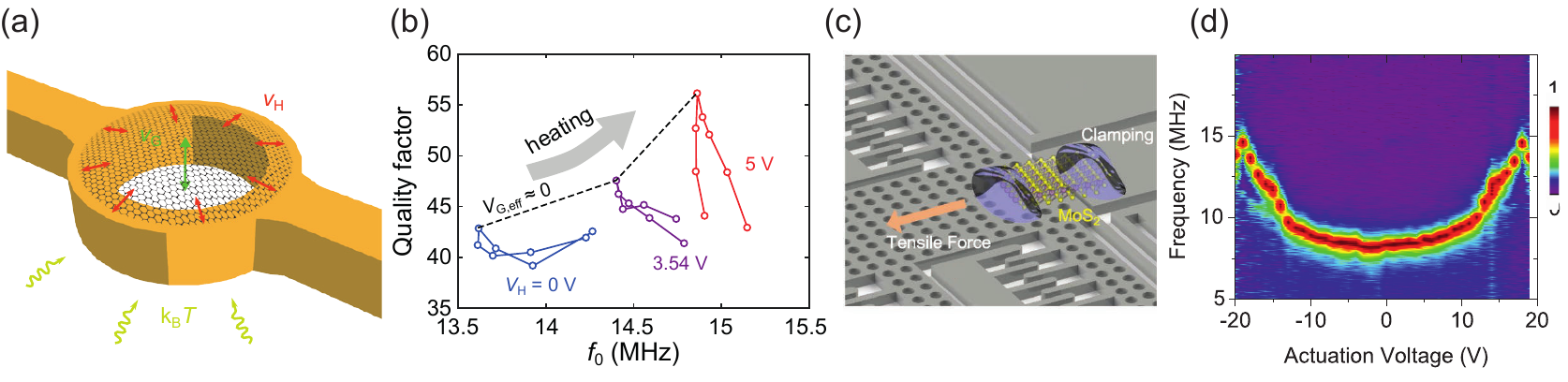}
    \caption{Engineering tension in 2D material resonators. (a) Schematic of a heater structure used to tension 2D material membranes. (b) Frequency versus quality factor plot at different heater voltages. (c) Schematic of a comb-drive actuator to tension 2D material membranes. (d) Resonance frequency tuning of a 3-layer MoS$_2$ resonator using a comb drive actuator. Figs. (a,b) are reprinted from Ref. \onlinecite{DavidovikjHeaters}  licensed under CC-BY-NC-ND. Figs. (c,d) are reprinted with permission from Ref. \onlinecite{xie2020straining}, copyright (2020) Wiley‐VCH GmbH.
    }
    \label{fig:tension}
\end{figure*}

A step beyond characterizing tension and its effect on the modal stiffness and resulting dynamics, is the ability to tune tension to change the dynamics of 2D material membranes. One strategy is to apply an out-of-plane force, by electrostatic actuation or gas pressure~\cite{nicholl2015, Nicholl2017}. An alternative method is to adjust the stress in the in-plane direction, which has the advantage that it maintains a flat membrane configuration. To this end, thermal heater substrates have been developed that consist of a metal ring on which a graphene membrane is suspended~\cite{DavidovikjHeaters}. Due to the positive thermal expansion of the substrate and the negative thermal expansion coefficient of graphene, tension is induced in the membrane when heating it by passing currents through the ring so that the membrane flattens (Fig.~\ref{fig:tension}(a)-(b)). Another more invasive approach to thermal tuning is to pass a large current through a suspended graphene membrane and heat up the membrane directly by Joule heating~\cite{ye2018electrothermally}. This leads to a nonuniform temperature and tension and only works for conductive materials, but does allow for much higher temperatures and therefore larger tuning range of resonance frequency. Similarly, cooling down the material changes the tension, which can be used to study material properties like thermal expansion (see Sec.~\ref{sec:texpansion}). Even more accurate control over tension is obtained using MEMS actuators (Fig.~\ref{fig:tension}(c)-(d)), which can tune 2D material resonance frequencies over a range of more that 10\%~\cite{verbiest2020tunable, xie2020straining}.

\subsubsection{Quality factor and dissipation}
\label{sec:Q}
\begin{figure*}[ht]
    \centering
    \includegraphics{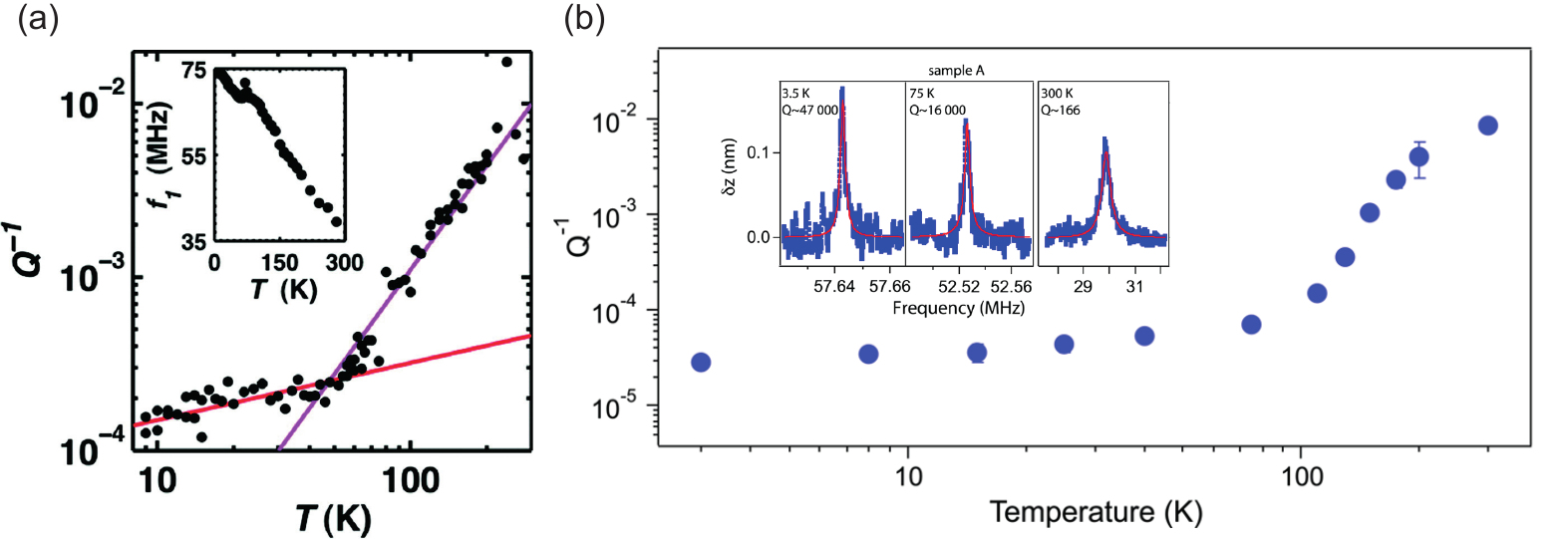}
    \caption{(a) Quality factor as a function of temperature for a single layer graphene resonator and (b) for a WSe$_2$ resonator. Fig. (a) is reprinted with permission from Ref. \onlinecite{Zande2012}, copyright (2010) American Chemical Society. Fig. (b) is reprinted with permission from Ref. \onlinecite{Morell2016}, copyright (2016) American Chemical Society.}
    \label{fig:Qfactor}
\end{figure*}
From experiments, the quality factor of a resonance mode $Q_i$, that is closely related to the damping coefficient $c_i=\sqrt{k_i m_i}/Q_i$, can be determined in a straightforward manner either from a frequency response fit by Eq.~(\ref{eq:FRF}) or from ring-down measurements~\cite{ronaldvanleeuwen2017,guttinger2017energy}. In a ring-down experiment the resonator is driven at resonance, after which the driving is stopped and the slow decay of the envelope of $q_i(t)$ is fit using Eq.~(\ref{eq:qringdown}). In particular when the $Q$-factor is very high, the ring-down measurement has the advantage that it is less sensitive to frequency drifts and fluctuations that can lead to spectral broadening~\cite{Barnard2012}. However, although the total losses can be determined from the measured $Q_i$, it is much more difficult to determine the microscopic mechanism that causes these losses, because of the large number of mechanisms that can contribute to damping.

The first measurements of single-layer graphene resonators showed quality factors of $Q=50-100$ at room temperature~\cite{Bunch2007}. In this work, the authors also hinted at the fact that the quality factor increases at lower temperatures, but the first systematic study of the temperature dependence of $Q$ was done by Chen \textit{et al.}~\cite{Chen2009} a few years later, where the authors observed an increase in $Q$ by a factor of more than 200 when cooling the resonators down to 5~K. Moreover, the trend of dissipation decrease was found to consist of two regimes: a $T^3$ dependence above 100~K and a $T^{0.3-0.4}$ dependence below this temperature. This trend is also observed in resonators from CVD graphene (see Fig. \ref{fig:Qfactor}(a)) \cite{Zande2012}. 

These measurements sparked two questions that led to a large number of studies in the years to follow: i) Why are the quality factors of graphene resonators so low at room temperature compared to, e.g. their diamond NEMS (nanoelectromechanical systems) counterparts? ii) Why does the quality factor increase so drastically with decreasing temperature and what constitutes the two distinct regimes? Interestingly, there was one known nanomechanical system where similar trends were observed: suspended carbon nanotubes~\cite{Verbridge2006high, Huttel2009carbon,sazonova2004tunable2}. It turned out later, however, that these observations were not limited to carbon-based NEMS, but were rather a characteristic of van-der-Waals nanomechanical resonators regardless of their chemical composition. $Q$-factors of the same magnitude have been observed in resonators made of transition-metal dichalcogenides (TMDC) (\ce{MoS2}~\cite{Castellanos2013, matis2017energy}, \ce{WSe2}~\cite{Morell2016} (Fig. \ref{fig:Qfactor}(b)), \ce{TaSe2}~\cite{Cartamil2015}, \ce{TaS2}~\cite{vsivskins2020magnetic}), \ce{hBN}~\cite{Zheng2017hexagonal,Cartamil2017}, b-P~\cite{Wang2015}, \ce{MPS3} antiferromagnets~\cite{vsivskins2020magnetic} (\ce{FePS3}, \ce{MnPS3}, \ce{NiPS3}) and even in other ultrathin materials, such as membranes of coordination polymers~\cite{lopez2018isoreticular} and complex oxides~\cite{davidovikj2019ultrathin}. The temperature dependence of the Q-factor of some of these resonators has also been measured, and has consistently shown a similar trend as graphene~\cite{Morell2016, matis2017energy,Cartamil2017,davidovikj2019ultrathin, vsivskins2020magnetic} (Fig.~\ref{fig:Qfactor}). 

An in-depth theoretical discussion on the different damping mechanisms in graphene can be found in Seo\'anez \textit{et al.}  ~ \cite{seoanez2007dissipation}. In the following we will outline the most relevant mechanisms that can limit the $Q$-factor and their temperature dependencies.

Studies have shown that membrane diameter and pre-tension are two parameters that are strongly correlated with the $Q$ factor of 2D resonators~\cite{barton2011high, Zande2012}.  In SiN membrane resonators, it is well-known that tension increases $Q$ by a mechanism called dissipation dilution~\cite{Ghadimi2018} and similar models were found to apply in 2D materials~\cite{Cartamil2015, Zalalutdinov2012}. Tension increase might also partly account for the increase in $Q$ with decreasing temperatures~\cite{Singh2010,vsivskins2020magnetic}. When the temperature decreases, the tension in the resonator increases, flattening the membrane and this may well be the main source of decreasing dissipation when lowering temperature. Tensioning the drum in a similar fashion at room temperature is challenging, as using a backgate deforms the resonator out-of-plane and introduces a strong electric field, which is known to deteriorate the $Q$ factor (Sec.~\ref{sec:elstinteractions})). Efforts have been made to establish in-plane tensioning at room/elevated temperatures using a piezo crystal~\cite{kramer2015strain} or a heater ring~\cite{DavidovikjHeaters} which resulted in a small increase of the $Q$ factor.

Besides tension and geometry, there is strong evidence that mechanical bending losses are important for determination of the $Q$-factor. The dynamic modulus of a material at a certain frequency can phenomenologically be represented~\cite{unterreithmeier2010damping, Zalalutdinov2012} by a complex number $E=E_1 + i E_2$, where $E_1$ and $E_2$ are the storage and loss modulus and their ratio is called the loss tangent $\tan \delta=\frac{E_2}{E_1}$. The effect of a nonzero loss modulus on the $Q$ factor of a resonator can be assessed by using\cite{Steeneken2007comsol} from Eq. (\ref{eq:qringdown}) that $Q_i=\frac{\Re \omega_i}{2\Im \omega_i}$ and substituting $E=E_1 + i E_2$ in Eq.~(\ref{eq:freqplatemem}) for the resonance frequency $\omega_i$. Since $\omega_{\rm mem,i}$ for a circular membrane (Eq.~(\ref{eq:resfreq})) does not depend on the elastic modulus, theoretically a perfectly ideal membrane has zero loss (and infinite $Q$), independent of the loss modulus of the material. The reason for this is that for strings and membranes in the linear regime the potential energy during resonance is stored in the direction change in the tensile force (from in-plane to out-of-plane) instead of in changes in the material's stress and strain. However, this is not true for a bending circular plate, since its resonance frequencies $\omega_{\rm plate,i}$ depend on $E$ (see Eq.~(\ref{eq:freqplate})) and the material  of the plate is experiencing a time variant strain during resonance. When both bending and membrane tension are taken into account (Sec.~\ref{sec:plateundertension}) to a good approximation~\cite{Castellanos2013},  $\omega_i^2=\omega_{\rm mem,i}^2+\omega^2_{\rm plate,i}$ and we find, for $E_2 \ll E_1$, for the fundamental mode an estimated $Q$-factor of:

\begin{equation}
\label{eq:Qdilution}
    Q_i = \frac{\Re \omega_i}{2\Im \omega_i} \approx \left( \frac{|\omega_{\rm mem,i}|^2}{|\omega_{\rm plate,i}|^2} + 1\right) \frac{E_1}{E_2}.
\end{equation}

This equation shows first of all that by increasing the tension in the material, the membrane resonance frequency $\omega_{\rm mem,i}$ increases and this results in a larger $Q$-factor, thus providing an illustration of the dissipation dilution mechanism~\cite{Ghadimi2018}. This dissipation dilution mechanism hinges on storing energy in 'lossless' membrane or string modes, thus reducing the relative contribution of other loss mechanisms to the $Q$-factor, which can also be defined as the ratio of 2$\pi$ times the stored energy divided by the energy loss per cycle. Secondly, Eq.~(\ref{eq:Qdilution}) shows that the $Q$-factor can also be increased by minimizing the material's loss modulus $E_2$ to reduce bending losses. Finally, reducing the thickness $h$ and increasing the radius $R$ of the membrane increases the ratio $|\omega_{\rm mem,i}|^2/|\omega_{\rm plate,i}|^2\propto \frac{R^2}{h^3}$, which increases $Q_i$.

A difficult question is: what mechanism does determine the loss modulus $E_2$ in 2D materials? A number of mechanisms is known to cause anelastic relaxation losses in solids\cite{nowick2012anelastic}, most of which might play a role in 2D materials. Information about the underlying mechanism can to some extent be obtained by measurements as a function of frequency and temperature. By testing the bulk material using dynamic mechanical analysis, $E_2$ can be estimated, although it might be different when thinned down to atomic thickness and can also be frequency dependent. 

An important anelastic loss mechanism affecting $E_2$ is thermoelastic damping~\cite{Zener1937} where the periodic compression and expansion of the material causes spatial temperature variations. When heat flows to equilibriate these temperature variations, the mechanical energy is lost. The degree of internal friction depends both on the thermoelastic properties of the material~\cite{jiang2014mos} and on the geometry of the resonator. The thermoelastic damping loss is proportional to the product of temperature and heat capacitance $c_v$ and a factor that depends on the geometry, mechanical frequency and thermal diffusivity. The recently observed large changes in loss close to phase transitions in 2D materials~\cite{vsivskins2020magnetic}, that approximately follow the trend $\frac{1}{Q} \propto c_v T$, support the idea that thermodynamic effects play a role in the $Q$-factor of 2D materials by demonstrating that large changes in specific heat are accompanied by large changes in $Q$.

Instead of damping by converting mechanical energy to heat, resonance losses can also occur when acoustic waves leave the membrane, transporting energy from the membrane resonator to the substrate via the edge of the drum, an effect that is called acoustic radiation loss, which is known to play a role in other NEMS and MEMS resonators~\cite{Steeneken2007comsol}. At the edge of the drum the 2D material often makes a kink due to edge adhesion. Acoustic waves might travel across this kink from the suspended to the unsuspended part of the 2D material~\cite{Dolleman2020B}, contributing to such acoustic radiation losses. In some cases the energy does not leave the resonance mode via the clamping points, but is transferred to other resonance modes of the same resonator via mode-coupling, which can lead to linear damping~\cite{midtvedt2014} but also to nonlinear damping as will be discussed in section~\ref{nonlineardynamics}.

Another mechanism related to the edge of the membrane, is adhesion loss, due to the repetitive adhesion and delamination of the material during resonance. This mechanism is usually dismissed in literature as a loss mechanism for 2D resonators because the energy density of a vibrating membrane at the edges is not sufficient to break the hydrogen bonds created between silanol groups (SiOH) at the interface and the exfoliated flake~\cite{seoanez2007dissipation}. The effect of the edge can be studied via the effect of membrane geometry on the Q-factor~\cite{Miller2017}.
 
Besides the aforementioned mechanisms, also the role of wrinkles, contamination and defects on the $Q$-factor cannot be ruled out, and a large reduction of $Q$ might be expected when a polymer contamination layer goes through the glass-transition. In general, more research is needed to obtain a full understanding of the loss mechanisms in 2D resonators and their temperature, material, frequency and geometry dependence.

\begin{figure*}[ht]
	\centering
	\includegraphics[scale=1]{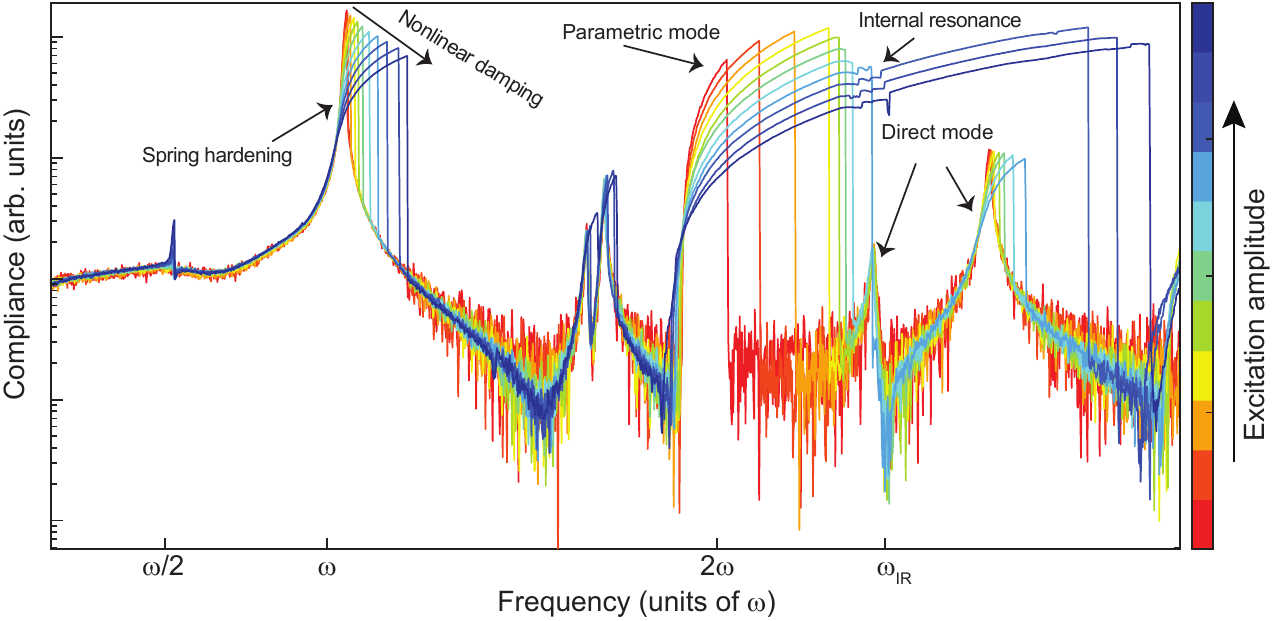}
	\caption{Measured nonlinear dynamic frequency response curves of a multi-layer graphene nanodrum (10~nm thick and 5~$\mu$m in diameter); from the red to the blue curves the drive level increases (color bar on the right hand side).  In a single set of measurements the Duffing response, nonlinear damping, parametric resonance and internal resonance (mode-coupling) are observed, by simultaneous application of both direct and parametric drive. The linear $x$-axis indicates the detection frequency in terms of the fundamental resonance frequency $\omega=\omega_1$, the logarithmic $y$-axis shows the measured mechanical compliance $|{\rm {FRF}(\omega)}|$. At a frequency $\omega_{\rm{IR}}$ an internal resonance between a direct and parametric mode is observed. Adapted from Ref.~\cite{kecskekler2020enhanced} licensed under CC BY 4.0.}
	\label{fig:nonlineardynamics}
\end{figure*}

\section{Nonlinear dynamics of 2D membranes: motion beyond the linear regime}
\label{nonlineardynamics}

Nonlinear dynamic effects are of paramount importance in 2D material resonators, since they precipitate already at excitation forces of only a few pN, and shrink the dynamic range over which the resonator's response is linear to span less than 2 orders of magnitudes (see Fig.~\ref{fig:introduction}). 2D material membranes exhibit a plethora of nonlinear dynamic phenomena that cannot be easily obtained in other mechanical systems. Many of these complex nonlinear dynamical phenomena can be present in the same device as shown in Fig.~\ref{fig:nonlineardynamics} for a multi-layer graphene nanodrum optothermally driven into resonance. In the first part of the following section, we will discuss how these complex frequency response curves arise from the nonlinear force terms of $F_{\rm nl,i}$ and $F_{\rm ext,i}$ in Eq.~(\ref{eq:motion}). Then we will discuss the various physical effects from which these force terms originate: geometric nonlinearity, nonlinear actuation forces, nonlinear damping, and mode coupling.

\subsection{Nonlinear dynamic phenomena}
\label{sec:nonlsoln} 

\subsubsection{Dynamics in the presence of nonlinear stiffness and damping}
\label{sec:nonlstiffness}
The most well-known nonlinear equation of motion in nonlinear structural dynamics  is the Duffing equation, which contains a nonlinear stiffness term of the form $F_{nl,i}=\gamma q_i^3$ resulting in this equation of motion: 

\begin{equation}\label{eq:Duffing}
    m_i \ddot{q_i} + c_i \dot{q_i} + k_i q_i +\gamma q_i^3= F_{\rm ext,i0} \cos(\omega t- \phi)+\tilde{F}.
\end{equation}
We will assume that the constant force term $\tilde{F}$ is zero for now. As a consequence of the Duffing term, the stiffness effectively becomes amplitude dependent, as can be seen by writing the total restoring force as $F_{k,i}=(k_i + \gamma q_i^2) q_i$. From this equation it follows that for positive $\gamma > 0$ the time averaged modal stiffness $k_{\rm eff,i}=k_i + \gamma \langle q_i^2\rangle $ will on average be larger than $k_i$. This effect is called spring hardening and leads to an increase in the resonance frequency $\omega_i \approx \sqrt{k_{\rm eff,i}/m_i}$ with increased amplitude. For $\gamma < 0$ the opposite effect, spring softening, occurs, resulting in a decrease in the resonance frequency at higher amplitudes.

\begin{figure}[ht]
    \centering
    \includegraphics[scale=1]{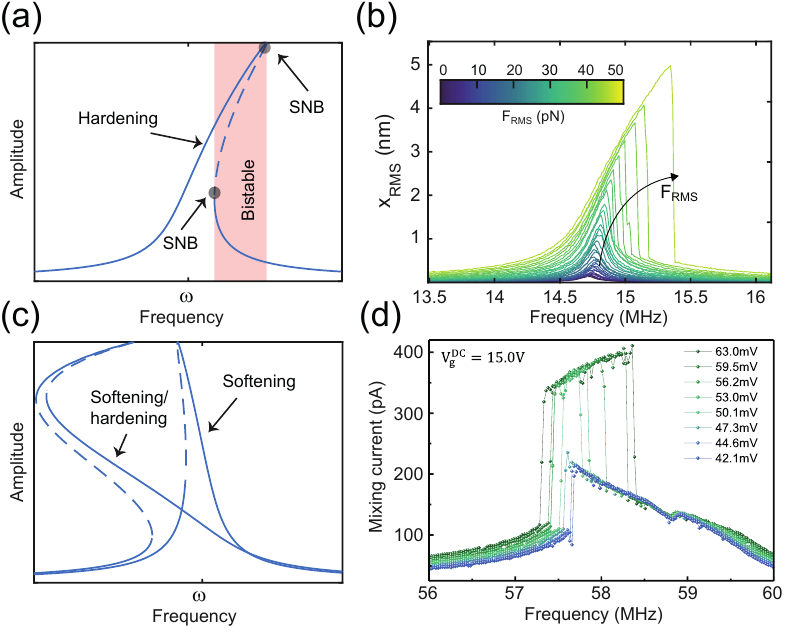}
    \caption{(a) Duffing hardening nonlinear response and (c) softening nonlinear response. The  response turns to hardening at large amplitude oscillations. The solid line in the two panels indicates stable solutions and dashed line is the unstable solution. SNB in the upper panel stands for Saddle-Node Bifurcation point. (b) Example of a Duffing hardening response measured in a graphene resonator. (d) Example of a measured softening response turning into hardening at large amplitudes of a graphene resonator.
    Fig. (b) is reprinted from Ref.~\onlinecite{Farbod2017} licensed under CC BY 4.0. Fig. (d) is reprinted with permission from Ref.~\onlinecite{Samanta2018}, copyright (2018) AIP Publishing LLC. 
    }
 \label{fig:hardening}
\end{figure}

Equation~(\ref{eq:Duffing}) can be solved analytically by approximating the solution by a function of the form  $q_i(t)\approx q_{i0} \sin \omega t$. By substituting this approximate solution in Eq.~(\ref{eq:Duffing}), balancing the fundamental harmonics ($\sin \omega t$, $\cos \omega t$) on both sides, and discarding higher-order harmonics (e.g. $\sin 3 \omega t$, $ \cos 3 \omega t$), the frequency response function reads: 
\begin{equation}\label{eq:nonlinear FRF}
    \left|\frac{q_{i0}}{F_0}\right|=\frac{1}{\sqrt {(k_i- m_i \omega^2+\frac{3 \gamma q_{i0}^2}{4})^2+ (\frac{\omega_i m_i \omega}{Q_i})^2}}.
\end{equation}
This function is plotted in Fig.~\ref{fig:hardening}(a). Comparing Eqs.~(\ref{eq:FRF}) and (\ref{eq:nonlinear FRF}), we note the presence of the Duffing constant $\gamma$ in the denominator of Eq.~(\ref{eq:nonlinear FRF}) that breaks the symmetry of the Lorentzian-like peak around $\omega=\omega_i$ and bends the frequency response curve. The bending direction of the response curve depends on the sign of $\gamma$ as discussed above. 

In an experiment, when the driving frequency $\omega$ is swept upward approaching a resonance frequency, spring hardening causes the amplitude to increase and the resonance frequency to shift to a higher frequency (Figs.~\ref{fig:hardening}(a--b)). This process continues until the driving frequency exceeds the resonance frequency. Beyond this point, the amplitude suddenly jumps to a lower value, which also causes the resonance frequency to reduce ($\omega_i \approx \sqrt{(k_i + \gamma \langle q_i^2\rangle )/m_i}$). When sweeping the drive frequency downward, a hysteresis cycle is formed since $\langle q_i^2\rangle $ remains small in the bi-stable region (Fig.~\ref{fig:hardening}(a)) until an upward jump towards higher amplitudes occurs near $\omega=\sqrt{k_i/m_i}$ . Both of these jumps occur at so-called saddle-node-bifurcation points (SNB) at which the system becomes unstable because the frequency response function (FRF) curve has an infinitely steep slope. At driving frequencies between the two jump frequencies there is a bi-stable region, that contains two stable solutions, with high and low amplitude, and an unstable solution that is indicated by a dashed line in Fig~\ref{fig:hardening}(a). This typical Duffing response has often been observed~\cite{Bunch2007,Chen2009,Chen2013,Farbod2017} in experimental studies of 2D material membranes at motion amplitudes of only a few nanometers. 

In the presence of externally applied stochastic forces, but also of intrinsic sources, like thermo-mechanical noise (Sec.~\ref{sec:brownian}), the bi-stable regime in a nonlinear resonator can enable stochastic switching between the two stable states (solutions) of the resonator. The noise might then help to amplify weak signals via a phenomenon called stochastic resonance~\cite{gammaitoni1998stochastic}. An advantage of graphene nanodrums is that they can achieve stochastic switching rates as high as a few kHz near room temperature~\cite{dolleman2019high}, hundred times faster than in conventional silicon resonators~\cite{venstra2013stochastic} at effective temperatures that are 3000 times lower, which could prove beneficial for enabling fast sensors based on stochastic resonance. 

It is also interesting to note that when a 2D material  membrane is deformed by a constant distributed pressure $p$ or {\it dc} electrostatic force, the Duffing term  shifts its resonance frequency~\cite{Banafsheh2017,Lee2019}.  To analyze this effect, one needs to calculate the resonance frequency of the membrane about the new equilibrium position induced by the constant pressure. For this purpose, the generalized coordinate $q_i$ is split into a static $q_{is}$ and a dynamic $q_{id}$ component: $q_i=q_{is}+q_{id}(t)$, with $|q_{id}| \ll |q_{is}|$. Inserting this expression in Eq.~(\ref{eq:Duffing}) gives:
\begin{equation} \label{eq:static}
    k_i q_{is}+\gamma_i q_{is}^3=\tilde{F},
\end{equation}
from which the static deflection $q_{is}$ due to $\tilde{F}$ can be calculated.  Once $q_{is}$ is known, the Duffing equation becomes:
\begin{multline}\label{eq:linearized}
   m_i \ddot{q_{id}} + c_i \dot{q_{id}} + k_i q_{id} + 3 \gamma_i q_{is}^2q_{id}+ 3 \gamma_i q_{is} q_{id}^2 \\ + \gamma_i q_{id}^3= F_{i0} \cos(\omega t- \phi).
\end{multline}
It can be seen from the linear terms in $q_{id}$ in Eq.~(\ref{eq:linearized}) that the resonance frequency about the deflected position becomes: 
\begin{equation}\label{eq:omegashift}
\omega_i=\sqrt{\frac{k_i+3\gamma_i q_{is}^2}{m_i}}.    
\end{equation}

Besides the extra linear term in Eq.~(\ref{eq:linearized}), a nonlinear term quadratic in $q_{id}$ arises due to $\tilde{F}$. This quadratic term is a consequence of the direction of the static force $\tilde{F}$ which breaks the symmetry~\cite{eichler2013symmetry} of the system, and can significantly alter the dynamics. The presence of quadratic nonlinearities can lead to a combined softening-hardening response~\cite{Samanta2018}, like the ones shown in Fig.~\ref{fig:hardening}(c--d).  Similar effects can occur when a Duffing system has an initial offset $q_{is}$ from the flat position due to other reasons than a static external force $\tilde{F}$. Such initial offsets can for instance arise from edge adhesion and uneven tension (wrinkling) during the fabrication process. 

\begin{figure}[ht]
    \centering
    \includegraphics[scale=0.8]{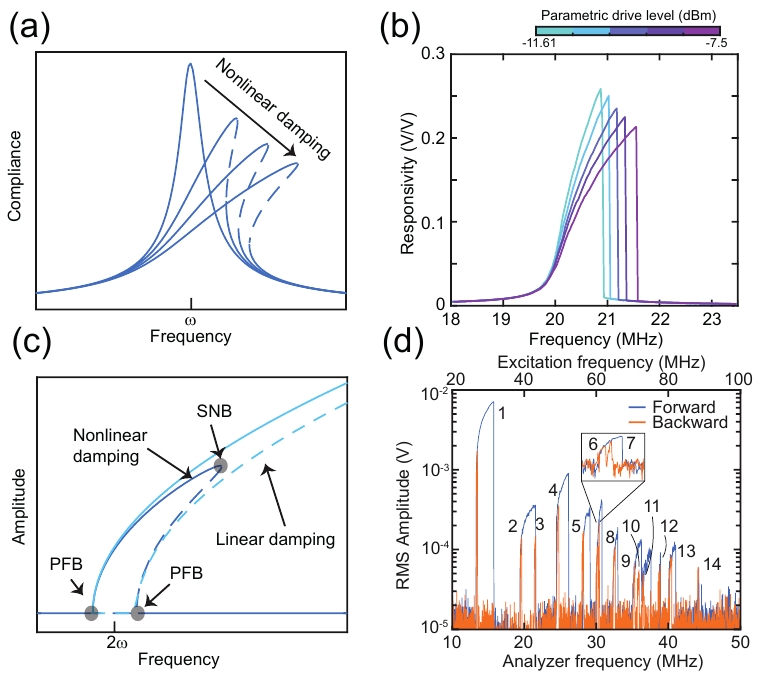}
    \caption{ (a) Effect of positive nonlinear damping on the Duffing response for increasing driving force in the direction of the arrow, resulting in a reduction in peak compliance ($|q_i/F_i|$). (b) Measured compliance as a function of driving power for a multilayer graphene resonator, showing the effect of positive nonlinear damping. (c) Parametric resonance in the presence and absence of nonlinear damping. SNB stands for Saddle-Node Bifurcation and PFB for Pitch-Fork Bifurcation. The solid lines in (a) and (c) indicate stable solutions and dashed lines the unstable solutions. (d) Measurement of parametric resonance in multiple mechanical modes in an opto-thermally excited single-layer graphene resonator. Fig. (b) is reprinted from Ref.~\cite{kecskekler2020enhanced} licensed under CC BY 4.0. Fig. (d) is reprinted from Ref.~\cite{Dolleman2018} licensed under CC BY 4.0.}
    \label{fig:parametric}
\end{figure}

The damping force in Eq.~(\ref{eq:motion}) can also be nonlinear, e.g. when  $F_{nl,i}=\eta_i q_i^2 \dot{q}_i$ with $\eta_i>0$, such that the effective dissipation coefficient $c_{{\rm eff},i}=c_i +\eta_i \langle q_i^2\rangle $ increases with an increase in the amplitude $q_i$. In practice, nonlinear damping can result from nonlinear stiffness, when a Duffing force term is combined with viscous delay, which can e.g. arise in materials where $\gamma_i$ is proportional to the complex Young's modulus $E=E_1+i E_2$ (see also Eq.~(\ref{eq:visco}) and section~\ref{sec:Q}).
For high-$Q$ resonators, this nonlinear damping term $\eta_i$, which in general can be both positive and negative, changes the compliance and speed of the resonator at high driving forces and amplitudes. As a result, by increasing the driving force, the compliance ($|q_i/F_{\rm ext,i0}|$) exhibits trends like the ones shown in Fig.~\ref{fig:nonlineardynamics} and Figs.~\ref{fig:parametric}(a--b), where with the increase in driving force, the Duffing resonance peak frequency increases due to spring hardening, while the peak amplitude decreases due to nonlinear damping.
Nonlinear damping can also strongly influence parametric resonance, which will be discussed in the next section. 

Before proceeding, we note that although analytical solutions can be found for several exemplary cases, in general obtaining them for nonlinear equations of motion is complicated, such that numerical techniques are often needed. A common technique in solving these equations is the use of numerical continuation schemes~\cite{nayfeh2008applied} that systematically trace the solutions of a system of differential equations in a space consisting of the state variables (displacements and velocities) and the parameter(s) (e.g. driving force, amplitude, driving frequency) over which continuation is performed and bifurcation points are identified. Several software tools are available for performing numerical continuation and bifurcation analysis including AUTO~\cite{doedel1997continuation} and Matcont~\cite{dhooge2003matcont}.   

\subsubsection{Parametric resonance}
\label{sec:parametricres}

The parametric forcing term $F_{{\rm p},i}(t)=-k_{{\rm p},i}(t) q_i$ (Sec.~\ref{parametricact}) in Eq.~(\ref{eq:motion}), can lead to interesting dynamics and provides an alternative to direct actuation via $F_{\rm ext,i}$ for driving and amplification of motion; it can even result in sustained oscillation or squeezing of noise. We consider the case that the parametric drive and resonator motion has a sinusoidal form $k_{{\rm p},i}(t)=k_{{\rm p},i0} \sin \omega_{{\rm p},i} t$ and $q_{i}(t)=q_{i,0} \sin \omega t$. Then, the parametric driving force $F_{{\rm p},i}(t)=-k_{{\rm p},i}q_i$ contains frequency components at $ |\omega_{{\rm p},i}\pm \omega|$. Therefore, when $\omega_{{\rm p},i} \approx 2 \omega_i$, it is found that $F_{{\rm p},i} $ contains a frequency contribution at $\omega_i$ that drives the motion of mode $i$ if the mode moves at its resonance frequency: $q_{i}(t)\approx q_{i,0} \sin \omega_i t$.

Parametric driving is often compared to a child on a swing, that periodically changes the length (and thus stiffness) of the swing. To maximize the kinetic and potential energy gain, the swing length should be maximum in the middle position and minimum at the end points. Since the swing passes the minimum position twice during one period of oscillation, the length should be modulated at twice the resonance frequency to amplify the amplitude of the swing. If the period of varying the stiffness is increased by an integer factor $n$, the energy gain mechanism is still synchronized with the motion (although less efficient), such that for lower frequencies $\omega_{{\rm p},i} \approx 2 \omega_i/n$ with $n=1,2,\dots$ being a positive integer, the system can still be driven parametrically~\cite{turner1998five, Seshia2016}.

The basic model that describes the nonlinear dynamics of membranes in the presence of parametric drive and a Duffing term is the Mathieu-Duffing equation:

\begin{equation}\label{eq:nonlinear-Mathieu}
    m_i \ddot{q_i} + c_i \dot{q_1} + (k_i+k_{{\rm p},i0} \cos(\omega_p t-\phi)) q_1 +\gamma q_1^3= 0.
\end{equation}
It can be shown from this equation, that at $\omega_p=2\omega_i$ the energy added per cycle to the system by parametric drive is $E_{\rm add}=\frac{\pi k_{{\rm p},i0} E_{\rm stored}}{k_i}$, where $E_{\rm stored}$ is the stored energy. If this added energy exceeds the dissipated energy per cycle ($E_{\rm diss}=\frac{2 \pi E_{\rm stored}}{Q_i}$), then for $\frac{k_{{\rm p},i0}}{k_i} > \frac{2}{Q_i}$ the energy supplied to the system will keep increasing indefinitely, unless it is limited by nonlinear damping effects that reduce the effective $Q$-factor. This condition is called parametric resonance~\cite{nayfeh2008nonlinear} or parametric oscillation. Parametric resonance can also occur at frequencies that do not exactly obey $\omega_p=2\omega_i$, although the further the separation of the driving frequency from this condition, the higher the drive levels $k_{{\rm p},i0}$ must be to reach parametric resonance as can be depicted in parametric (in)stability maps~\cite{Dolleman2018} (see Eq.~(\ref{eq:parametric})).

To obtain the frequency response curve of the Mathieu-Duffing oscillator for the fundamental (or principal) parametric resonance  ($\omega_p\approx 2 \omega_i$) and highlight the difference with Eq.~(\ref{eq:Duffing}), we again assume the solution to be harmonic as a first approximation, but since the principal resonance occurs at $\omega_i \approx \frac{1}{2} \omega_p$, we change our assumed solution to $q_i \approx q_{i,0} \sin(\frac{1}{2}\omega_p t)$. By following an analysis similar to the one performed for the Duffing resonator, i.e., balancing the fundamental harmonics and discarding higher order ones after inserting the solution in Eq.~(\ref{eq:nonlinear-Mathieu}), we obtain the trivial solution $q_{i,0}=0$ and the non-trivial solutions, of which 1 is stable and the other unstable:
\begin{equation} \label{eq:parametric} 
    q^2_{i,0}=\frac{4 k_i}{3 \gamma} \left(\frac{\omega_p^2}{4 \omega_i^2}-1\pm \frac{\omega^2_i}{2} \sqrt{\frac{k_{{\rm p},i0}^2}{k_i^2}- \left(\frac{\omega_p}{\omega_i Q_i}\right)^2}\right).
\end{equation}
    
A number of solutions of this equation at different driving forces are shown in Fig.~\ref{fig:parametric}(c). Equation~(\ref{eq:parametric}) also shows that parametric resonance at $\omega_p= 2 \omega_i$ exists only if $k_{{\rm p},i0}> \frac{2 k_i}{Q_i}$.  At parametric resonance, the solution branches with $q_i=0$ lose stability through the so-called pitch-fork bifurcations (PFB). These bifurcation points are the points where the trivial ($q_i=0$) and non-trivial solutions meet in the frequency amplitude response (see Fig.~\ref{fig:parametric}(c). Parametric resonances can be observed at different driving frequencies in the same structure, either for different values~\cite{turner1998five, Seshia2016} of the integer $n$ with $\omega_{{\rm p},i} \approx 2 \omega_i/n$, or when different eigenmodes of the structure are excited parametrically via the same modulation parameter, such as shown in Fig.~\ref{fig:parametric}(d)~\cite{Dolleman2018}.
2D materials can exhibit record number of parametric resonances when the parametric drive $k_{{\rm p},i0}$ is gradually increased by opto-thermal tension modulation~\cite{Dolleman2018}; a phenomenon that has not been observed in macro-mechanical systems due to the high dissipation and the difficulty of modulating the stiffness.

Equation~(\ref{eq:parametric}) predicts that the amplitude of the resonator will tend to infinity when sweeping the driving frequency $\omega_p$ upward. However, in experiments it usually drops down to the low-amplitude solution, after having followed the solution (Eq.~(\ref{eq:parametric})) up to a certain point. Nonlinear damping can account for this drop in amplitude, since it effectively causes a decrease of $Q_i$ with increasing amplitude until the relation $k_{{\rm p},i0}> \frac{2 k_i}{Q_i}$ does not hold anymore as shown in Fig.~\ref{fig:parametric}(d). Parametric resonance can therefore be used as a very sensitive technique for characterizing nonlinear damping~\cite{Dolleman2018, kecskekler2020enhanced}.

At driving levels that are not sufficient for parametric oscillation, $k_{{\rm p},i0}< \frac{2 k_i}{Q_i}$, the parametric term in Eq.~(\ref{eq:motion}) can result in parametric gain (amplification) and noise squeezing~\cite{rugar1991mechanical, bothner2020cavity}. Parametric gain is the enhancement of a sinusoidal direct driving force $F_{\rm ext,i}$ by supplementing it with a parametric drive. The parametric gain factor $G=|q_{{\rm p}i}/q_i|$ is the ratio between the amplitude of the resonator in the presence of parametric drive ($|q_{{\rm p}i}|$) and in the absence of parametric drive ($|q_i|$), and depends on the phase difference between the parametric and direct forcing terms. For certain phases the parametric gain is larger than 1, but for other phases the gain is reduced below 1, with a minimum gain of $G=0.5$, such that it reduces the resonator's amplitude ($G<1$). If the signal $F_{\rm ext,i}$ is due to noise, with random phase, adding a parametric drive will therefore reduce this noise for certain phases and amplify it for other phases. This effect (for $G<1$) is called noise squeezing~\cite{Singh2018} and can be useful, when for example the position of a resonator needs to be stabilized. It should be noted that since velocity is 90 degrees out of phase with position, the phase dependent gain caused by a parametric drive will either amplify velocity while reducing position amplitude, or vice-versa, so it is impossible to squeeze both position and velocity noise simultaneously.

\subsubsection{Mode-coupling and internal resonance}

Since the eigenmodes are orthogonal, there are no linear coupling terms between the equations with different mode number \textit{i} in Eq.~(\ref{eq:motion}) and therefore mode coupling can only occur via nonlinear terms in the EOM~\cite{venstra2010}. In particular, mode coupling is caused by forces that are generated on one mode by the motion of another and corresponds to the terms in $F_{nl,i}$ in Eq.~(\ref{eq:motion}) that involve products of the generalized coordinates, like $q_iq_j$, with $j \neq i$. Up to cubic order $F_{nl,i}$ can thus be written as~\cite{Muravyov2003, amabili2008nonlinear}:
\begin{equation} \label{eq:fnl}
    F_{nl,i}=\sum_{j=1}^N \sum_{k=j}^N \alpha ^{(i)}_{jk} q_j q_k + \sum_{j=1}^N \sum_{k=j}^N \sum_{l=k}^N \gamma^{(i)}_{jkl} q_j q_k q_l ,
\end{equation}
where $N$ is the number of degrees-of-freedom, and the coefficients $\alpha^{(i)}_{jk}$ and $\gamma^{(i)}_{jkl}$ are the quadratic and  cubic coupling terms that depend on the geometry, elasticity, and curvature of the membrane, but can also originate from e.g. electrostatic or optical forces, resulting in mode coupling between 2D materials and SiN membranes~\cite{singh2019giant} or cavity modes~\cite{DeAlba2016}, as will be discussed later. Besides the nonlinear stiffness mode coupling terms in Eq.~(\ref{eq:fnl}), there is a similar set of nonlinear damping mode coupling terms like $F_{nl,i}=\phi^{(i)}_{jk} q_i \dot{q}_j$, which has received little attention up to now. 
Mode coupling effects can be studied by driving both modes near their resonance frequency simultaneously, or by exciting one mode and study its ringdown~\cite{leeuwen2014, guttinger2017energy} while monitoring the effects of interactions with other modes. However, most commonly these coupling effects are studied~\cite{van_der_Avoort_2010, kecskekler2020enhanced, Samanta2015, Mathew2016, nathamgari2019nonlinear} by driving one mode $j$ directly or parametrically with an external sinusoidal force $F_{\rm ext,j}=F_{\rm ext,j0} \sin (\omega t)$ while tuning its resonance frequency until interactions with other modes become noticeable. These interactions can be understood as follows: at sufficiently high driving force and when driving the mode close to its resonance frequency, the motion $q_j(t)=q_{j,0} \sin \omega_j t$ reaches large amplitudes, and as a consequence, mode $i$ experiences a mode-coupling force, e.g. of the form $\alpha^{(i)}_{ij} q_i q_{j,0} \sin{(\omega_j t)} $. Interestingly, it is seen that when $\omega_j=2 \omega_i$, the force has the same form of a stiffness modulated at $2 \omega_i$ and thus resembles parametric driving (Sec.~\ref{sec:parametricres}), but since it is caused by the resonator itself instead of an external modulation, it is called auto-parametric excitation~\cite{van_der_Avoort_2010}. More generally, this kind of resonant excitation of a mode by driving another mode at resonance is called internal resonance, and can occur not only at $\omega_i/\omega_j=1/2$ but also at other ratios, where the condition $\omega_i/\omega_j=n/m$ is also called a (n:m) internal resonance~\cite{nayfeh2008nonlinear}.

It is interesting to note that for intermodal coupling, the quadratic and cubic nonlinear restoring force terms of Eq.~(\ref{eq:fnl}) have contributions that depend on the type of internal resonance that exists between the modes. For instance, in case of two-to-one (2:1) internal resonance~\cite{kecskekler2020enhanced} the dominant coupling terms in two-degree-of-freedom system are: $F_{nl,1}= \alpha_{12}^{(1)} q_1 q_2$ and $F_{nl,2}=\alpha_{11}^{(2)} q_1^2$, and for three-to-one~\cite{guttinger2017energy} (3:1)  
$F_{nl,1}= \gamma_{112}^{(1)} q_1^2 q_2$ and $F_{nl,2}=\gamma_{111}^{(2)} q_1^3 $. This is a consequence of mixing since e.g. driving at $\omega_2 = 3 \omega_1$ is generated by the $q_1^3$ term and driving at $\omega_1 = \omega_2-2 \omega_1$ is generated by the $q_1^2 q_2$ term. 

When two mechanical modes of the resonator interact in the vicinity of internal resonance, part of the mechanical energy is transferred to vibrations at other frequencies, which can be seen as a kind of nonlinear damping~\cite{midtvedt2014,guttinger2017energy, kecskekler2020enhanced}, since in an uncoupled (orthogonal) undamped system of eigenmodes, each mode is expected to maintain its energy forever, whereas coupling can lead to a distribution (equipartition) of energy over all modes. It has therefore been suggested that the nonlinear terms in the equation of motion can play a role in this equipartition~\cite{midtvedt2014, dauxois2005fermi}. An experimental example of this type of damping by mode coupling, which will be discussed in more detail in section~\ref{sec:nldamping} can be seen in Fig.~\ref{fig:nonlineardynamics}, where near frequency $\omega_{\rm{IR}}$, the peak amplitude of the parametric mode reduces.

\subsection{Physical origin of nonlinear effects}
\label{sec:orignonl}

After having discussed the different types of phenomena that can be caused by the nonlinearities in the equation of motion, we now discuss the physics and mechanics from which these nonlinear terms originate. 
We can distinguish two types of origins for the nonlinear coefficients in Eq.~(\ref{eq:motion}), the mechanical nonlinearities in $F_{nl,i}$ and the nonlinearities induced by the actuation terms $F_{\rm ext,i}$ or $k_{{\rm p},i}$. Both material nonlinearities and geometric nonlinearities can contribute to the mechanical nonlinearities. For capturing the nonlinearities in actuation, the physics behind the actuation mechanism needs to be analysed. Finally, the output signal can be distorted by nonlinearities in the readout, however, since these nonlinearities do not affect the dynamics of the device, they can be calibrated and corrected for in a rather straightforward manner~\cite{DollemanCalibration} and will not be discussed in detail here.

\subsubsection{Mechanically induced nonlinear terms} 
\label{sec:geomnonl}
Geometric nonlinearity is a primary cause of the nonlinear dynamic response of free-standing 2D materials. Its effect on a suspended 2D membrane under tensile stress can be intuitively understood by considering, as an example, a straight string at zero tension, that is deflected in the middle by a distance $x$ from its equilibrium position, thus forming a triangular shape. The length change of the string increases proportionally to $x^2$ for small $x$ according to Pythagoras' theorem, and the tension force in the string increases proportional to this length increase. Moreover, after multiplication by another factor $x$, that is obtained by taking the tension force component perpendicular to the string, which is approximately proportional to the small angle by which the string is deflected, it is found that the geometric nonlinear force required to deflect the string is a cubic function of the center deflection $x$, i.e. $F_{\rm geom,nl} \propto x^3$. As a consequence of the geometric nonlinearity, the tension in the string increases, and therefore the system exhibits spring hardening such that its dynamics is described by the Duffing equation with $\gamma>0$ (Sec.~\ref{sec:nonlstiffness}).

Geometric nonlinear terms are often difficult to compute analytically, however numerical~\cite{Farbod2017} and finite element model~\cite{Muravyov2003} based methodologies have been developed to determine the coefficients in Eq.~(\ref{eq:fnl}). Some of these methodologies are discussed in appendix \ref{AppendixC}. Once calculated, a fit of the nonlinear frequency response function, can provide high-frequency information about device and material properties, such as its Young's modulus~\cite{Farbod2017} and  loss tangent~\cite{Dolleman2018} that cannot be obtained with quasi-static techniques like atomic force microscopy. An interesting question is if these methodologies might be used to study differences between the static and dynamic material properties. Geometric nonlinearities can also result in nonlinear damping in a viscous material that will be discussed in section~\ref{sec:nldamping}. In the presence of large rotations, inertia nonlinearities in the mass term of the equation of motion might also play a role~\cite{villanueva2013nonlinearity}, however, in membranes this effect is usually negligible.
Besides geometric nonlinearities, also material nonlinearities, caused by nonlinearities in the stress-strain curve of the material, can induce nonlinear terms in the equation of motion. Up to now, no experimental evidence of the observation of effects of material nonlinearities in 2D materials on their nonlinear dynamics has been reported to our knowledge.

\subsubsection{Actuation induced nonlinear terms}
\label{sec:actnonlin}
All types of actuation forces $F_{\rm ext,i}$, including electrostatic forces~\cite{DeAlba2016, Banafsheh2017} and opto-thermal forces~\cite{Samer2017,Dolleman2018,dolleman2019high, inoue2017resonance}, are nonlinear to a certain degree. In general the effect of nonlinear actuation terms can be analyzed by performing a series expansion of the force around the equilibrium position. As an example one can expand the electrostatic actuation force discussed in section~\ref{sec:elstact}, Eq.~(\ref{eq:elpressure2}) around $w=q_i=0$ for $q_i\ll 1$ to obtain: 
\begin{equation}\label{eq:nonlinear-taylor}
\begin{split}
    F_{\rm ext}= \frac{\varepsilon_0 A}{2} \left(V_{dc}+V_{ac} \cos(\omega t)\right)^2 \\
    \left(\frac{1}{g^2}+\frac{2 q_i}{g^3}+\frac{3 q_i^2}{g^4}+\frac{4 q_i^3}{g^5} \right).
\end{split}
\end{equation}
From this equation it can be seen that the different powers of $q_i$ in the actuation force add to the different mechanical linear and nonlinear terms on the left side of the equation of motion, Eq.~(\ref{eq:motion}). In this case the term proportional to $2 q_i/g$ is a softening term that reduces the linear stiffness, and the terms $3 q_i^2 /g^4$ and $4 q_i^3/g^5$ are quadratic and cubic spring softening terms that can be used for tuning nonlinear stiffness of 2D materials~\cite{Samanta2018,dash2019optical}. In addition to these nonlinear terms, it is noted that the static actuation term $1/g^2$, in combination with the geometric nonlinear terms, can induce nonlinearities as discussed in Eq.~(\ref{eq:static}). For other types of actuation forces a similar analysis of nonlinearities via actuation terms can be performed. 

Tuning of the nonlinear terms in the mechanical equation of motion via the actuation force can be useful. For instance, it was shown that by adjusting the cavity depth $g$ and the {\it dc} bias voltage, the electrostatic softening by the electrostatic force can compensate for the hardening geometric nonlinearity and thus increase the dynamic range over which 2D material resonators operate linearly~\cite{Parmar2015}. On the other hand, the nonlinear terms in the actuation force do complicate the analysis and control over the intrinsic nonlinearities in 2D material resonators.

\subsubsection{Nonlinear damping terms} 
\label{sec:nldamping}
As has been shown in Fig.~\ref{fig:nonlineardynamics} and discussed in section~\ref{sec:parametricres}, signatures of nonlinear damping have been observed in frequency response and ringdown measurements of 2D material resonators~\cite{Eichler2011, guttinger2017energy, Dolleman2018,kecskekler2020enhanced}. Figures \ref{fig:parametric}(a),(b) show how nonlinear damping manifests itself in direct and parametrically driven resonators, and how it can be modelled by a term of the form $F_{nl,i}=\eta_i q^2_i \dot{q}_i$ in which $\eta_i$ is the the nonlinear damping coefficient. 

Although the presence of nonlinear damping has been detected in 2D materials, its origin is still a subject of debate,  and different physical phenomena  are believed to lie at its root including a combination of geometric nonlinearity and viscoelasticity~\cite{Dolleman2018,amabili2019}, coupling between the flextural modes and the in-plane phonons~\cite{Croy2012}, Akhiezer damping~\cite{atalaya2016nonlinear}, and nonlinear mode coupling~\cite{midtvedt2014, guttinger2017energy,shoshani2017, kecskekler2020enhanced}, of which we will discuss the first and last in more detail. 

An anelastic (viscous) material can be described by a complex Young's modulus $E=E_1+i E_2$ (see section \ref{sec:Q}). Since the geomtrically induced Duffing term $\gamma$ of 2D material membranes explicitly depends on the Young's modulus of the material (see Eq.~(\ref{eq:coeffs}c)), the real part of the nonlinear stiffness near resonance for harmonic motion $q_i=q_0 \exp {i \omega t}$ becomes~\cite{Dolleman2018,amabili2016nonlinear}:
\begin{equation} \label{eq:visco}
\Re \gamma q_i^3= \frac{\pi h}{f(\nu) R^2} (E_1 q_i^3+E_2 \frac{q_i^2 \dot q_i}{\omega}),
\end{equation}
with $f(\nu)=(1.27-0.97 \nu-0.27 \nu^2)$, which directly yields a nonlinear damping term in the equation of motion proportional to the loss modulus $E_2$ . Although this type of damping in viscous membranes in the presence of geometric nonlinearities is thought to be important in 2D materials, to our knowledge Eq.~(\ref{eq:visco}) has not been tested experimentally.

Nonlinear damping via mode-coupling~\cite{midtvedt2014, guttinger2017energy, kecskekler2020enhanced} is based on the notion that energy can leave the driven mode via the coupling terms in Eq.~(\ref{eq:fnl}). This occurs in particular near internal resonances where the driven mode actuates another mode auto-parametrically, as is supported by experimental evidence~\cite{guttinger2017energy, kecskekler2020enhanced}. Strictly, the energy is not lost because it is not converted into heat, but in mechanical motion of another mode. Under certain circumstances the energy can return into the driven mode, an effect found in the work of Fermi, Pasta and Ulam~\cite{berman2005fermi}. However, practically, the energy usually does not return, such that the coupling contributes to nonlinear damping of the main mode at high amplitudes during driven motion~\cite{kecskekler2020enhanced} or during ringdown~\cite{guttinger2017energy}, which might even be tuned by adjusting the driving amplitude.
Nonlinear damping terms might also contribute to our understanding of linear damping~\cite{midtvedt2014}, heat transfer and equipartition between coupled modes. For instance, for two mechanical modes with generalized coordinates $q_1$ and $q_2$ and modal stiffnesses $k_1$ and $k_2$, nonlinear damping forces $F_{d12}=-\eta_{12} q_2^2 \dot{q_1}$ can arise, resembling Eq.~(\ref{eq:visco}). At a finite temperature $T$ equipartition holds, such that the motion of mode 2 is given by $\langle q_2^2\rangle =k_{\rm B} T/k_2$. This could result in a temperature dependent average damping force on mode 1 $F_{d12,av}=-\eta_{12} \langle k_{\rm B} T/k_2\rangle \dot{q_1}$ .

\section{Physical Interactions}
\label{sec:physint}
The dynamics of suspended 2D materials is affected both by the internal membrane physics and by external processes that affect the membrane's surface and edges. As such, the membrane's motion is not only sensitive to the internal thermal, electric and magnetic processes in the membrane, but also to external forces from gases, liquids, charge and electromagnetic fields. 
A deeper understanding of the effects of these physical interactions on the mechanical motion and internal and external membrane processes can provide new methods for material characterization of atomically thin membranes, while at the same time offering a platform for enabling novel and improved environmental and force sensing applications, as has recently been reviewed~\cite{lemme2020nanoelectromechanical}. 

In this section we review the effect of both the internal and external physical interactions on the dynamics of 2D materials. Some of the basic mechanisms have already been introduced in earlier sections, since these interactions also provide routes for actuating and detecting the motion of the membranes. We now provide a more in-depth discussion of the underlying physics, the possibilities for using interactions for material characterisation, and their importance for sensing external forces from the environment.

\subsection{Thermodynamic and electromagnetic interactions}
\label{thermalcharacterization}
 
Heat in 2D materials can be stored in phononic lattice vibrations, as well as in electronic and magnetic degrees of freedom. There are not many techniques available to characterize these thermodynamic properties in suspended ultrathin membranes. In this subsection we will present several methodologies that utilize the coupling between membrane dynamics and thermodynamics for characterizing thermal and electromagnetic properties, that are difficult to obtain otherwise.

\subsubsection{Thermal characterization via opto-thermal response}
\label{sec:fdepthresp}
The high-frequency temperature modulation by a modulated laser does not only provide a route for contactless actuation of suspended 2D materials (Sec. ~ \ref{sec:thermalact}), but can also be exploited to learn more about their thermal properties. 
When heating a 2D membrane with a power-modulated laser at frequency $\omega$, the absorbed laser power, $\mathcal{P}(t)=\mathcal{P}_0 e^{i\omega t}$, causes its temperature $T(t)=T_0 e^{i\omega t}$ to be modulated at an amplitude $T_0$ which can be expressed in the frequency domain as $T_0=\frac{R_{\rm th} \mathcal{P}_0}{i\omega \tau_{\rm th} + 1}$. Here $R_{\mathrm{th}}$ and $C_{\mathrm{th}}$ are the membrane's thermal resistance and heat capacity respectively, and $\tau_{\rm th} = R_{\rm th}C_{\mathrm{th}}$ is the membrane's characteristic thermal time constant. The resulting effective thermal expansion force is $F_{\mathrm{ext}} = \alpha_{\rm eff} T$, where  $\alpha_{\rm eff}$ is an effective thermal expansion coefficient of the device (see Sec. \ref{sec:thermalact}). Its frequency dependence can be written as:
\begin{equation}\label{eq:Fextthermal1}
    F_{\mathrm{ext}} = \alpha_{\rm eff} R_{th} \mathcal{P}_0 \frac{1}{1+\omega^2\tau_{\rm th}^2} - \alpha_{\rm eff} R_{\rm th} \mathcal{P}_0 \frac{ i \omega \tau_{\rm th}}{1+\omega^2 \tau_{\rm th}^2}.
\end{equation}

\begin{figure}[ht]
    \centering
    \includegraphics{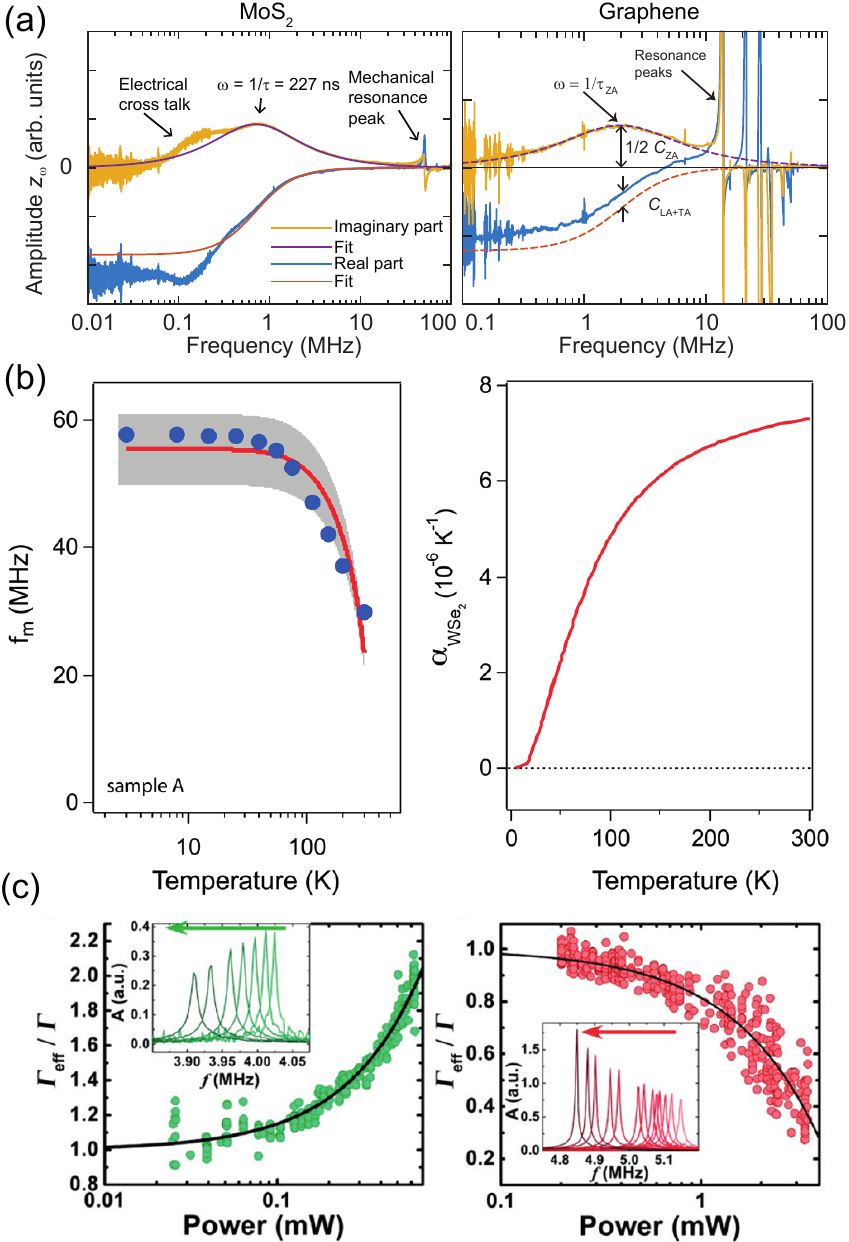}
    \caption{(a) Complex amplitude of the mechanical motion of MoS$_2$ and graphene as a function of the frequency of the photothermal heating. (b) Left figure: resonance frequency of a WSe$_2$ resonator as a function of temperature, compared to a theoretical prediction of the thermal expansion coefficient (red line and right figure). (c) Tuning of the linewidth of a single layer graphene resonator as a function of incident laser power. The left plot uses a laser with a wavelength of 568~nm, while the right plot used a wavelength of 633~nm. The sign of the linewidth change $\Delta \Gamma$ changes because the optical gradient $\nabla \mathcal{P}$ has a different sign for both wavelengths.
    Fig. (a) (left) reprinted with permission from Ref.~\onlinecite{DollemanMoS2}, copyright 2018 American Physical Society. Fig. (a) (right) reprinted from Ref.~\onlinecite{Dolleman2020A} licensed under CC BY 4.0. Fig. (b) is reprinted with permission from Ref.~\onlinecite{Morell2016}, copyright (2016) American Chemical Society. Fig. (c) is reprinted with permission from Ref.~\onlinecite{Barton2012}, copyright (2012) American Chemical Society.
    }
    \label{fig:thermalex}
\end{figure}

This frequency dependence of the real (in-phase) and imaginary (out-of phase) parts of the force $F_{\mathrm{ext,i}}$ can be characterised via the motion $q_i$, using the relation $q_{i} = F_{\mathrm{ext,i}}/k_i$, which holds far below the fundamental resonance frequency.
Equation~(\ref{eq:Fextthermal1}) matches experiments well and fits can be used to obtain the thermal time-constant $\tau_{\rm th}$ of the membrane~\cite{DollemanThermal, DollemanMoS2}. This is illustrated in Fig.~\ref{fig:thermalex}(a), which shows an example of the measured real and imaginary part of the frequency dependent amplitude in a suspended single-layer MoS$_2$ drum with a fit to Eq.~(\ref{eq:Fextthermal1}). The imaginary part shows an extremum at $\omega = 1/\tau_{\rm th}$ = 1/227 ns. By combining the thermal delay time $\tau_{\rm th}$ with calculated values of $C_{\rm th}$, it is possible to determine the thermal conductivity $1/R_{\rm th}$, which was found to be 24.7~W/mK for 8~$\mu$m diameter MoS$_2$ drums~\cite{DollemanMoS2}, in accordance with estimates from literature. Thus, dynamic characterisation of the thermal time constant of 2D membranes can be used as a contactless tool to probe their thermal properties.

Recently, experimental evidence~\cite{Dolleman2020A} was found for the theoretical hypothesis~\cite{Vallabhaneni2016} that for graphene Eq.~(\ref{eq:Fextthermal1}) is incomplete, because the out-of-plane flexural acoustic phonons are effectively decoupled from the in-plane (longitudinal and transverse) acoustic phonons. This causes the in-plane and out-of-plane phononic baths to have different thermal time constants. 
Since the time constants corresponding to in-plane phonons are estimated to be much smaller than that of the out-of-plane flexural phonons, the slow time-constant $\tau_{\rm th}$ measured in Ref.~\cite{DollemanThermal} is attributed to flexural phonons in combination with phonon boundary scattering effects at the boundary of the suspended drum~\cite{Dolleman2020B}. Evidence for heat transport by fast in-plane phonons, with time-constants below the experimental detection limit, was obtained from observations of an experimental offset (see right panel of Fig. \ref{fig:thermalex}(a)), in comparison to the real part of Eq.~(\ref{eq:Fextthermal1}).
 
 \subsubsection{On-resonance thermal characterisation}
An alternative methodology for determining the membrane's thermal properties is by characterizing the effect of the gradient in the laser intensity on the line width and frequency of the resonance peak. To do this, one exploits the optical gradient of the electric field in the Fabry-Perot cavity formed between the membrane and the substrate (Sec.~\ref{optreadout}), which causes a position dependent optical absorption $\mathcal{P}(q_i) = (\mathcal{P}_0+ \frac{d\mathcal{P}}{dq_i}q_i)$. Because the thermal expansion force is delayed with respect to the power absorption, part of the resulting gradient in the opto-thermal force $dF_{\rm ext}/dq_i \propto d\mathcal{P}/dq_i$ in Eq.~(\ref{eq:Fextthermal1}) will be in phase with the velocity $\dot{q}_{i}$, such that both the squared resonance frequency and linewidth change by respectively $\Delta \omega_i^2=-k_{\rm fb}/m_i$ and $\Delta \Gamma_i = -c_{\rm fb}/(2 \pi m_i)$ (see Sec.~\ref{sec:feedback}, Eqs.~(\ref{eq:omegatuning}), (\ref{eq:qtuning})). This effect was first observed on single-layer graphene resonators with optical readout of the motion~\cite{Barton2012}, as shown in Fig.~\ref{fig:thermalex}(c). The damping can become larger or smaller when changing the laser wavelength, which can be attributed to the sign of the optical gradient.  

By combining the effect of opto-thermal backaction on the resonance frequency and damping, the thermal time-constant can be determined~\cite{Morell2019} with this equation:
\begin{equation}
\label{eq:backaction3}
    \tau_{\rm th} = \frac{\Delta \Gamma_i}{\Delta \omega_i^2} = \frac{1}{\omega_i} \frac{\Im{d F_{\rm ext}/dq_i}}{\Re{dF_{\rm ext}/dq_i}}.
\end{equation}
The expression on the right side of the equation is valid more generally and can also be applied to extract electrical or gas permeation time constants (Secs.~\ref{sec:elstinteractions}, \ref{sec:gasint}, Fig.~\ref{fig:physicalinteractions}). The appealing feature of this equation is that the prefactors of Eq.~(\ref{eq:Fextthermal1}) drop out by taking the ratio. Since uncertainties increase when one has to distinguish very small shifts in line width or resonance frequency, which can occur according to Eq.~(\ref{eq:Fextthermal1})  and Eq.~(\ref{eq:backaction3}) if $\omega \tau_{\rm th} \gg 1$ or $\omega \tau_{\rm th} \ll 1$, this methodology works most accurately when $1/\tau_{\rm th}$ is of the same order of magnitude as the resonance frequency and not much smaller than the intrinsic line width of the resonance peak. 

Thus, there are two complementary methods for determining the thermal time constant: the fitting procedure outlined in Sec.~\ref{sec:fdepthresp} that works well at frequencies far below $\omega_i$, and Eq.~(\ref{eq:backaction3}) that complements this technique by providing the possibility to study the thermal time constant  at higher frequencies, for which $1/\tau_{\rm th}$ is of the order of the resonance frequency. 

\subsubsection{Thermal expansion characterisation}
\label{sec:texpansion}
Finally, besides the presented high-frequency temperature modulation techniques, it is also possible to learn about the thermal properties of the membrane material by characterizing changes in resonance frequency due to a static temperature change $\Delta T$. Due to thermal expansion, the tension in the membrane changes due to thermal strain $\epsilon$:
\begin{equation}\label{eq:tensiontemp}
    n(T) = \frac{E h}{1-\nu} \left[ \epsilon(T_0)-\int_{T_0}^T \alpha(T') dT'\right].
\end{equation}
Singh {\it et al.}~\cite{Singh2010} used this concept to measure the temperature dependent thermal expansion coefficient of graphene by tracking the resonance frequency as a function of the environmental temperature, utilizing Eq.~(\ref{eq:tensiontemp}) in combination with $\omega_{\rm mem,i} \propto \sqrt{n(T)}$. This methodology has also been applied to other 2D materials such as WSe$_2$ (Fig.~\ref{fig:thermalex}(b)) \cite{Morell2016}. When applying Eq.~(\ref{eq:tensiontemp}), corrections have to be made for the thermal expansion of the substrate. Vice-versa, after having performed an independent calibration measurement of $\omega_i(T)$, the resonance frequency can also be used to sense the membrane temperature, for instance if the membrane is locally heated by a laser with power $P$, under the assumption that differences in temperature distributions across the membrane can be neglected. The obtained temperature can be used to calculate the thermal resistance $R_{th} = \Delta T/P$, and then calculate the thermal conductivity from the geometry~\cite{Morell2019, islam2018anisotropic}.  
By determining $R_{th}$ with a static measurement or from literature estimates, and $\tau_{th}$ with a dynamic measurement, the heat capacity can be estimated using $C_{th} = \tau_{th}/R_{th}$, as was done~\cite{Morell2019} for MoSe$_2$ and  for~\cite{Dolleman2018} MoS$_2$.
As shown in the previous sections, the dynamics of suspended 2D materials coupled to an optical field that acts as a heat source, has proven itself as a versatile probe to study the thermal properties of these materials. In the future, these dynamical characterisation methods are expected to further contribute to an improved understanding of heat transport in ultra thin materials. 

\subsubsection{Coupling to electronic and magnetic phases}
\label{sec:stateofmatter}
\begin{figure}[ht]
    \centering
    \includegraphics{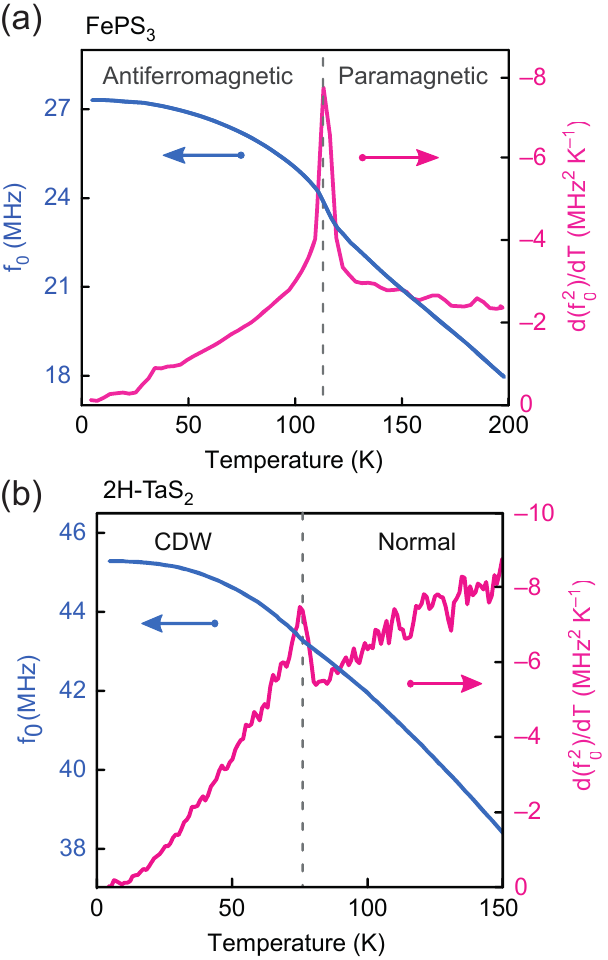}
    \caption{Measured resonance frequency as a function of temperature (solid blue line) and line—temperature derivative of $f_0^2$ (solid magenta line) of a (a) \ce{FePS3} and a (b) 2H-TaS$_2$ nanodrum showing prominent features at the phase transition of the materials. Reprinted from Ref.~\cite{vsivskins2020magnetic} licensed under CC BY 4.0.}
    \label{fig:phasetransitions}
\end{figure}

Interestingly, clear changes in the slope and curvature of the $\omega_i(T)$ curve were recently observed~\cite{vsivskins2020magnetic} when crossing the antiferromagnetic Ne\'el temperature in the thin layered 2D materials FePS$_3$, MnPS$_3$, and NiPS$_3$ (Fig.~\ref{fig:phasetransitions}(a)). These were, similarly as in the previous section, attributed to changes in the thermal expansion coefficient. In fact, the thermal expansion coefficient also holds information on a more fundamental material parameter, the specific heat $c_V$, which, according to thermodynamic models, is proportional to the thermal expansion coefficient and shows a discontinuity (jump) at second order phase transitions, that can account for the observed changes in $\omega_i(T)$ near the phase transition temperature. This demonstrates that the dynamics of 2D materials can be used to probe phase transitions which are difficult to study by conventional methods, because ultra thin materials do not respond strongly to conventional electronic and magnetic probes. 

To demonstrate the potential of the technique, in two studies~\cite{vsivskins2020magnetic,jiang2020exchange} a significant effect of strain on the magnetic order in suspended 2D materials was observed by applying electrostatic force on the membrane while monitoring its mechanical resonance frequency. Since anomalies in the specific heat are quite universal signatures of phase transitions, the methodology is not only applicable to probe magnetic order, but also to electronic phase transitions, as was demonstrated by probing the charge density wave transition\cite{lee2021study} in 2H-TaS$_2$ (Fig.~\ref{fig:phasetransitions}(b)). Furthermore, it was observed that the phase transition has a large effect on the quality factor, which was also attributed to the change in the specific heat and its effect on the energy dissipation mechanisms like thermo-elastic damping~\cite{vsivskins2020magnetic}. 

Similar to the previously discussed opto-thermal actuation, the actuation mechanism can also be used to probe electronic, magnetic and optical properties of 2D materials via other routes. For example, in the electronic band structure of 2D materials like MoS$_2$, the K and K' valleys form an electronic binary system, whose symmetry is broken if strong spin-orbit coupling is present. It was shown~\cite{li2019valley} that this effect can be used to actuate the vibration of single-layer MoS$_2$ in a magnetic field gradient induced by the substrate.

\subsection{External interactions}
In this section we discuss how external electrostatic fields and the gaseous environment of 2D material membranes can influence their dynamics, and in particular the $Q$-factor and resonance frequency.

\begin{figure}[ht]
    \centering
    \includegraphics{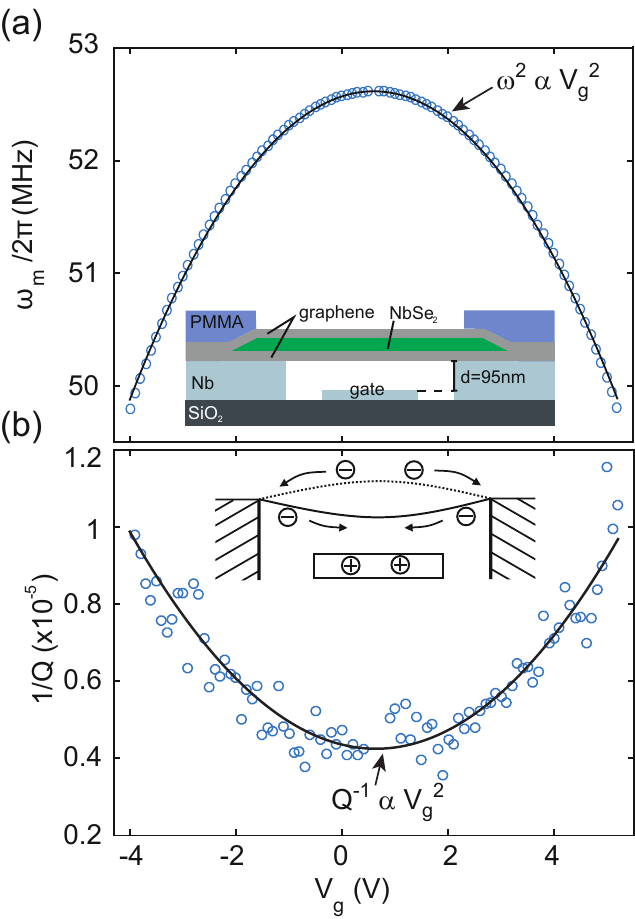}
    \caption{Effect of an external electric field on the resonance of suspended 2D materials. (a) Resonance frequency of a graphene/NbSe$_2$ heterostructure membrane as a function of gate voltage, showing a quadratic dependence as predicted by Eq.~\ref{eq:Ffit}. (b) Quality factor of the same membrane as (a) as a function of gate voltage, showing a quadratic dependence on it in agreement with Eq.~\ref{eq:qfactorcapacitance}. Reprinted with permission from Ref.~\onlinecite{will2017high}, copyright 2017 American Chemical Society.}
    \label{fig:electrostatic}
\end{figure}

\subsubsection{Electrostatic interactions}
\label{sec:elstinteractions}

Suspended 2D materials are often coupled to an electric field, especially when driven electrostatically (see Sec.~\ref{sec:elstact}). For a perfectly conducting membrane, the electrostatic force is in-phase with the applied voltage. However, in practice, 2D materials have a substantial resistance $R$ due to their low thickness, which in combination with the gate capacitance $C$ results in an electrical $RC$-circuit that causes the voltage and electrostatic force on the membrane to be delayed by a characteristic charge relaxation time $\tau_{el}=R C$, with respect to the externally applied voltage. 
Similar to the case of on-resonant thermal interaction, this leads to a time-delayed position dependent part of the electrostatic force with real and imaginary parts, that modify stiffness and damping terms~\cite{Song2012} in the equation of motion, respectively (Sec.~\ref{sec:feedback}). In the limit $1/\omega_i \ll \tau_{\rm el}$, which is applicable for a good electrical conductor such as graphene, the resonance frequency and quality factor can be written as:
\begin{equation}\label{eq:Ffit}
 \omega_i^2 = \omega_{\mathrm{i,int}}^2- \frac{ \varepsilon_0 V_{\mathrm{dc}}^2}{\rho h} \frac{1}{g^3},
\end{equation}
\begin{equation}\label{eq:qfactorcapacitance}
    Q_i^{-1} =
    Q^{-1}_{\mathrm{i,int}} + \frac{\varepsilon_0  V_{\mathrm{dc}}^2}{ \rho h \omega_i}  \frac{\tau_{\rm el}}{g^3}, 
\end{equation}
where $\omega_{\mathrm{i,int}}$ and $Q_{\mathrm{int}}$ are the intrinsic resonance frequency and quality factor when $V_{\mathrm{dc}} = 0$. Applying a {\it dc} voltage to the back gate thus lowers the resonance frequency (spring softening) and at the same time introduces additional damping and $Q$-factor reduction, an example of this can be seen in Figs.~\ref{fig:electrostatic}(a) and (b). This effect has been reported in several works~\cite{Morell2016,song2014graphene,Song2012,Weber2014,Weber2016,guttinger2017energy,will2017high} and the voltage dependence given by Eqs.~(\ref{eq:Ffit}) and (\ref{eq:qfactorcapacitance}) can be used to extract the mass and $\tau_{\rm el}$.

Equation~(\ref{eq:Ffit}) is obtained by assuming a constant capacitance, however, in certain situations this is not valid. The electrostatic force coupled to the membrane results in an effective stiffness $k_{\rm eff} = \frac{1}{2} V_{\rm DC}^2 \mathrm{d}^2C/\mathrm{d}z^2 $. In 2D materials, quantum capacitance effects can be important, which directly alters $k_{\rm eff}$ through the "$\mathrm{d}^2C/\mathrm{d}z^2$" term and this results in a resonance frequency shift. This effect has been modeled and measured in detail by \citeauthor{chen2016modulation} \cite{chen2016modulation}.

Besides the electrostatic softening in Eq.~(\ref{eq:Ffit}), electrostatic forces can also cause an electrostatic tensioning effect, where tension is proportional to the square of the force perpendicular to the membrane that causes a static deflection $q_{is}\propto V_{dc}^2$, and a tension $n\propto q_{is}^2$, such that approximately $\omega_i^2\propto n \propto V_{dc}^4$. The combination of this effect with electrostatic softening Eq.~(\ref{eq:Ffit}) leads to typical W-shaped or U-shaped graphs~\cite{Chen2009, KaiMingHu2020} when plotting the resonance frequency versus the applied gate voltage $V_{\rm dc}$, with the exact shape depending on parameters like pre-tension.

For applications in the optomechanics community one wants to apply high $V_{dc}$ to achieve high frequency tuning, without significantly reducing the quality factor. Therefore, according to Eqs.~(\ref{eq:Ffit}) and (\ref{eq:qfactorcapacitance}), it is of interest to minimize $\tau_{\rm el}$. The most successful approach towards this is to use heterostructures of 2D materials to improve the electron mobility and reduce resistive dissipation~\cite{will2017high}. 
 
\subsubsection{Interaction with gas molecules}
\label{sec:gasint}

When an unsealed graphene drum moves at low frequencies in a gas at an ambient pressure $P_a$, the gas molecules will enter and leave the gap region with height $g$, such that the pressure stays constant. However, this situation changes when the membrane moves at a high resonance frequency, because in that case viscous forces effectively prevent the gas from flowing back and forth during an oscillation period.
If the resonance frequency is much higher than the characteristic time ($\tau_{\rm gas}$) it takes the gas to escape the cavity, the gas is compressed by the motion of the membrane (the squeeze-film effect), which results in a stiffness and corresponding resonance frequency increase given by: 

\begin{equation}\label{eq:sqz_freqshift}
    \omega_{i,{\rm sq}}^2 = \omega_i^2 + \frac{P_{\rm a}}{g \rho h}
\end{equation}

\begin{figure*}[ht]
    \centering
    \includegraphics{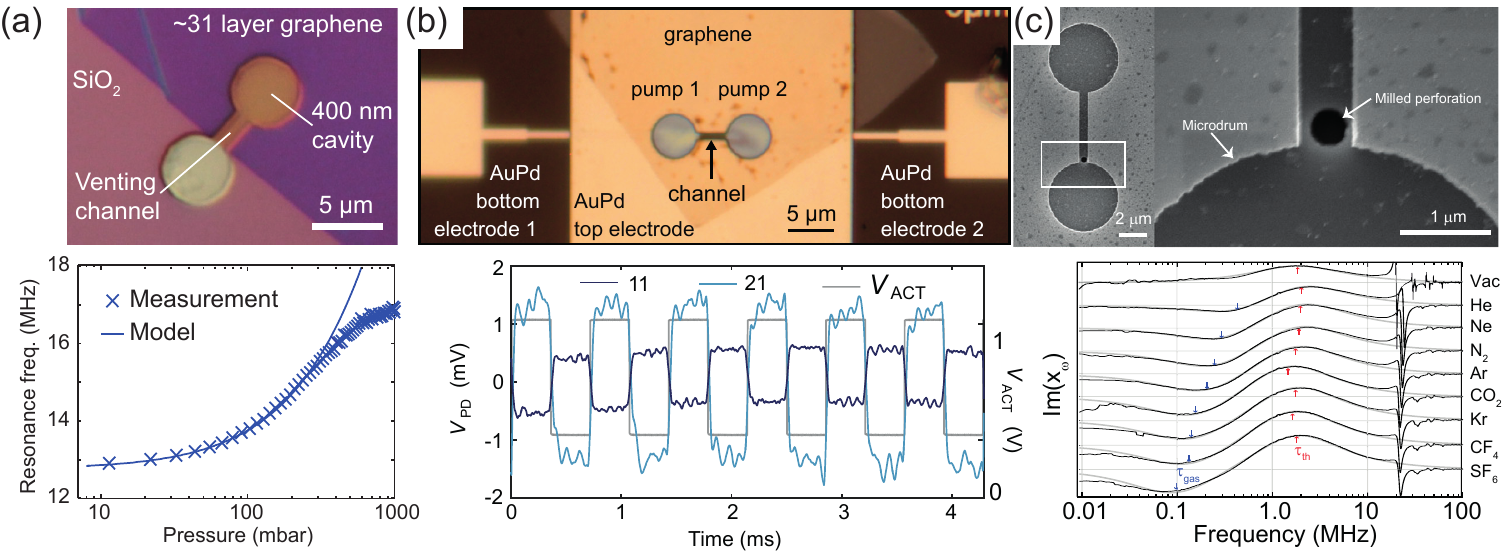}
    \caption{(a) Optical image of a 31 layer graphene squeeze-film pressure sensor and its measured pressure dependent resonance frequency~\cite{Dolleman2016A}. (b) Pneumatic coupling between two graphene nanodrums~\cite{DavidovikjIEEEpumps}. (c) Measurement of $\tau_{\mathrm{gas}}$ and $\tau_{\rm th}$ in a graphene drum with a milled nanopore~\cite{roslon2020graphene}.
    Fig. (a) reprinted with permission from Ref.~\cite{Dolleman2016A}, copyright 2015 American Chemical Society. Fig. (b) reprinted with permission from Ref.~\onlinecite{DavidovikjIEEEpumps}, copyright 2018 IEEE. Fig. (c) is reprinted from Ref.~\onlinecite{roslon2020graphene} licensed under CC BY 4.0.
    }
    \label{fig:squeezefilmanddamping}
\end{figure*}

The squeeze-film effect has been demonstrated using multi-layer graphene devices~\cite{Dolleman2016A} as illustrated in Figure~\ref{fig:squeezefilmanddamping}(a). The device consists of a dumbbell shaped cavity which is 400~nm deep, and a graphene flake (31 layers) that only covers half of this dumbbell. This half covering ensures the existence of a venting channel that keeps the average pressure in the cavity equal to that of the environment. The resonance frequency rises to higher values as the pressure is increased, as expected from Eq.~(\ref{eq:sqz_freqshift}), that is is fitted to the data using only the resonance frequency at vacuum $\omega_i$ as a fit parameter, and assuming that $\omega_i \tau_{\mathrm{gas}} \gg 1$. Excellent agreement with the experimental data is found up to 200 mbar, but a significantly lower stiffness is observed above this pressure, which is likely due to the breakdown of the assumption that gas does not escape above this pressure. 

Thin films of gas have also been used to pneumatically couple two graphene resonators by connecting them via a channel~\cite{DavidovikjPumps} (see Fig.~\ref{fig:squeezefilmanddamping}(b)). In this figure, the drum on the left is actuated, pushing gas through the channel and actuating the other drum on the right which responds with opposite sign. By observing the delay between the response of the opposite drum, the gas flow induced by these graphene pumps can be studied, making it an interesting platform to study gas flow and thermodynamics at these length scales. 

The pressure relaxation time, $\tau_{\mathrm{gas}}$, can also be directly measured by studying the vibration of a membrane with a nanopore perforation below the resonance frequency~\cite{roslon2020graphene}. 
Similar to the opto-thermal response, an extremum in the imaginary part of the complex amplitude is expected when $\omega \tau_{\mathrm{gas}} =1$, which can be seen in the measurement shown in Fig.~\ref{fig:squeezefilmanddamping}(c). The blue arrows indicate this extremum due to gas permeation, the frequency of which changes as different gases are brought in the environment. In addition, the red arrows indicate the extremum due to the thermal time constant discussed in section~\ref{sec:fdepthresp}), which occurs due to the opto-thermal actuation and does not change position when different gasses are used. The gas dependence of $\tau_{\mathrm{gas}}$ agrees well with a model for effusion through the milled nanopore, an effect that can potentially enable new types of gas sensors.    

\subsection{Single relaxation time models}
\begin{figure*}[ht]
    \centering
    \includegraphics[scale = 0.9]{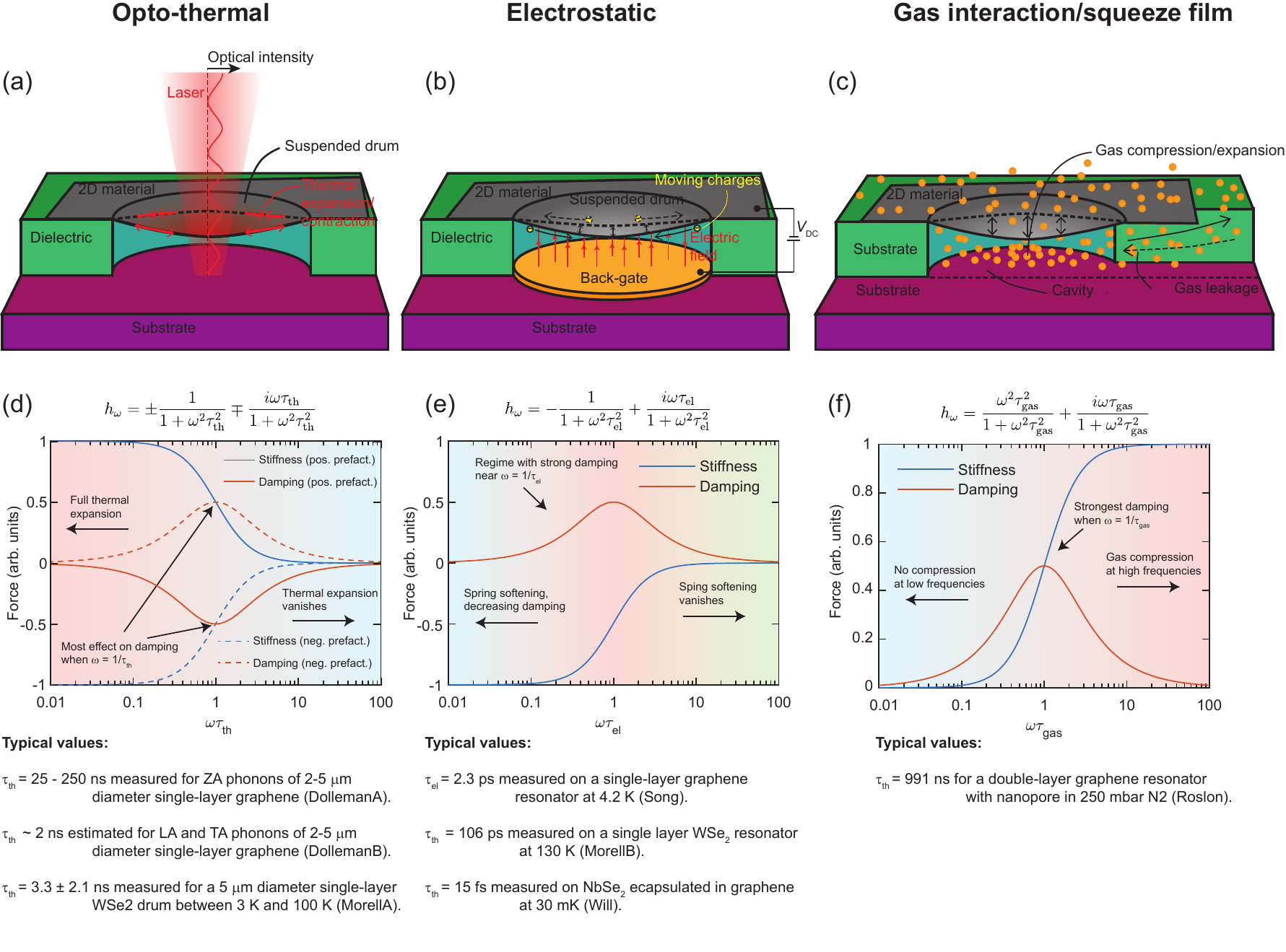}
    \caption{Overview of physical interactions that result in position and velocity-dependent forces on the membrane, discussed in detail in appendix \ref{AppendixD}. (a) Opto-thermal interaction. The position dependence of the force occurs due to the optical gradient, the thermal diffusion in the membrane causes a phase delay of the force that alters the damping. (b) Electrostatic interaction. Due to the capacitor configuration, charges are stored on the membrane, resulting in a stiffness force. The diffusion of these charges results in a phase shift of the force, resulting in damping. (c) Gas interaction, shown here is the squeeze-film effect where the membrane is suspended over a thin gap. Due to the membrane motion, gas is trapped resulting in a stiffness force. The lateral gas flow causes a phase delay that damps the motion of the membrane. (d) $h_{\omega}$ in the case of opto-thermal interaction and a plot of the real and imaginary part of $h_{\omega}$ as a function of $\omega \tau_{\rm th}$. Some typical values of $\tau_{\rm th}$ from literature (DollemanA~\cite{DollemanThermal,Dolleman2020A}, DollemanB~\cite{Dolleman2020B}, MorellA~\cite{Morell2019}) are cited below the figure. See also Figs. \ref{fig:thermalex}(a) and (c) for experimental results on opto-thermal interactions. (e) $h_{\omega}$ in the case of electrostatic interaction. Typical values from the literature (Song~\cite{Song2012}, MorellB~\cite{Morell2016}, Will~\cite{will2017high}) are calculated by taking the reported effective resistance and multiplying by the geometric capacitance. See also Figs. \ref{fig:electrostatic}(a--b) for experimental results on this interaction. (f) $h_{\omega}$ in the case of the squeeze-film effect, typical measured values (Roslon~\cite{roslon2020graphene}) can also be seen in Fig. \ref{fig:squeezefilmanddamping}(c). Note that either fits from Eq.~(\ref{eq:Fextthermal1}) or Eq.~(\ref{eq:backaction3}) are used to obtain the reported time constants.}
    \label{fig:physicalinteractions}
\end{figure*}

In subsections \ref{sec:fdepthresp}, \ref{sec:elstinteractions} and \ref{sec:gasint}, we have shown that the interactions of 2D material membranes with their environment can well be modelled by a single relaxation time model, with relaxation times like $\tau_{\rm th}$, $\tau_{\rm el}$, $\tau_{\rm gas}$. In this subsection we shortly highlight the commonalities between these model and underlying physics. A more extensive discussion of single relaxation time models can be found in appendix \ref{AppendixD}. In Figure~\ref{fig:physicalinteractions} we summarize the dynamics that is observed when 2D materials resonators interact with temperature, electrical $RC$-circuits and gases. In each case, the relaxation time emerges due to a delay between the membrane motion and the force action on the membrane. In the case of opto-thermal interaction this occurs due to heat diffusion, in the case of electrostatic interaction due to Joule dissipation and in the case of gas interaction due to friction.  Figures \ref{fig:physicalinteractions}(a--c) illustrate these mechanisms and the real and imaginary parts of the frequency dependent interaction forces and transfer functions $h_\omega$ (Fig. \ref{fig:physicalinteractions}(d--f)) that can be accurately measured by the dynamic motion of 2D material membranes using the methods described in this section. The similarities between these interactions can be found in their dependence on the frequency of the oscillation $\omega$. As shown in Figs. \ref{fig:physicalinteractions}(d--f), the effect of damping is the largest when $\omega = 1/\tau$ and furthermore a transition between stiffness regimes occurs at this frequency. In Fig. \ref{fig:physicalinteractions} we also cite typical values of the time constant $\tau$, this reveals the main difference between the different types of interaction as they occur on quite different timescales. Although single-relaxation time models are often only approximately valid, since multiple timescales can play a role, we note that in practice they provide quite a good model of the observed effects of interactions on membrane dynamics. Moreover the models can be extended to analyze more intricate and new types of physics, characterized via the motion of 2D material membranes.

\section{Discussion and Outlook}
Experimental investigations of the dynamics of 2D materials have progressed from phenomenological studies of the motion of 2D materials to studies that utilize the high-frequency motion of 2D materials as a probe for gaining a deeper and more quantitative understanding of the physics of this class of materials. 
 In the future, we expect that 2D membranes will continue to evolve as a tool to study the physics and properties of 2D materials. Let us summarize the areas where we anticipate that the dynamics of 2D membranes can contribute to advance our understanding of 2D materials. First, mechanical techniques can contribute to the study of electronic and magnetic phases of matter in 2D materials that are insulating or are difficult to couple to an electronic lead (for example due to Schottky barrier formation), such that electronic characterisation methods fail. Second, the methodologies presented in Sec.~\ref{sec:physint} can contribute to studies of transient heat transport which happens in the nanosecond range and is difficult to study by other means. Third, further study of electrostatically coupled 2D materials can help to understand what ultimately limits the ultrafast relaxation times of these systems. Finally, we expect that smart engineering of mechanical resonators from heterostructures of 2D materials, can enable deeper insight and control over interactions at interfaces, which can bring significant improvements in device performance and lead to new functionalities~\cite{lemme2020nanoelectromechanical}. 

We conclude the review with some open research questions, highlighting promising directions for the study of the dynamics of 2D materials.

\begin{itemize}

\item \textbf{Unravelling the Q-factor of 2D resonators.} Although many effects have been linked to dissipation in 2D resonators, a quantitative picture of the different components contributing to damping, and their temperature dependence is still missing. Clamping, surface, contamination, bending and thermo-elastic losses all limit the Q-factor of 2D materials, whereas dissipation dilution by high-tension increases Q. How can these individual factors and their dependence on geometry, contamination, material and temperature be distinguished and quantified?

\item \textbf{Understanding the nonlinear coefficients in the equation of motion.} The nonlinear (cubic) spring constant in the motion of 2D materials can be used to extract their Young's modulus. However, many of the nonlinear coupling and dissipation terms have not been quantified yet. Can these coefficients be extracted and be used to characterize geometric nonlinearities and material properties such as the specific heat, thermal expansion coefficient or viscoelasticity?

\item \textbf{Frequency-dependent material properties.} Many studies focusing on static properties of 2D materials have been carried out. Some of these properties, like the Young's modulus, can host a strong imaginary component which makes the measurement dependent on the frequency at which the materials are characterized. Are properties like Young’s modulus and loss tangent frequency dependent, and if so why?

\item \textbf{Dynamic characterisation of wrinkles.} Morphological imperfections have been haunting the field of 2D nanomechanics since its inception. 
Wrinkles affect almost all aspects of the dynamics of 2D materials, including their vibrational mode shapes, their resonance frequency and their thermal properties. Can methodologies be developed to detect and characterize static and dynamic wrinkles in 2D resonators, such that the relation between wrinkles and membrane dynamics can be unravelled?

\item \textbf{Measurement and control over phonons in suspended 2D membranes.} Although initial studies have revealed that phononic and thermal transport in 2D membranes is significantly different from that in bulk materials, is it possible to actuate and detect phonons in 2D materials more accurately to elucidate their role in sound and heat transport?

\item \textbf{Nonlinear dynamic sensing with 2D membranes.} The limited dynamic range of 2D material membranes puts a barrier on the sensing performance of these devices, since at low amplitudes their dynamics is affected by thermomechanical noise and at the upper end by nonlinearities (Fig. \ref{fig:introduction}(a)). Can 2D resonant sensors be operated in the nonlinear range effectively, and can their nonlinearities be used for increasing sensitivity?

\item \textbf{Extending 2D resonators as probes of condensed matter physics.} Initial evidence of the detection of phase transitions and thermodynamics has shown the potential of 2D resonators for probing non-mechanical effects. Could these techniques be improved in accuracy and scope to facilitate quantitative characterisation of material properties?

\end{itemize}

\section{Acknowledgements}
The authors acknowledge the fruitful discussions and collaborations they have had with colleagues on the topics outlined in this review. The research leading to these results has received funding from the European Union’s Horizon 2020 research and innovation programme under Grant Agreement Nos. 785219 and 881603 Graphene Flagship. F.A. acknowledges financial support from the European Union’s Horizon 2020 research and innovation programme under Grant Agreement 802093 (ERC starting grant ENIGMA) and 966720 (ERC PoC GRAPHFITI). R.J.D. acknowledges funding by the Deutsche Forschungsgemeinschaft (DFG, German Research Foundation) under grant agreement No. STA 1146/12-1. 
\newpage
\newpage
\appendix

\numberwithin{equation}{section}

\section{Appendix A: Dynamic regimes of motion}
\label{AppendixA}

In Fig.~\ref{fig:introduction}(a) the different ranges of motion for circular graphene membranes are shown, indicating that when scaling down the graphene membrane radius, the nonlinear and thermal motion become increasingly important. In this appendix we present the equations used to estimate the motion amplitudes that separate the different regimes of motion in Fig.~\ref{fig:introduction}(a). To determine the boundaries between the ranges, we use typical parameters for a graphene membrane: a constant pretension $n_0$=0.03 N/m, density $\rho = 2260$ kg/m$^3$, graphene thickness $h$ = 0.335 nm, Young's modulus $Y$=10$^{12}$ N/m$^2$ and Poisson's ratio $\nu$=0.16.

Starting from the smallest scale, the ultimate detection limit for motion detection by interferometric readout is~\cite{LaHaye74} the standard quantum limit\ $\langle z^2_{\rm SQL}\rangle^{1/2} = \sqrt{\hbar/(2 m \omega_i)}$ which, for a fundamental resonance frequency $f_1=\omega_1/2\pi$= 30 MHz, of a 5 micron diameter graphene membrane, gives an imprecision of $\langle z^2_{\rm SQL}\rangle^{1/2}$=0.3~pm, where we use that the fundamental mode has a modal mass~\cite{Hauer2013} $m_{1}$ of 0.27 times the total mass $\pi R^2 h \rho$ according to equation~(\ref{meff10}). 
Brownian, or thermal motion of the membrane corresponds, according to the equipartition theorem, to an energy  $k_{\rm B} T$ per mode and a root mean squared motion amplitude $\langle z^2_{\rm th}\rangle^{1/2}$ given by:
\begin{equation}
\label{Bmotion}
    \langle z^2_{\rm th}\rangle^{1/2} = \sqrt{k_{\rm B} T/(m_{i} \omega_i^2)}.
\end{equation}
At room temperature ($T=$293~K) and for a resonance frequency of $\omega_1/(2\pi)$=30~MHz, this results in a position standard deviation of  $ \langle z^2_{\rm th}\rangle^{1/2}$=0.17~nm, which is significantly higher than the quantum motion. Therefore quantum motion effects can be neglected at room temperature.

It is of interest to note that when scaling down the drum size, while keeping its mechanical pretension $n_0$ constant, the amplitude of the thermal Brownian fluctuations remains constant because the mass increases proportional to the square of the radius $R$, but the resonance frequency $\omega_1$ decreases inversely proportional to $R$ (Eq.~(\ref{eq:resfreq})), such that the product $m_{i} \omega_i^2$ in Eq.~(\ref{Bmotion}) stays constant.

Since nonlinearities exist at any membrane amplitude, there is not a sharp boundary between linear and nonlinear regimes. Here, we define the nonlinear regime as the amplitude region for which the nonlinear component of the membrane elastic force exceeds 10\% of the linear membrane force. To estimate this point, we utilize the equation~\cite{vinci1996mechanical} that relates the center deflection $w_c$ of a circular membrane to the applied uniform pressure $P$ and the related force $F_p=\pi R^2 P$:
\begin{equation}
    \label{eq:nonlinstrain}
    F_p=k_1 w_c + k_3 w_c^3 = 4 n_0 w_c + \frac{8 Y h w_c^3}{3 R^2 (1- \nu)} .
\end{equation}
Thus, the system is found to be in the nonlinear regime for center amplitudes that exceed $w_{c,nl}$ as found from this equation:
\begin{equation}
    w_{c,nl} = \sqrt{0.1 \frac{k_1}{k_3}}=  \sqrt{0.1 \frac{3 n_0 R^2 (1-\nu)}{2 Y h}}.
\end{equation}
If the amplitude of the membrane becomes even larger, it can rupture. For this yield strain we take a conservative value of breaking strain~\cite{Goldsche2018}: $\epsilon_b$=1.2 \%. By rewriting Eq.~(\ref{eq:nonlinstrain}) in terms of an effective deflection induced tension, $n_{\rm eff}=\frac{2 Y h w_c^2}{3 R^2 (1- \nu)}$ such that $F_p= 4 (n_0 + n_{\rm eff}) w_c$, and using $n_{\rm eff,b}=Y h \epsilon_b/(1-\nu)$, we obtain as rough estimate for the breaking strain amplitude $w_{c,b}$:
\begin{equation}
    w_{c,b}=\sqrt{\frac{3}{2} \epsilon_b R^2}.
\end{equation}
Using these equations and the above-mentioned constants we find $ \langle z^2_{\rm th}\rangle^{1/2}$ = 0.17~nm for all drum diameters. For the drum radii of 50 nm, 2.5 $\mu$m and 100 $\mu$m, we find $w_{c,nl}$=0.17~nm, 8.4~nm and 340 nm and $w_{c,b}$=7.0 nm, 349 nm and 14~$\mu$m respectively as indicated in Fig.~\ref{fig:introduction}.

\section{Appendix B: Resonant motion of circular membranes and plates}
\label{AppendixB}

In this appendix we review the linear theory governing the undamped resonances of a circular drum resonator, deriving the modal masses $m_i$, stiffnesses $k_i$ and resonance frequencies $\omega_i$, in membrane, plate and intermediate regime.
\subsection{Membrane dynamics}
\label{sec:membranedynamics}
 If the membrane is very thin, the pre-tension dominates the restoring force and we will assume this pre-tension is uniform over the membrane surface. For the out-of-plane deflection $w$ of the membrane in cylindrical coordinates $r$, $\theta$ we can write the following equation of motion:
\begin{equation}
\frac{\partial^2 w}{\partial r^2} + \frac{1}{r^2}\frac{\partial^2 w}{\partial \theta^2} + \frac{1}{r} \frac{\partial w}{\partial r} = \frac{\rho h}{n_0} \frac{\partial^2 w}{\partial t^2}, 
\end{equation}

\begin{figure}[ht]
    \centering
    \includegraphics{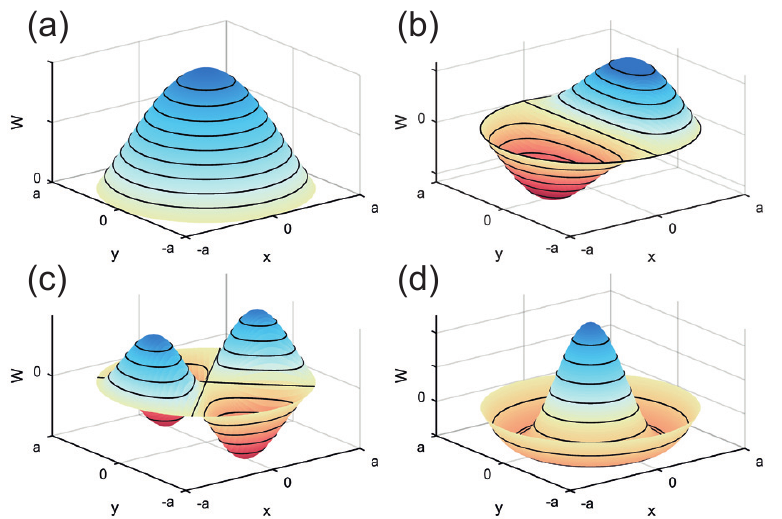}
    \caption{The four lowest mode shapes of a circular membrane. Reprinted from Ref.~\cite{dolleman2018phdthesis}.}
    \label{fig:modeshape}
\end{figure}

The solution to this equation gives the resonance frequency and eigenmodes: 
\begin{equation}\label{eq:resfreq}
\omega_{\rm mem,\alpha n} = \frac{\gamma_{\alpha n}}{R} \sqrt{\frac{n_0}{\rho h}},
\end{equation}
\begin{equation}
W^{(1)}_{\alpha n} (r,\theta) = R(r) T(\theta) = J_\alpha (\gamma_{\alpha n} r/R) \cos{\alpha  \theta},
\end{equation}
\begin{equation}
W^{(2)}_{\alpha n} (r,\theta) = J_\alpha  (\gamma_{\alpha n} r/R) \sin{m \theta},
\end{equation}
where $\alpha = 0,1,2,...$ and $n = 1, 2, 3,...$. The constants $\gamma_{\alpha n}$ are found by solving the frequency equation: 
  \begin{equation}
 J_\alpha (\gamma_{\alpha n}) = 0,
 \end{equation}
which for the first four modes gives: $\gamma_{01} = 2.405$, $\gamma_{11} = 3.832$, $\gamma_{21} = 5.135$ and $\gamma_{02} = 5.520$. The first four modes of a circular membrane are shown in Fig.~\ref{fig:modeshape}.

An important property of vibrational modes is that their shape is fixed, therefore if one knows the deflection of a certain point of the membrane at a resonance, one automatically knows the deflection of all the other parts of the membrane. As discussed in Sec. \ref{sec:eqom}, a mode is therefore a single degree of freedom that can describe the motion by a single generalized coordinate $q_i$. It is this property that allows one to formulate the vibration at a resonance as an equation of motion with a single degree of freedom:
\begin{equation}
 m_{i} \ddot{q}_i + k_{i} q_i = F_{i},
\end{equation}
where $m_{i}$ is the effective (or modal) mass, $k_{i}$ is the modal stiffness and $F_{\mathrm{eff}}$ the effective force, where the modal mass, stiffness and force depend on the vibrational mode under consideration and the choice of generalized coordinate. We choose as generalized coordinate the maximum deflection of the membrane. In that case, for the fundamental mode (Fig.~\ref{fig:modeshape}a) the modal mass of a circular membrane compared to its real total mass $m_{\mathrm{tot}}$ is:

\begin{equation}
\label{meff10}
    \frac{ m_{(0,1)}}{m_{\mathrm{tot}}} = 0.2695,
\end{equation}
while for the second mode (1,1) in Fig.~\ref{fig:modeshape}(b) this ratio is:
\begin{equation}
    \frac{ m_{(1,1)} }{m_{\mathrm{tot}}} = 0.2369.
\end{equation}
An overview of modal masses and how to calculate them for various geometries can be found in Ref.~\cite{Hauer2013}. Once the modal mass is determined, the modal stiffness follows directly from Eq.~(\ref{eq:resfreq}):
\begin{equation}\label{eq:keff}
k_{\alpha n}= \frac{\gamma_{\alpha n}^2}{a^2} {\frac{n_0}{\rho h}} m_{\alpha n}.
\end{equation}

\subsection{Plate dynamics}
\label{sec:platedynamics}
If the membrane becomes thicker or if the tension is small, tension effects can be neglected and the restoring force in the membrane is generated by the internal bending moments in the material. This is called the plate limit. Taking into account this bending rigidity, the equation that governs the dynamics of the circular plate becomes:
\begin{equation}\label{eq:platefull}
    \rho h \frac{\partial^2 w(r,\theta,t)}{\partial t^2} + D \nabla_r^2 \nabla_r^2 w(r,\theta,t)  = 0,
\end{equation}
where $\nabla_r^2 = \frac{\partial^2}{\partial r^2} + \frac{1}{r} \frac{\partial}{\partial r} + \frac{1}{r^2} \frac{\partial^2}{\partial \theta^2}$ and $D$ is the bending rigidity:
\begin{equation}\label{eq:bending}
    D = \frac{Eh^3}{12(1-\nu^2)}.
\end{equation}
Equation (\ref{eq:bending}) is valid for a continuous uniform material, however for a 2D material that assumption breaks down in the direction perpendicular tot the plane. Nevertheless, single-layer 2D materials also have a bending rigidity~\cite{Lindahl2012}, but Eq.~(\ref{eq:bending}) is not valid anymore, because the bending rigidity is dominated by the electron orbitals that resist bending, such that $D$ needs to be calculated by other means~\cite{Banafsheh2018}. Theory even suggests that for single-layer materials the bending rigidity can be temperature dependent~\cite{Roldan2011}.
Equation~(\ref{eq:platefull}) can also be solved by separation of variables and this leads to the following eigenmodes and resonance frequencies:
\begin{equation}
\begin{split}
   W^{(1)}_{\alpha n} (r,\theta) = (J_{\alpha} (\lambda_{\alpha n} r) I_{\alpha}(\lambda_{\alpha n} R) - \\ J_{\alpha} (\lambda_{\alpha n} R) I_{\alpha}(\lambda_{\alpha n} r)) \cos(\alpha \theta),
   \end{split}
\end{equation}
\begin{equation}
\begin{split}
   W^{(2)}_{\alpha n} (r,\theta) = (J_{\alpha} (\lambda_{\alpha n} r) I_{\alpha}(\lambda_{\alpha n} R) - \\ J_{\alpha} (\lambda_{\alpha n} R) I_{\alpha}(\lambda_{\alpha n} r)) \sin(\alpha \theta),
   \end{split}
\end{equation}
\begin{equation}
\label{eq:freqplate}
    \omega_{\rm plate,\alpha n} = \lambda_{\alpha n} \sqrt{D}{\rho h},
\end{equation}
where $\lambda_{\alpha n}$ is the solution to the frequency equation:
\begin{equation}
    I_{\alpha} (\lambda_{\alpha n} R)  J_{\alpha-1} (\lambda_{\alpha n} R) - J_{\alpha} (\lambda_{\alpha n} R)  I_{\alpha-1} (\lambda_{\alpha n} R) = 0.
\end{equation}
The first four roots of this equation are: $\lambda_{01} R = 3.196$, $\lambda_{11} R = 4.611$, $\lambda_{21} R = 5.906$ and $\lambda_{02} R = 6.306$.

\subsection{Plates under tension}
\label{sec:plateundertension}
When increasing the thickness of suspended 2D materials, their dynamics can be described by the membrane equation of motion (Sec. \ref{sec:membranedynamics}) in the single-layer limit and transforms~\cite{Castellanos2013} to plate dynamics (Sec. \ref{sec:platedynamics}) for multi-layer membranes where bending rigidity quickly becomes dominant, since the bending rigidity $D \propto h^3$. In the intermediate regime  both the bending rigidity and the tension contribute to the restoring force, resulting in the equation of motion: 
\begin{equation}
D \nabla_r^2 \nabla_r^2 w(r,\theta,t) + n_0 \nabla_r^2 w(r,\theta,t) = {\rho h} \frac{\partial^2 w}{\partial t^2}. 
\end{equation}
The resonance frequency is:
\begin{equation}\label{eq:freqplatemem}
   \omega_{\rm tot,\alpha n} = \sqrt{\zeta_{\alpha n}^2\frac{n_0}{\rho h} + \zeta_{\alpha n}^4 \frac{D}{\rho h}}. 
\end{equation}
The procedure to determine $\zeta_{\alpha n}$, the resonance frequencies and mode shapes can be found in Refs.~\cite{DollemanMScthesis, Wah1962}. $\zeta_{\alpha n}$ only depends on the dimensionless number $R^2 \frac{n_0}{D}$. This also gives a useful test to determine whether the lower modes of a circular membrane can be described using membrane or plate theory. 
 
Since the evaluation of the resonance frequency is quite complicated in this case, often~\cite{Castellanos2013} the approximation $\omega_{\rm tot,\alpha n}^2 \approx \omega_{\mathrm{mem,\alpha n}}^2 + \omega_{\mathrm{plate,\alpha n}}^2$ is used.

\section{Appendix C: Derivation of the nonlinear reduced order model of a circular drum}
\label{AppendixC}
After having considered the linear dynamics in the previous appendix, we show here how a nonlinear reduced order model for the fundamental mode of a circular membrane can be derived, leading to the Duffing equation. The resulting analytic expression for the nonlinear Duffing term was used in Ref.\cite{Farbod2017} to extract the Young's modulus of graphene from the experimental nonlinear frequency response curves. The demonstrated procedure can be extended to include other modes and evaluate mode-couplings as well, and variants of it have been implemented with FEM techniques\cite{Muravyov2003}. 

The Lagrange equations of a tensioned circular 2D membrane are given by
\begin{equation}\label{eq:lagrange}
\frac {d}{dt} (\frac{\partial T_k}{\partial \dot{q_{i}}})-\frac{\partial T}{\partial q_{i}}+\frac{\partial U}{\partial q_{i}}=\frac{\partial W}{\partial q_{i}}   , \quad  i=1,2...,\bar{N},
\end{equation}
in which $T_k$ is the kinetic energy, $U$ is the potential energy, $W$ is the virtual work done by external forces, and $q_i$ are the generalized coordinates that are used to define the motion of the drum.  
The potential energy of a circular drum for an isotropic material is
\begin{equation}\label{eq:potential}
\begin{split}
U=\int_{0}^{2 \pi}\int_{0}^{R} \frac{ Eh}{2(1-\nu^2)} \Big(\epsilon_{rr}^2+\epsilon_{\theta \theta}^2\\
+2\nu \epsilon_{rr} \epsilon_{\theta \theta}+\frac{1-\nu}{2}\gamma_{r \theta}^2\Big) r dr d\theta ,
\end{split}
\end{equation}
where $E$ is the Young's modulus, $\nu$ is the Poisson's ratio, $h$ is the thickness and $R$ is the radius of the drum. Moreover, $\epsilon_{rr}$, $\epsilon_{\theta \theta}$ are the normal strains and $\gamma_{r \theta}$ is the shear strain that can be written in terms of the drum's deflection as follows~\cite{amabili2008nonlinear}:
\begin{subequations}\label{eq:vk1}
\begin{equation}
\epsilon_{rr}=\frac{\partial u}{\partial r}+\frac{1}{2}\Big(\frac{\partial w}{\partial r}\Big)^2,
\end{equation}
\begin{equation}
\epsilon_{\theta \theta}=\frac{\partial v}{r \partial \theta}+\frac{u}{r}+\frac{1}{2}\Big(\frac{\partial w}{r \partial \theta}\Big)^2,
\end{equation}
\begin{equation}
\gamma_{r \theta}=\frac{\partial v} {\partial r}-\frac{v}{r}+\frac{\partial u}{r \partial \theta}+\Big(\frac{\partial w}{\partial r}\Big)\Big(\frac{\partial w}{r \partial \theta}\Big).
\end{equation}
\end{subequations}
In Eq.~(\ref{eq:vk1}) $u$, $v$, and $w$ are the radial, tangential and transverse membrane deflections. The quadratic terms in Eq.~(\ref{eq:vk1}) are commonly known as von K\'arm\'an nonlinear terms and shall be included for studying the nonlinear dynamic response of the drum. 
 
 The kinetic energy of the drum can be calculated by neglecting in-plane inertia ($\dot u=\dot v=0$), because the in-plane resonance frequencies are much higher than the considered out-of-plane dynamics and is found to be:
 \begin{equation} \label{eq:kinetic}
T_k=\frac{1}{2}\rho h \int_{0}^{2 \pi}\int_{0}^{R} \dot{w} r dr d\theta,
\end{equation}
where the overdot indicates differentiation with respect to time, $t$.  For a membrane subjected to transverse distributed pressure consisting of a constant part $p_0$ and a harmonic component $F \cos(\omega t-\phi)$, the virtual work done can be calculated as:
 \begin {equation}\label{eq:work}
W=\int_{0}^{2 \pi}\int_{0}^{R} \Big(\frac{F \cos(\omega t-\phi)}{\pi R^2}+p_0 \Big) w r dr d\theta.
\end{equation}
To obtain the governing equations of motion, one would first need to discretize Eqs.~(\ref{eq:potential}),(\ref{eq:kinetic}), and (\ref{eq:work}) using a set of admissible functions that satisfy the boundary conditions. 2D material membranes stamped on top of cavities have fixed boundary conditions due to the strong adhesion force that exists between the substrate and the membrane. For this type of boundary condition $u=v=w=0$ at $r=R$. 
Moreover, for axi-symmetric deflection/oscillations, strain-displacement relations (Eq.~(\ref{eq:vk1})) can be further simplified since for this type of motion $v=0$, and $\partial u/\partial \theta =\partial v/\partial \theta=\partial w/\partial \theta=0$. Thus, for a drum driven into its first resonance mode, one can use the following set of functions for discretization:
\begin{subequations}\label{eq:discrete}
\begin{equation}
w=q_{1}(t) J_{0} \Big(\gamma_{01}\frac{r}{R}\Big) , 
\end{equation}
\begin{equation}
 u= {\frac{n_{0} (1-\nu)}{E h} r}+ r (R-r) \sum_{k=2}^{\bar{N}} q_{k}(t) r^{k-2}.
\end{equation}
\end{subequations}
Eventually inserting the discretized $U$, $T$ and $W$ in Lagrange equations leads to a system of nonlinear equations comprising a single differential equation associated with the out-of-plane generalized coordinate $q_1$ and $\bar{N}-1$ algebraic equations in terms of the in-plane generalized coordinates $q_2,...,q_{\bar{N}}$. By solving the $\bar{N}-1$ algebraic equations in terms of $q_1$, and adding a damping term, it is then possible to reduce the number of equations to the following $2^{nd}$ order, Duffing type (Eq.~(\ref{eq:Duffing})) differential equation describing the out-of-plane dynamics of the fundamental mode:
\begin{equation}\label{eq:Duffing2}
    m_1 \ddot{q_1} + c_1 \dot{q_1} + k_1 q_1 +\gamma q_1^3= F_{\rm ext,1} \cos(\omega t- \phi)+\tilde{F}_1.
\end{equation}

According to the described procedure, the parameters in this equation are obtained as:
 \begin{eqnarray} \label{eq:coeffs}
     m_1&=&0.2695 \rho h \pi R^2, \\
     k_1&=&4.8967 n_0, \\ 
     \gamma&=&\frac{\pi E h}{(1.27 -0.97\nu-0.27 \nu^2) R^2 }, \\
        F_{\rm ext,1}&=&0.432 F, \\
            \tilde{F}_1&=&1.3567 p_0 R^2.
 \end{eqnarray}
Here, $\gamma$ is the cubic spring constant or Duffing constant. In section~\ref{sec:nonlsoln} the role of this constant on the nonlinear phenomena that are commonly observed in 2D material membranes is highlighted.

\section{Appendix D: Single relaxation-time models}
\label{AppendixD}

In this appendix the single relaxation-time models presented in section \ref{sec:physint} and Fig. \ref{fig:physicalinteractions} are discussed in more detail.

Many physical phenomena resemble the physics of an electrical $RC$ circuit, consisting of a capacitor in parallel to a resistor. When a current $I_{\rm in}(t)=I_{\rm in}(\omega)e^{i\omega t}$ is run through this circuit, the voltage $V_{\rm out}$ across the capacitor is governed by the following differential equation:

\begin{equation}
    I_{\rm in}(t) = \frac{V_{\rm out}}{R} + C \frac{{\rm d}V_{\rm out}}{{\rm d} t} = \frac{V_{\rm out}(\omega)}{R} (1 + i\omega RC) e^{i \omega t},
\end{equation}
which can be rewritten as:
\begin{equation}
    V_{\rm out}(\omega)=\beta h_\omega I_{\rm in}(\omega)=\beta I_{\rm in}(\omega) \frac{1}{1 + i\omega \tau},
\end{equation}
where the transfer~\cite{metzger2008optical} function $h_\omega=\frac{1}{1+i\omega \tau}$ captures the time delay $\tau = RC$ between current and voltage, indicative of the time taken to charge the capacitor and $\beta=R$ is a characteristic transduction constant. Since this relaxation time is fully described by the time constant $\tau$, it is called a single relaxation time model.

In the dynamics of 2D material membranes we encounter 3 types of systems that can be described as single relaxation time models. The first is an optothermally actuated system with a heat capacitance $C_{th}$ and thermal resistance $R_{th}$, that is quite analogous to the electrical system we just discussed. In this case we find a thermal expansion force of the form:
\begin{equation}
\label{eq:FextSRT}
    F_{\rm ext,th}=\alpha_{\rm eff} T_{\rm out}(\omega)=\beta \mathcal{P}_{\rm in}(\omega) \frac{1}{1 + i\omega \tau_{th}},
\end{equation}
where $\mathcal{P}(t)$ is the modulated optical power absorbed by the membrane, $T_{\rm out}(\omega)$ is the membrane temperature and $\tau_{th}=R_{th} C_{th}$ is the thermal time-constant.

When the force is position dependent, the real and imaginary parts of $F_{\rm ext}$ in Eq.~\ref{eq:FextSRT} can alter the effective stiffness and damping coefficient as discussed in section \ref{sec:feedback}. These changes in stiffness and damping can also be used to determine the value of $\tau$, like in Eq.~(\ref{eq:backaction3}).

The circuit of an electrostatically actuated membrane consists of a capacitor (the membrane) in series with a resistor (the interconnect). The circuit acts as a voltage divider, where only the part of the input voltage $V_{in}$ across the capacitor contributes to the electrostatic force. For such a voltage divider it is straightforward to show that we get a single relaxation time model of the form:
\begin{equation}
    F_{\rm ext,el}=\beta_1 V_{\rm out}(\omega)=\beta_2 {V}_{\rm in}(\omega) \frac{1}{1 + i\omega \tau_{el}},
\end{equation}
Where $V_{\rm in}$ is the applied {\it ac} voltage, and the constants $\beta$ depend on the voltage, electrical component values and geometry. $\tau_{el}=R C$ is the electrical time-constant, which is the product of the electrical resistance and capacitance of the electrostatically actuated membrane.

The third type of relaxation time model is used to capture the squeeze-film effect, where the pressure difference across the drum reduces both due to gas flow out of the enclosing cavity with time-constant $\tau_{\rm gas}$ at a rate proportional to the pressure difference $\Delta p$, and due to membrane motion with a speed $\dot{q}_1$, as indicated by the following differential equation~\cite{roslon2020graphene}:
\begin{equation}
       \frac{{\rm d}\Delta p}{{\rm d}t} = -\frac{1}{\tau_\text{gas}} \Delta p + \gamma  \frac{{\rm d}q_1}{{\rm d}t},
 \end{equation}
 where $\gamma$ is a proportionality constant.
For a sinusoidal pressure variation $\Delta p(t)=\Delta p(\omega) e^{i\omega t}$, this becomes:
 \begin{equation}
 \label{eq:deltaP}
        \Delta p(\omega)  = \frac{i\omega \tau_{\rm gas}}{1  + i \omega \tau_\text{gas}} \gamma  q_1(\omega),
 \end{equation}
where we again obtain a single relaxation time model of a slightly different form between the pressure and membrane position. The pressure across the membrane drives the dynamics of the drum, and can be coupled to the mass spring system, ignoring the intrinsic damping of the resonator:
 \begin{equation}
    \label{eq:eomdeltaP}
      -m_1 \omega^2  q_1 + k_1 q_1 =  \beta \Delta p.
 \end{equation}
If we substitute Eq.~(\ref{eq:deltaP}) into this equation, and rewrite $\beta$ and $\gamma$ using the well known solution in the high-frequency limit ($\omega \tau_\text{gas}$) and using $\omega^2_{1}=k_1/m_1$, we obtain:
\begin{equation}
    \omega^2 = \omega_1^2  + \frac{p}{g_0 \rho h}  \frac{\omega^2 \tau_\text{gas}^2}{\omega^2 \tau_\text{gas}^2 +1}+   i\frac{p}{g_0 \rho h}  \frac{\omega \tau_\text{gas}}{\omega^2 \tau_\text{gas}^2 +1} .
\end{equation}
 For high frequencies, $\omega \tau_\text{gas} > 1$ the resonance frequency becomes Eq.~\ref{eq:sqz_freqshift}. When $\omega \tau_\text{gas} < 1$, the membrane is not oscillating fast enough to compress the gas, and the stiffness reduces. The imaginary part describes the damping, and this is maximum when $\omega \tau_\text{gas} = 1$. 

\bibliography{References}

\end{document}